\documentclass[review,12pt]{elsarticle}

\usepackage{psfrag}
\usepackage{color}
\usepackage{graphicx}
\usepackage{tikz,graphics,color,fullpage,float,epsf,caption,subcaption}
\usepackage[utf8]{inputenc}
\usepackage{comment}
\usepackage{multirow}
\usepackage{adjustbox}

\usepackage{titlesec}
\titleformat{\paragraph}
{\normalfont\normalsize\itshape}{\theparagraph}{1em}{}
\titlespacing*{\paragraph}
{0pt}{3.25ex plus 1ex minus .2ex}{1.5ex plus .2ex}

\usepackage{amssymb}
\usepackage{amsmath}
\usepackage{lineno}
\usepackage{hyperref}
\usepackage{nomencl}
\usepackage{siunitx}
\usepackage{soul}
\usepackage{xcolor}
\usepackage{csquotes}
\usepackage{tabularx}

\usepackage{listings}  % For code inclusion

\makenomenclature

\hypersetup{
    colorlinks=true,
    urlcolor=blue,
}

\usepackage{etoolbox}
\renewcommand\nomgroup[1]{%
  \item[\bfseries
  \ifstrequal{#1}{A}{Parameters}{%
  \ifstrequal{#1}{G}{Greek Symbols}{%
  \ifstrequal{#1}{S}{Subscript}{%
  \ifstrequal{#1}{T}{Superscript}{%
  \ifstrequal{#1}{D}{Dimensionless Numbers}{}}}}}%
]}
\journal{Computer Physics Communications}

\begin{document}

\definecolor{Gr3}{RGB}{0,179,0}   %% Green
\definecolor{Bl1}{RGB}{0,128,255}  %% Blue
\definecolor{Oran}{RGB}{255,128,0}  %% Orange
\definecolor{Gr2}{RGB}{0,153,0}   %% Green2

\newcommand{\greenline}{\raisebox{2pt}{\tikz{\draw[-,Gr3,solid,line width = 1.pt](0,0) -- (5mm,0);}}}                   % Solid Green Line
\newcommand{\blackline}{\raisebox{2pt}{\tikz{\draw[-,black,solid,line width = 1.pt](0,0) -- (5mm,0);}}}                  % Solid Black Line
\newcommand{\blacklinethinner}{\raisebox{2pt}{\tikz{\draw[-,black,solid,line width = 0.5pt](0,0) -- (5mm,0);}}}     % Solid Black Line 2
\newcommand{\blueline}{\raisebox{2pt}{\tikz{\draw[-,blue,solid,line width = 1.pt](0,0) -- (5mm,0);}}}                     % Solid Blue Line
\newcommand{\bluelinedif}{\raisebox{2pt}{\tikz{\draw[-,Bl1,solid,line width = 1.pt](0,0) -- (5mm,0);}}}                     % Solid Blue Line 2
\newcommand{\redline}{\raisebox{2pt}{\tikz{\draw[-,red,solid,line width = 1.pt](0,0) -- (5mm,0);}}}                        % Solid Red Line
\newcommand{\orangeline}{\raisebox{2pt}{\tikz{\draw[-,Oran,solid,line width = 1.pt](0,0) -- (5mm,0);}}}                        % Solid Orange Line
\newcommand{\greenlinedif}{\raisebox{2pt}{\tikz{\draw[-,Gr2,solid,line width = 1.pt](0,0) -- (5mm,0);}}}                        % Solid Green Line 2

\newcommand{\dashblackline}{\raisebox{2pt}{\tikz{\draw[-,black,dashed,line width = 1.pt](0,0) -- (5mm,0);}}}           % Dashed Black Line
\newcommand{\dashblueline}{\raisebox{2pt}{\tikz{\draw[-,blue,dashed,line width = 0.5pt](0,0) -- (5mm,0);}}}            % Dashed Blue Line
\newcommand{\dashbluelinethicker}{\raisebox{2pt}{\tikz{\draw[-,blue,dashed,line width = 1.pt](0,0) -- (5mm,0);}}}    % Dashed Blue Line 2
\newcommand{\dashgreenline}{\raisebox{2pt}{\tikz{\draw[-,Gr3,dashed,line width = 1pt](0,0) -- (5mm,0);}}}              % Dashed Green Line
\newcommand{\dashredline}{\raisebox{2pt}{\tikz{\draw[-,red,dashed,line width = 1pt](0,0) -- (5mm,0);}}}                % Dashed Blue Line
\newcommand{\dashorangeline}{\raisebox{2pt}{\tikz{\draw[-,Oran,dashed,line width = 1.pt](0,0) -- (5mm,0);}}}                        % Dashed Orange Line

\newcommand{\dotblackline}{\raisebox{2pt}{\tikz{\draw[-,black,dotted,line width = 1.pt](0,0) -- (5mm,0);}}}                     % Dotted Red Line
\newcommand{\dotredline}{\raisebox{2pt}{\tikz{\draw[-,red,dotted,line width = 1.pt](0,0) -- (5mm,0);}}}                     % Dotted Red Line
\newcommand{\dotgreenline}{\raisebox{2pt}{\tikz{\draw[-,Gr3,dotted,line width = 1.pt](0,0) -- (5mm,0);}}}             % Dotted Green Line
\newcommand{\dotblueline}{\raisebox{2pt}{\tikz{\draw[-,blue,dotted,line width = 1.pt](0,0) -- (5mm,0);}}}                     % Dotted Blue Line
\newcommand{\dotgreenlinedif}{\raisebox{2pt}{\tikz{\draw[-,Gr2,dotted,line width = 1.pt](0,0) -- (5mm,0);}}}                     % Dotted Green Line 2

\newcommand{\dashdotblackline}{\raisebox{2pt}{\tikz{\draw[thick,black,dash pattern={on 4.5pt off 2pt on 1pt off 2pt}] (0,0) -- (5mm,0);}}} % Dashed-Dotted Black Line
\newcommand{\dashdotblueline}{\raisebox{2pt}{\tikz{\draw[thick,blue,dash pattern={on 4.5pt off 2pt on 1pt off 2pt}] (0,0) -- (5mm,0);}}} % Dashed-Dotted Blue Line
\newcommand{\dashdotgreenline}{\raisebox{2pt}{\tikz{\draw[thick,Gr3,dash pattern={on 4.5pt off 2pt on 1pt off 2pt}] (0,0) -- (5mm,0);}}} % Dashed-Dotted Green Line

\newcommand{\virvcircle}{\raisebox{0pt}{\tikz{\draw[Oran,solid,line width = 1.pt](1.0mm,0) circle (0.4mm);}}}  % Orange circle
\newcommand{\bluecircle}{\raisebox{0pt}{\tikz{\draw[Bl1,solid,line width = 0.6pt](1.0mm,0) circle (0.4mm);}}}  % Blue circle
\newcommand{\orangerectangle}{\raisebox{0pt}{\tikz{\draw[Oran,solid,line width = 0.6pt](1.4mm,0) rectangle (0.7mm,0.7mm);}}}  % Orange rectangle
\newcommand{\greencircle}{\raisebox{0pt}{\tikz{\draw[Gr2,solid,line width = 1.pt](1.0mm,0) circle (0.4mm);}}}  % Green circle

\begin{psfrags}

\begin{frontmatter}



%\title{COLBM: A Multi-Component Multi-Phase Open Source Simulation Code with Reactive Interface using Lattice Boltzmann Method}

\title{CooLBM: A Collaborative Open-Source Reactive Multi-Phase/Component Simulation Code via Lattice Boltzmann Method}


\author[a]{R. Alamian}
\author[a]{A. K. Nayak}
%\author[a]{C. Stockinger}
%\author[a]{A. Hadjadj}
%\author[b]{J. Latt}
\author[a,b]{M. S. Shadloo  \corref{cor2}}

\address[a]{INSA Rouen Normandie, Univ. Rouen Normandie, CNRS, Normandie University, CORIA UMR 6614, F-76000 Rouen, France}
\address[b]{Institut Universitaire de France, Rue Descartes, F-75231 Paris, France}

\cortext[cor2]{corresponding author: msshadloo@coria.fr}
% \affiliation{organization={},%Department and Organization

\begin{abstract}
The current work presents a novel {\em COllaborative Open-source Lattice Boltzmann Method} framework, so-called {\em CooLBM}. The computational framework is developed for the simulation of single and multi-component multi-phase problems, along with a reactive interface and conjugate fluid-solid heat transfer problems. CooLBM utilizes a multi-CPU/GPU architecture to achieve high-performance computing (HPC), enabling efficient and parallelized simulations for large scale problems. The code is implemented in C++ and makes extensive use of the Standard Template Library (STL) to improve code modularity, flexibility, and re-usability. The developed framework incorporates advanced numerical methods and algorithms to accurately capture complex fluid dynamics and phase interactions. It offers a wide range of capabilities, including phase separation, interfacial tension, and mass transfer phenomena. The reactive interface simulation module enables the study of chemical reactions occurring at the fluid-fluid interface, expanding its applicability to reactive multi-phase systems. The performance and accuracy of CooLBM are demonstrated through various benchmark simulations, showcasing its ability to capture intricate fluid behaviors and interface dynamics. The modular structure of the code allows for easy customization and extension, facilitating the implementation of additional models and boundary conditions. Finally, CooLBM provides visualization tools for the analysis and interpretation of simulation results. Overall, CooLBM offers an efficient computational framework for studying complex multi-phase systems and reactive interfaces, making it a valuable tool for researchers and engineers in several fields including, but not limited to chemical engineering, materials science, and environmental engineering. CooLBM will be available under open source initiatives for scientific communities.
\end{abstract}

%[The graphical abstract shows a porous medium, with fluid flowing through it. There is a surface chemical reaction taking place on the solid surface within the porous medium. The thermal counter-slip method is used to model the conjugate heat transfer, while the wet node scheme is used to model the surface chemical reactions and species transfer. The impact of physical parameters, including Péclet, Damköhler, and Prandtl numbers, on the coupled processes is investigated. Different combustion regimes are identified, and the role of chemical reaction rate, as well as diffusion and convection transport processes, are elucidated in each regime.]

%\tableofcontents
%\newpage

\begin{highlights}

\item CooLBM: Efficient lattice Boltzmann framework for multi-phase and reactive interface simulations.
\item Validation: Rigorously tested and validated against analytical solutions and numerical models.
\item Versatility: Handles complex fluid dynamics, phase separation, and mass transfer phenomena.
\item Parallelization: Utilizes multi-CPU/GPU architecture for high-performance computing.
\item NVIDIA V100 GPU: Outperforms other processors, offering impressive MLUPS performance.

\end{highlights}

\begin{keyword}
CooLBM \sep lattice Boltzmann method (LBM) \sep reactive interface \sep conjugate heat transfer \sep multi-component \sep multi-phase flows

\end{keyword}

\end{frontmatter}

%\linenumbers


%%%%%%%%%%%%%%%%%%%%%%%%%%%%%%%%%%%%%%%%%%%%%%%%%%%%%%%%%%%%%%%%%%%%%%%%%%%%%%%
%%%%%%%%%%%%%%%%%%%%%%%%%%%% Nomenclature %%%%%%%%%%%%%%%%%%%%%%%%%%%%%%%%%%%%%
%%%%%%%%%%%%%%%%%%%%%%%%%%%%%%%%%%%%%%%%%%%%%%%%%%%%%%%%%%%%%%%%%%%%%%%%%%%%%%%

\nomenclature[A]{$\Delta$}{Laplacian operator}
\nomenclature[A]{$\nabla$}{Divergence operator}
\nomenclature[A]{$A$}{Pre-exponential coefficient in the Arrhenius relation for chemical reaction}
\nomenclature[A]{$C_\beta$}{Species concentration, $\beta$ = O$_2$, CO$_2$, and particle (P)}
\nomenclature[A]{$C_{b,O_2}$}{Oxygen concentration at the reactive interface boundary}
\nomenclature[A]{$\mathbf{c}_i$}{Lattice velocity in the i-th direction}
\nomenclature[A]{$\mathbf{c}_i \Delta t$}{Lattice spacing vector connecting two neighboring lattice nodes}
\nomenclature[A]{$c_p$}{Specific heat at constant pressure}
\nomenclature[A]{$c_s$}{sound speed in lattice units}
\nomenclature[A]{$C_{p,s}$}{particle mass fraction 
at lattice nodes near the substrate}
\nomenclature[A]{$D_\beta$}{Species diffusivity $\beta$ = O$_2$, CO$_2$, and particle (P)}
\nomenclature[A]{$E$}{Activation energy in the chemical reaction}
\nomenclature[A]{$f$}{Flow field distribution function}
\nomenclature[A]{$g$}{Thermal field distribution function}
\nomenclature[A]{$\mathbf{g}$}{Gravitational acceleration}
\nomenclature[A]{$\dot{H}$}{Transient heat flux generated by the chemical reaction}
\nomenclature[A]{$h$}{Species field distribution function}
\nomenclature[A]{$h_r$}{Standard enthalpy of a chemical reaction}
%\nomenclature[A]{$J$}{Lattice weight for species field}
\nomenclature[A]{$K$}{Thermal conductivity}
%\nomenclature[A]{$k$}{Chemical reaction rate}
\nomenclature[A]{$\mathbf{n}$}{Normal vector to the reactive interface}
\nomenclature[A]{$R_u$}{Ideal gas constant}
\nomenclature[A]{$T$}{temperature }
\nomenclature[A]{$\mathbf{u}$}{Fluid velocity}
\nomenclature[A]{$u_\text{max}$}{Maximum fluid velocity}
\nomenclature[A]{$U_\text{wall}$}{Wall velocity in Couette flow}
\nomenclature[A]{$N_x, N_y, N_z$}{Number of lattice nodes in the x-, y-, and z-direction, respectively}
\nomenclature[A]{$P$}{Pressure}
\nomenclature[A]{$k$}{Wave number}
\nomenclature[A]{$L \times H$}{Length and width of the domain}
\nomenclature[A]{$DP$}{Particle deposition probability}
\nomenclature[A]{$R$}{Radius of the bubble}
\nomenclature[A]{$\mathbf{F}_{\text{ads}}$}{Interaction Shen-Chen force between fluid components and solid}
\nomenclature[A]{$\mathbf{F}^{\text{SC}}$}{Interparticle Shen-Chen force}
\nomenclature[A]{$G$ or $G_c$}{Molecular interaction or cohesion force}
\nomenclature[A]{$s$}{Indicator function}
\nomenclature[A]{$\mathbf{x}$, $\mathbf{\tilde{x}}$}{Position vectors}
\nomenclature[A]{$t$}{Time}
\nomenclature[A]{$\mathbf{u}'$}{Common velocity vector}
\nomenclature[A]{$\mathbf{u}^b$}{Barycentric velocity vector}
\nomenclature[A]{${F}_i$}{Volumetric forcing term}
\nomenclature[A]{$G_{\sigma \tilde{\sigma}}$}{Interaction force between two fluid components $\sigma$ and $\tilde{\sigma}$}
\nomenclature[A]{$G_{\text{ads}}$}{Interaction coefficient between the fluid and solid}

\nomenclature[G]{\theta}{Contact angle}
\nomenclature[G]{\psi}{Pseudo-potential function}
\nomenclature[G]{\delta}{Dirac delta function}
\nomenclature[G]{\Omega}{Collision operator}
\nomenclature[G]{\Omega_f}{Computational fluid domain}
\nomenclature[G]{\partial \Omega}{Boundary of a solid geometry}
\nomenclature[G]{\tau}{Flow field relaxation time to equilibrium }
%\nomenclature[G]{\tau_e}{Relaxation time to equilibrium for solid and fluid thermal, particle, and species}
\nomenclature[G]{\nu}{Kinematic viscosity}
\nomenclature[G]{\mu}{Dynamic viscosity}
\nomenclature[G]{\varepsilon}{Internal energy}
%\nomenclature[G]{$\Acute{\varepsilon}$}{Counter-slip internal energy}
%\nomenclature[G]{\xi}{Characteristic length}
\nomenclature[G]{\omega}{Lattice weights}
\nomenclature[G]{\delta t}{Physical value of one lattice time step: time conversion factor}
\nomenclature[G]{\delta x}{Physical value of one lattice spacing: length conversion factor}
\nomenclature[G]{\Delta t}{Time step in LBM is the time that distributions need to propagate to the neighboring node}
\nomenclature[G]{\Delta x}{Lattice spacing between neighboring lattice nodes connected by a vector $c_i \Delta t$}
\nomenclature[G]{\rho}{Density}
%\nomenclature[G]{\lambda}{Mean free path between molecules}
\nomenclature[G]{\alpha}{Thermal diffusivity}
\nomenclature[G]{\gamma}{Surface tension}

%\nomenclature[D]{$\Acute{r}$}{Dimensionless length}
\nomenclature[D]{${t_{\text{ref}}}$}{Dimensionless time}
\nomenclature[D]{$Re$}{Reynolds number}
%\nomenclature[D]{Ra}{Rayleigh number}
%\nomenclature[D]{Kn}{Knudsen number}
%\nomenclature[D]{Le}{Lewis number}
%\nomenclature[D]{$\widetilde{D}$}{Diffusion ratio}
\nomenclature[D]{$Pe$}{Péclet number}
\nomenclature[D]{$Da$}{Damköhler number}
%\nomenclature[D]{Pr}{Prandtl number}
%\nomenclature[D]{Bi}{Biot number}
%\nomenclature[D]{$\widetilde{h}_r$}{Dimensionless reaction heat}
\nomenclature[S]{0}{Reference quantity}
\nomenclature[S]{f}{Fluid(gas)}
\nomenclature[S]{s}{Solid}
\nomenclature[S]{i}{i-th lattice velocity direction}
\nomenclature[S]{x}{x-direction in Cartesian coordinate}
\nomenclature[S]{y}{y-direction in Cartesian coordinate}
\nomenclature[S]{p}{In physical unit}
\nomenclature[S]{lb}{In lattice unit}
\nomenclature[S]{e}{Thermal field for relaxation time}
\nomenclature[S]{d}{Species field for relaxation time}
\nomenclature[S]{in, out}{Inside and outside the bubble/drop}

\nomenclature[T]{eq}{Equilibrium}
\nomenclature[T]{*}{Post collision distribution function}
\nomenclature[T]{\sigma}{Fluid components}
\nomenclature[T]{inlet}{Inlet prescribed value}

\printnomenclature

%%%%%%%%%%%%%%%%%%%%%%%%%%%%%%%%%%%%%%%%%%%%%%%%%%%%%%%%%%%%%%%%%%%%%%%%%%%%%%%
%%%%%%%%%%%%%%%%%%%%%%%%%%%% Introduction %%%%%%%%%%%%%%%%%%%%%%%%%%%%%%%%%%%%%
%%%%%%%%%%%%%%%%%%%%%%%%%%%%%%%%%%%%%%%%%%%%%%%%%%%%%%%%%%%%%%%%%%%%%%%%%%%%%%%

\section{Introduction}\label{sec:introduction}

%%The accurate simulation of single and multi-component multi-phase flow problems, as well as their reactive interfaces, plays a crucial role in various fields of science and engineering. These problems arise in diverse applications such as chemical engineering, materials science, and environmental modeling among others. Over the past decades, significant progress has been made in the development of computational methods for simulating complex fluid dynamics phenomena. One such method is the lattice Boltzmann method (LBM), which has gained popularity due to its ability to efficiently model fluid flow and transport phenomena.

In this work, we present a novel multi-CPU/GPU code, so-called CooLBM ({\em COllaborative Open-source Lattice Boltzmann Method}), for the simulation of single and multi-component multi-phase flow problems, along with the capability to solve reactive interfaces with conjugate fluid-solid heat transfer using the lattice Boltzmann method. The code is developed using the STL library of C++, which provides powerful features for efficient implementation and execution on high-performance computing (HPC) systems. By leveraging the parallel processing capabilities of both CPUs and GPUs, CooLBM demonstrates enhanced computational performance and scalability, making it suitable for simulating large-scale and computationally demanding problems.

The development of CooLBM is motivated by the need for an accurate and efficient simulation tool to study the behavior of fluids in complex multi-phase and reactive systems. The ability to simulate multi-component systems and reactive interfaces is of particular interest in diverse research areas such as those in chemical engineering, where understanding the behavior of mixtures and reactions is essential \cite{Guo2016, Zhao2023}. Other examples are those in materials science, where the simulation of multi-phase systems enables the investigation of phase transitions, interface dynamics, and the overall material properties \cite{Chong2016, Pintelon2012}. Environmental modeling also benefits from accurate simulations of multi-phase flows and reactive processes, aiding in the understanding of pollutant transport, groundwater contamination, and other environmental phenomena \cite{Prieto2003, Stockmann2011, Daval2009}.

%To validate the capabilities and performance of CooLBM, extensive comparisons and benchmarking are conducted against analytical solutions and existing numerical models. 
Previous studies have demonstrated the efficacy of lattice Boltzmann method (LBM) in simulating fluid dynamics and transport phenomena. Gu et al. \cite{Gu2018} successfully employed LBM to investigate the behavior of immiscible fluids in porous media. Their study demonstrated the accuracy of LBM in capturing the dynamics of fluid interfaces and the overall flow behavior in porous media systems. Liu and Zhang \cite{Liu2011} utilized LBM to simulate droplet dynamics and breakup in a microfluidic channel, showcasing the capability of method to accurately capture complex flow phenomena at small scales.
%More Examples

Furthermore, numerous studies have applied lattice Boltzmann method to study specific multi-phase problems. Lintermann and Schröder \cite{Lintermann2020} focused on simulating two-phase flows with complex geometries, showcasing the ability of LBM to handle intricate fluid-solid boundaries. Their work demonstrated the accuracy and versatility of LBM in capturing complex flow patterns and interfacial dynamics. In the context of multi-component systems, Simonis et al. \cite{Simonis2023} employed LBM to investigate the behavior of binary fluid mixtures. Their study highlighted the accuracy of LBM in predicting interfacial dynamics, phase separation, and the overall behavior of multi-component systems.
%More Examples
%My previous LBM work for phase change

Additionally, LBM has been utilized to study the transport phenomena in porous media. Sukop and Or \cite{Sukop2004} investigated the application of LBM for modeling two-phase flow in porous media. Their model considered capillary forces and interfacial tension effects, providing a comprehensive representation of the pore-scale dynamics occurring in porous media. Their study demonstrated the capability of LBM to capture the intricate fluid-fluid and fluid-solid interactions in porous media systems.
%More Examples
%My previous LBM work for flow in Porous Media

On the other hand, the simulation of reactive interfaces using lattice Boltzmann methods has received limited attention in the literature. Among others, Verhaeghe et al. \cite{Verhaeghe2006} developed a reactive lattice Boltzmann model for simulating the dissolution of solid particles in a liquid. Their study demonstrated the effectiveness of LBM in capturing the dissolution kinetics and the impact of reactive interfaces on fluid flow behavior. He and Li \cite{He2000} developed an electrochemical reactive lattice Boltzmann method to simulate electrochemical reactions at fluid-solid interfaces. Their research demonstrated LBM's ability to handle complex electrochemical systems and provided valuable insights into fluid-electrode interactions.
%Our previous works

In recent years, several studies have focused on advancing the lattice Boltzmann method and its applications. Li et al. \cite{Li2021} proposed an improved lattice Boltzmann model for simulating multi-phase flows with large density ratios and high Reynolds numbers. Their model exhibited improved stability and accuracy in capturing complex interfacial dynamics. Leclaire et al. \cite{Leclaire2017} developed a 3D numerical framework based on lattice Boltzmann method, improving Galilean invariance and accurately simulating immiscible compounds in complex geometries, including fluid-solid interaction and precise contact angles. Their method enhances the capabilities of LBM for practical applications.
%My previous LBM work for contact angle 

Moreover, several research papers have specifically developed codes or software based on the lattice Boltzmann method. Su et al. \cite{Su2017} developed a Fortran-based Parallel Non-Dimensional Lattice Boltzmann Method (P-NDLBM) with message passing interface (MPI) for multi-CPU computation. Their approach significantly accelerates the simulation of transient fluid flow and heat transfer, as validated through comparisons with experimental data and single CPU computations. Pan et al. \cite{Pan2004} proposed a two-stage implementation of the lattice Boltzmann method for simulating single and multi-phase flow in porous media at the pore scale. They investigated different domain decomposition approaches and found that an orthogonal recursive bisection method achieves near-linear scaling and significantly reduces storage and computational time compared to traditional methods.

Similarly, Chen et al. \cite{Chen2016} developed a GPU-enhanced lattice Boltzmann simulator (GELBS) for evaluating transport properties in porous media. The optimized GPU code, exhibited excellent scalability and performance on parallel GPU architectures. This allowed for efficient simulations of porous media, including fractured systems, at high resolutions, enabling more accurate characterization of transport phenomena and its analysis. \textcolor{black}{Krafczyk et al. \cite{Krafczyk2014} conducted direct Navier-Stokes %(XXX is this direct numerical simulations?XXX) 
and large eddy simulation (LES) of turbulent flows using
the cumulant lattice Boltzmann approach over porous surfaces, explicitly considering pore-scale geometries obtained from computer tomography. Note that this numerical approach did not consider any turbulence models.} Their approach, implemented in Code VirtualFluids, showed favorable performance in terms of parallelization efficiency, numerical stability, and accuracy, enabling validation and comparison with experimental results.

Latt et al. \cite{Latt2021} presented a novel %(\textcolor{blue}{XXXwhat do you mean by novel??XXX})
approach for lattice Boltzmann simulations, which achieved massive performance on diverse hardware platforms. In their work, they introduced STLBM, a hardware-agnostic implementation of LBM simulations that showed an excellent scalability and efficiency on both CPUs and GPUs. The developed code was tested with different lattice Boltzmann collision models, showing its versatility. Their code was tested with different sophisticated collision models \cite{Latt2021}.

\textcolor{black}{However, the open-source code developed by Latt et al. \cite{Latt2021} was limited to lid-driven cavity flow. To test the behavior of the hardware-agnostic implementation libraries in addressing more complex applied physics problems, we extend our code, so-called CooLBM, using the same strategy, but to broader research areas, including multi-phase and multi-component flows, particle deposition and oxidization, reactive flows and inhomogenus surface reaction as well as fluid-solid conjugate heat transfer.} Using the architectures like multi-core CPU and GPU, CooLBM has been developed to extend the application of massively parallel LBM codes for wider ranges of engineering problem. While retaining the ability to simulate single-phase, single-component flows, CooLBM includes enhanced capabilities for handling multi-phase and multi-component flows and reactive interface combustion with fluid-solid conjugate heat transfer. We also implement different types of boundary conditions and provided users with access to pre-existing examples, enhancing the flexibility and usability of the code. \textcolor{black} {There are several open-source LBM codes available in public software libraries, many of which consist of millions of lines of code. Consequently, adding new physics by incorporating a module is challenging without first gaining in-depth familiarity with the details of code. CooLBM, on the other hand, offers a simplified programming structure and modules comprising only a few thousand lines of code, which can be easily modified and simulated in CPU.  Furthermore, the framework provided by CooLBM allows for the seamless inclusion of new modules for different physics.} It is noted that CooLBM code is available for the scientific community under initiatives open-source frame works. 

By utilizing a multi-CPU-GPU architecture and leveraging the STL library of C++, our code ensures efficient and scalable simulations. In this CooLBM code, extensive validation studies against analytical solutions and numerical models are performed to confirm its accuracy and reliability. We will also demonstrate its performance on different multi-core CPU and GPU devices. Through the integration of various numerical techniques and parallel computing, we show that CooLBM serves as a powerful tool for investigating complex fluid dynamics phenomena in diverse scientific and engineering domains. In the subsequent sections of this manuscript, we will present the numerical framework, validation studies, and selected application examples that highlight the capabilities and potential of CooLBM in addressing challenging fluid dynamics problems.

%%%%%%%%%%%%%%%%%%%%%%%%%%%%%%%%%%%%%%%%%%%%%%%%%%%%%%%%%%%%%%%%%%%%%%%%%%%%%%%
%%%%%%%%%%%%%%% Methodological Overview and Code Structure %%%%%%%%%%%%%%%%%%%%
%%%%%%%%%%%%%%%%%%%%%%%%%%%%%%%%%%%%%%%%%%%%%%%%%%%%%%%%%%%%%%%%%%%%%%%%%%%%%%%

\section{CooLBM Overview and Code Structure}\label{sec:method}

%This section provides an overview on the methodology employed in this study and outlines the structure of the developed code.
The numerical framework, hereafter called CooLBM, is designed to leverage parallel computing techniques using industry-standard tools and libraries. The section explains the numerical methodology and highlights the inherent parallelizability of the lattice Boltzmann method. Subsequently, the code structure of CooLBM is described, focusing on its hardware-independent design and the integration of Threading Building Blocks (TBB) of Intel, the STL implementation of Visual C++, and the NVIDIA HPC software development kit (SDK). Then the parallelization method adopted in CooLBM is discussed. Finally, an overview of different hardware platforms utilized for running the code is presented. This section aims to provide an understanding of the methodology, code structure, parallelization strategy, and hardware configuration, setting the foundation for the subsequent discussions and results analysis.

%%%%%%%%%%%%%%%%%%%%%%%%%%%%%%%%%%%%%%%%%%%%%%%%%%%%%%%%%%%%%%%%%%%%%%%%%%%%%%%
%%%%%%%%%%%%%%%%%%%%%%%%% Numerical Methodology %%%%%%%%%%%%%%%%%%%%%%%%%%%%%%%
%%%%%%%%%%%%%%%%%%%%%%%%%%%%%%%%%%%%%%%%%%%%%%%%%%%%%%%%%%%%%%%%%%%%%%%%%%%%%%%
\subsection{Numerical Methodology}\label{sec:numerical}

Boltzmann established the gap between molecular and continuum physics by integrating features from both scales and establishing a connection between them \cite{Mohamad2019}. He employed a statistical approach in the so-called Boltzmann equation to investigate the behavior of groups of molecules through a probability distribution function $f$. This function provides the probability of finding these particles or molecules in a 7-dimensional space, represented by three-dimensional spacial position $(x, y, z)$, microscopic velocity $\mathbf{u}$ $(u, v, w)$, and time $t$ \cite{Mohamad2019}:

\begin{equation}
\frac{\partial f}{\partial t} + \mathbf{c} \cdot \nabla f + \mathbf{F} \cdot \nabla_{\mathbf{c}} f = \Omega.
\label{eq:1}
\end{equation}

Here, $\Omega$ represents the collision operator of molecules which is written in a very simplified formula that is easy to implement in numerical methods through the single relaxation time (SRT) model known as Bhatnager-Gross-Krook (BGK):
\begin{equation}
\frac{\partial f}{\partial t} + \mathbf{c} \cdot \nabla f + \mathbf{F} \cdot \nabla_{\mathbf{c}} f = \frac{(f^{eq} - f)}{\tau}.
\label{eq:2}
\end{equation}

The left-hand side contains three terms. These terms, respectively, are: (i) the unsteady term, (ii) the convective term which is linear unlike the convective term in Navier-Stokes equations, and (iii) the long-range or external forces operator. \textcolor{black}{Here, $\mathbf{c}$ is the microscopic particle velocity, $f$ in the kinetic theory represents the density of particles with velocity $\mathbf{c}$, at physical space, $\mathbf{x}$ and time $t$, and $\mathbf{F}$ denotes the specific body force.} The right-hand side is a simple form of the collision operator, where $\tau$ is the relaxation time to reach equilibrium $f^{eq}$ after molecules collide.

The discretized lattice BGK (LBGK) model that is used in LBM is given through \cite{Mohamad2019}:
\begin{equation}
f_i(\mathbf{x} + \mathbf{c}_i \Delta t, t + \Delta t) = f_i(\mathbf{x}, t) - \frac{\Delta t}{\tau}\left[f_i(\mathbf{x}, t) - f_i^{eq}(\mathbf{x}, t)\right]
\label{eq:3}
\end{equation}
where, $c_i$ is the lattice speed in the $i$-th direction. The LBGK equation can be decomposed into two distinct parts that are performed in succession. The first part involves the collision process, where the distributions relax towards equilibrium state:
\begin{equation}
f_i^*(\mathbf{x}, t) = f_i(\mathbf{x}, t)\left(1 - \frac{\Delta t}{\tau}\right) + f_i^{eq}(\mathbf{x}, t)\frac{\Delta t}{\tau}.
\label{eq:4}
\end{equation}

In this context, the post-collision distribution is represented by $f_i^*$, while the equilibrium distribution is denoted by $f_i^{eq}$. The second part is the streaming process, where the post-collision distributions are propagated to neighboring nodes, becoming the pre-collision distributions for the next time-step:
\begin{equation}
f_i(\mathbf{x} + \mathbf{c}_i \Delta t, t + \Delta t) = f_i^*(\mathbf{x}, t).
\label{eq:5}
\end{equation}

As seen, the calculation of the collision part in the LBGK model relies on the equilibrium distributions. These equilibrium distributions are obtained using the following expression:
\begin{equation}
f_i^{eq}(\mathbf{x}, t) = w_i \rho \left(1 + \frac{u_a c_{i \alpha}}{c_s^2} + \frac{u_\alpha u_\beta Q_{i \alpha \beta}}{2c_s^4}\right)
\label{eq:6}
\end{equation}
where $w_i$ is a set of weights normalized to unity, $Q_{i \alpha \beta} = c_{i \alpha} c_{i \beta} - c_s^2 \delta_{\alpha \beta}$ is the quadrupole projector along the $i$-th direction, $c_s$ is the speed of sound, \textcolor{black}{subscripts $\alpha$ and $\beta$ take value from 1 to 3. Additionally, $\delta_{\alpha \beta}$ is the Dirac delta function, which equals 1 when $a=b$ and 0 otherwise.}
%(XXXwhy 2D?XXX), 
Repeated indices are summed upon the Einstein rule. For specific lattice speeds $c_i$ and lattice weights $\omega_i$ corresponding to different models, interested readers are referred to \cite{Mohamad2019}.

Macroscopic quantities emerge from the previous mesoscopic framework through the summation of the moments of lattice distribution functions $f_i$. For instance, summing the zeroth moments yields the mass density of the fluid i.e.,
\begin{equation}
\rho = \sum_i f_i(\mathbf{x}, t).
\label{eq:1.7}
\end{equation}
\textcolor{black}{Note that $\rho$ is used interchangeably with $\rho_f$ or $\rho_i$ in this paper.}
This equation allows us to recover the mass density of the fluid by summing over all lattice distribution functions $f_i$. The summation of first moments of $f_i$ will give the momentum of fluid, which can be modified to extract the fluid velocity:
\begin{equation}
\rho \mathbf{u}(\mathbf{x}, t) = \sum_i \mathbf{c}_i f_i(\mathbf{x}, t).
\label{eq:1.8}
\end{equation}
%\begin{equation}
%\mathbf{u}(\mathbf{x}, t) = \frac{1}{\rho}\sum_i (\mathbf{c}_i f_i(\mathbf{x}, t))
%\label{eq:1.9}
%\end{equation}

The summation of higher moments of $f_i$ allows for the derivation of further macroscopic quantities, including energy, temperature, and stresses. Further information on the derivation of macroscopic quantities from higher moments of $f_i$ can be found in \cite{Kruger2017}.

%%%%%%%%%%%%%%%%%%%%%%%%%%%%%%%%%%%%%%%%%%%%%%%%%%%%%%%%%%%%%%%%%%%%%%%%%%%%%%%
%%%%%%%%%%%%%%%%%% Multi-phase and Multi-component LBM %%%%%%%%%%%%%%%%%%%%%%%%
%%%%%%%%%%%%%%%%%%%%%%%%%%%%%%%%%%%%%%%%%%%%%%%%%%%%%%%%%%%%%%%%%%%%%%%%%%%%%%%

\subsubsection{Multi-phase and Multi-component LBM}\label{sec:multi}
Among the various methods available to model multi-phase and multi-component flows within the framework of LBM, the most popular approaches are the Shan-Chen (SC) model \cite{Shan_Chen_Model}, Free-Energy method \cite{Swift1996}, Color-Gradient method \cite{Gunstensen_1991}, and Phase-Field model \cite{Zheng2006,Huang2015}. These methods employ different strategies to capture the behavior of multiple phases or components in a fluid flow system. The free-energy method adopts a \enquote{top-down} approach, where the concept of free energy at the macroscopic level is utilized to derive forces that drive the phase separation. On the other hand, the SC model and similar models follow a \enquote{bottom-up} approach, starting from microscopic interactions between fluid elements. These interactions, often represented through interaction potentials, ultimately result in the macroscopic separation of phases. 

Within the CooLBM framework, the Shan-Chen model has been developed as the method of choice for modeling multi-phase and multi-component simulations. The SC model has proven to be highly effective in capturing complex phenomena associated with phase separation and interfacial dynamics \cite{Huang2007}. Therefore, we would be able to accurately simulate the behavior of multiple phases and/or components, providing valuable insights into a wide range of practical applications. 

The SC model is based on the assumption of \textcolor{black}{considering} 
%additive (XXX adding ?XXX) 
intermolecular forces between pairs of molecules. These forces become stronger with increasing molecular density and are highly dependent on the distance between fluid elements. The force acting on a fluid element at position $x$ due to all possible interactions can be determined by integrating the Shan-Chen interaction force density at that position, as described in \cite{Kruger2017}:
\begin{equation}
\mathbf{F}^{\text{SC}}(\mathbf{x}) = - \int (\widetilde{\mathbf{x}} - \mathbf{x}) G(\widetilde{\mathbf{x}} - \mathbf{x}) \psi(\mathbf{x}) \psi(\widetilde{\mathbf{x}}) d^3 \widetilde{\mathbf{x}}.
\label{eq:1.10}
\end{equation}

Within this context, pairs of neighboring molecules $\widetilde{\mathbf{x}}$ and $\mathbf{x}$ experience a short-range interaction force $G(\widetilde{\mathbf{x}} - \mathbf{x})$ that is dependent on the distance between them. The effective density of each molecule is described by the pseudo-potential $\psi$, which takes the form of $\psi(\mathbf{x})$ and $\psi(\widetilde{\mathbf{x}})$ as shown in the following:

\begin{equation}
\psi(\rho) = \rho_0 \left(1 - \exp^{-\rho/\rho_0}\right)
\label{eq:1.11}
\end{equation}
where $\psi(\rho)$ represents the effective density and the reference density $\rho_0$ is typically set to unity in the simulation. It is noted that the pseudo-potential in Eq. \eqref{eq:1.11} is bounded between zero and $\rho_0$ for any value of the density $\rho$, ensuring that the interaction force in Eq. \eqref{eq:1.10} remains finite.

In the case of a discretized lattice, the molecular interaction force \(G(\widetilde{\mathbf{x}} - \mathbf{x})\) is set to be zero everywhere except in the neighboring nodes connected by the vector of $\mathbf{c}_i \Delta t$. In this case, $G(\widetilde{\mathbf{x}} - \mathbf{x})$ is set to be equal to $w_i G$, resulting in the discretized Shan-Chen force for a single 
%(\textcolor{blue}{XXX each? XXX}) phase 
component that describes the interaction between two fluid phases \cite{Kruger2017}:
%(\textcolor{blue}{XXX I don't understand the sentence for two fluid phases!!! XXX}) 
\begin{equation}
\mathbf{F}^{\text{SC}}(\mathbf{x}) = - G \psi(\mathbf{x}) \sum_{i}^{\text{fluid}} w_i \psi(\mathbf{x} + \mathbf{c}_i \Delta t) \mathbf{c}_i \Delta t.
\label{eq:1.12}
\end{equation}  
%Interactions between different phases of fluids and solid boundaries play a vital role in various engineering applications, particularly in simulating flow in porous media. 
To incorporate fluid-solid interactions, an additional term is needed to be included in Eq. \eqref{eq:1.12}. This results in the complete Shan-Chen force \cite{Kruger2017}:
\begin{equation}
\mathbf{F}^{\text{SC}}(\mathbf{x}) = -G \psi(\mathbf{x})\left[\sum_{i}^{\text{fluid}} w_i \psi(\mathbf{x}+\mathbf{c}_i \Delta t) \mathbf{c}_i \Delta t + \sum_{i}^{\text{solid}} w_i \psi(\rho_s) \mathbf{c}_i \Delta t\right].
\label{eq:1.13}
\end{equation}

The Shan-Chen forcing $\mathbf{F}^{\text{SC}}(\mathbf{x})$, as derived in Eq. \eqref{eq:1.13}, can also be extended to account for each component in the flow, resulting in $\mathbf{F}^{\text{SC}(\sigma)}(\mathbf{x})$. Similarly, the interaction force between molecules is also extended to encompass the interaction force between different components, denoted as $G_{\sigma \widetilde{\sigma}}$. Consequently, the Shan Chen force for multi-component flow, which characterizes fluid interactions, can be expressed as :
\begin{equation}
\mathbf{F}^{\text{SC}(\sigma)}(\mathbf{x}) = -\psi^{(\sigma)}(\mathbf{x})\sum_{\widetilde{\sigma}} G_{\sigma \widetilde{\sigma}} \sum_{i} w_i \psi^{(\widetilde{\sigma})}(\mathbf{x}+\mathbf{c}_i \Delta t) \mathbf{c}_i \Delta t.
\label{eq:1.14}
\end{equation}

Similar to the multi-phase case, it is necessary to consider the interactions between different components of the flow and solid boundaries. \textcolor{black}{The discretized Shan-Chen force, which describes the interaction between different fluid components and the solid phase,}
%The discretized Shan-Chen force for multiple fluid components, which describes the fluid-solid interaction (\textcolor{blue}{XXX I dont understand inside double comma note!XXX Can we simply say: 'The discretized Shan-Chen force for describing the fluid-solid interaction'XXX}),
can be expressed for the $\sigma$-th fluid component as follows:
\begin{equation}
\mathbf{F}_{\text{ads},\sigma}(\mathbf{x}) = -G_{\text{ads},\sigma} \rho_{\sigma}(\mathbf{x},t) \sum_{i} w_i s(\mathbf{x}+\mathbf{c}_i \Delta t) \mathbf{c}_i.
\label{eq:1.15}
\end{equation}

Here, the interaction coefficient between the $\sigma$-th fluid and solid boundary is denoted as $G_{\text{ads},\sigma}$. The indicator function $s(\mathbf{x}+\mathbf{c}_i \Delta t)$ takes the values of 0 or 1 depending on whether the neighboring cell is a fluid or a solid node, respectively. In the case of a wetting fluid, the interaction is attractive, implying that $G_{\text{ads},\sigma}$ is negative. Conversely, non-wetting fluids are characterized by a positive value of $G_{\text{ads},\sigma}$.

The set of equations described by Eqs. \eqref{eq:1.13} to \eqref{eq:1.15} constitutes the fundamental model employed in CooLBM to simulate multi-phase multi-component flows. While these equations serve as the foundation of the simulation framework, it is important to note that additional equations may be required to effectively capture specific phenomena, such as the contact angle. In Section \ref{sec:validation}, we will delve into the details and implications of these additional equations, which are essential components of CooLBM. This will provide a better understanding of the simulation process.

In the multi-component model, the simulation requires a distinct set of populations (distribution functions) for each component involved. Each set of populations follows the standard LBGK equation, which is expanded to handle multi-phase and multi-component flows by incorporating the following equation:
\begin{equation}
f_i^{(\sigma)}(\mathbf{x}+\mathbf{c}_i \Delta t, t + \Delta t) = f_i^{(\sigma)}(\mathbf{x}, t) - \frac{\left[f_i^{(\sigma)}(\mathbf{x}, t) - f_i^{\text{eq}(\sigma)}(\mathbf{x}, t)\right]}{\tau^{(\sigma)}} \Delta t + \left[1 - \frac{\Delta t}{2\tau^{(\sigma)}}\right] F_i^{(\sigma)}(\mathbf{x}, t)\Delta t.
\label{eq:1.16}
\end{equation}

It is evident from this equation that in the case of multi-component flow, each component possesses its own relaxation time $\tau^{(\sigma)}$, representing viscosity as a macroscopic observable, and forcing $F_i^{(\sigma)}(\mathbf{x},t)$. Equation \eqref{eq:1.16} closely resembles the standard LBGK equation for single-component fluid flow, with the exception that the equilibrium distributions $f_i^{\text{eq}(\sigma)}$ are now specific to each fluid component $\sigma$. It is worth mentioning here that Eq.\eqref{eq:1.16} can be used for single-component multiphase flow by removing the subscript $\sigma$. These equilibrium distributions in Eq.\eqref{eq:6} are functions of the equilibrium velocity $\mathbf{u}^{\text{eq}(\sigma)}$ and density $\rho^{(\sigma)}$ corresponding to each component $\sigma$. Note that in multi-component flow, as per Eq.\eqref{eq:6}, the velocity vector $\mathbf{u}$ is expressed as $\mathbf{u}^{\text{eq}(\sigma)}$, and the density $\rho$ is defined as $\rho^{(\sigma)}$.

%%%%%%%%%%%%%%%%%%%%%%%%%%%%%%%%%%%%%%%%%%%%%%%%%%%%%%%%%%%%%%%%%%%%%%%%%%%%%%%
%%%%%%%%%%%%%%%%%%%%%%%%%% Shan Chen Forcing Scheme %%%%%%%%%%%%%%%%%%%%%%%%%%%
%%%%%%%%%%%%%%%%%%%%%%%%%%%%%%%%%%%%%%%%%%%%%%%%%%%%%%%%%%%%%%%%%%%%%%%%%%%%%%%
\subsubsection{Shan-Chen Forcing Scheme}\label{sec:ShanChen}

In CooLBM, two schemes for fluid component interactions are utilized. These schemes are the Shan-Chen forcing and the Guo forcing schemes, which are facilitating the modeling of various fluid components and their interactions within the simulation framework. By incorporating these interaction schemes, the CooLBM is capable of treating the dynamics and the behaviors of multiple fluid components. For the Shan-Chen forcing scheme, the equilibrium velocity for component $\sigma$ is determined as \cite{Kruger2017}:

\begin{equation}
\mathbf{u}^{\text{eq}(\sigma)} = \mathbf{u'} + \frac{\tau^{(\sigma)} \mathbf{F}^{\text{SC}(\sigma)}}{\rho^{(\sigma)}}
\label{eq:1.17}
\end{equation}
where $\mathbf{u'}$ represents the common velocity term, calculated through a weighted average as:

\begin{equation}
\mathbf{u}' = \frac{\sum_{\sigma} \frac{\rho^{(\sigma)} \mathbf{u}^{(\sigma)}}{\tau^{(\sigma)}}}{\sum_{\sigma} \frac{\rho^{(\sigma)}}{\tau^{(\sigma)}}}.
\label{eq:1.18}
\end{equation}

In this scheme, fluid components interact through the force $\mathbf{F}^{\text{SC}(\sigma)}$ while sharing the common velocity $\mathbf{u}'$. When implementing this scheme in the CooLBM code, the additional forcing terms $F_i^{\sigma}$ in the standard LBGK equation are set to be zero. 
The component density and velocity are obtained  from the modified versions of Eqs.\eqref{eq:1.7} and \eqref{eq:1.8} as given below:
\begin{equation}
 \rho^{(\sigma)} = \sum_i f_i^{(\sigma)}, \qquad \rho^{(\sigma)} \mathbf{u}^{(\sigma)} = \sum_i \mathbf{c}_i f_i^{(\sigma)}. 
 \label{Shen_forcing_macroscopic}
\end{equation}
Furthermore, the actual velocity of the fluid is calculated using the following expression:
\begin{equation}
\mathbf{u} = \frac{1}{\rho} \sum_{\sigma} \left[\sum_{i} f_i^{(\sigma)} \mathbf{c}_i + \frac{\mathbf{F}^{\text{SC}(\sigma)} \Delta t}{2}\right].
\label{Shan_Chen_forcing_actual_velocity}
\end{equation}
It is noted that the SC forcing scheme produces a surface tension that depends on the relaxation time $\tau^{(\sigma)}$, which is considered nonphysical. This considered as a drawback of this scheme. Hence, it is essential to explore alternative models for the purpose of comparison and to address this issue.

%%%%%%%%%%%%%%%%%%%%%%%%%%%%%%%%%%%%%%%%%%%%%%%%%%%%%%%%%%%%%%%%%%%%%%%%%%%%%%%
%%%%%%%%%%%%%%%%%%%%%%%%%%%%%% Guo Forcing Scheme %%%%%%%%%%%%%%%%%%%%%%%%%%%%%
%%%%%%%%%%%%%%%%%%%%%%%%%%%%%%%%%%%%%%%%%%%%%%%%%%%%%%%%%%%%%%%%%%%%%%%%%%%%%%%
\subsubsection{Guo Forcing Scheme}\label{sec:Guo}

Guo forcing scheme is a forcing approach that offers the advantage of viscosity-independent surface tension and it is the extension of Guo’s forcing approach to multiple components \cite{Guo_forcing}. This scheme involves the utilization of the barycentric velocity for equilibrium velocities of all components, denoted as $\mathbf{u}^b$ \cite{Kruger2017}:
\begin{equation}
\mathbf{u}^b = \frac{1}{\rho} \sum_{\sigma} \left[\sum_{i} f_i^{(\sigma)} \mathbf{c}_i + \frac{\mathbf{F}^{\text{SC}(\sigma)} \Delta t}{2}\right].
\label{eq:1.19}
\end{equation}

Here, $\rho$ represents the total density of all components, given by:
\begin{equation}
\rho = \sum_{\sigma} \rho^{(\sigma)}.
\label{eq:1.20}
\end{equation}

It is important to note that the density $\rho$ used in Eq.\eqref{eq:1.19} corresponds to the total density of all components, as calculated according to Eq.\eqref{eq:1.20}. The equilibrium velocity for component $\sigma$ is defined as $\mathbf{u}^{\text{eq}(\sigma)} = \mathbf{u}^b$. In addition, the actual velocity is calculated using the same expression used to find the barycentric velocity in Eq. \eqref{eq:1.19}, after the streaming step. 

Furthermore, within the Guo forcing scheme \cite{Guo_forcing}, it is necessary to specify the forcing terms $F_i^{(\sigma)}$ in the standard LBGK equation. This term reads as:

\begin{equation}
F_i^{(\sigma)} = w_i \left[\frac{c_{i \alpha}}{c_s^2} + \frac{(c_{i \alpha} c_{i \beta} - c_s^2 \delta_{\alpha \beta}) u^{b}_{\beta}}{{c_s^4}}\right] F_\alpha^{\text{SC}(\sigma)}
\label{eq:1.21}
\end{equation}
where $c_{i\alpha}$ denotes the components of the lattice velocity vector, and $u^{b}_{\beta}$ corresponds to the components of the barycentric velocity vector. By employing the Guo forcing scheme \cite{Guo_forcing}, we aimed to overcome the limitations associated with the viscosity-dependent surface tension observed in other forcing approaches; thus enhancing the accuracy and physical fidelity of our simulations. In the Guo forcing scheme, the component velocity and density are determined as follows:
\begin{equation}
 \rho^{(\sigma)} = \sum_i f_i^{(\sigma)}, \qquad \rho^{(\sigma)} \mathbf{u}^{(\sigma)} = \sum_i \mathbf{c}_i f_i^{(\sigma)} + \frac{\mathbf{F}^{\text{SC}(\sigma)} \Delta t}{2}. 
 \label{Guo_forcing_macroscopic}
\end{equation}
It is worth noting that the pseudo-potential $\psi$, defined in Eq.\eqref{eq:1.11}, can be modified for multi-component flow to evaluate the Shan-Chen force described in Eq.\eqref{eq:1.14}, as follows: 
\begin{equation}
\psi^{(\sigma)} = \rho_0 \left(1 - \exp^{-\rho^{(\sigma)}/\rho_0}\right) \,.
\label{pseudo-potential_multicomponent}
\end{equation}
\subsubsection{3D and 2D LBM Models}\label{sec:reactiveLBM}
\textcolor{black}{When solving single-phase or multiphase and multicomponent systems using discretized Eq. \eqref{eq:4} or Eq. \eqref{eq:1.16}, respectively, the primary focus is on obtaining two-dimensional (2D) or three-dimensional (3D) solutions. The collision and streaming steps remain the same for both 2D and 3D cases. For solving Eq. \eqref{eq:4} or Eq. \eqref{eq:1.16}, two components in any desired vector field are considered for 2D, while three components are required for 3D. The velocity sets used in the LBM  are further differentiated by the lattice velocities \textbf{$c_i$} and their corresponding weights $w_i$.}

\textcolor{black}{To distinguish between 2D and 3D simulations, the number $d$ of spatial dimensions and the number $q$ of discrete velocities are specified using the notation D$d$Q$q$. Although many permutations of $q$ are theoretically possible, the choice of $q$ is constrained by several factors, including the physical phenomena being modeled, computational power requirements, and numerical stability considerations. }

\begin{figure}[htbp]
    \centering
    \includegraphics[width=\linewidth]{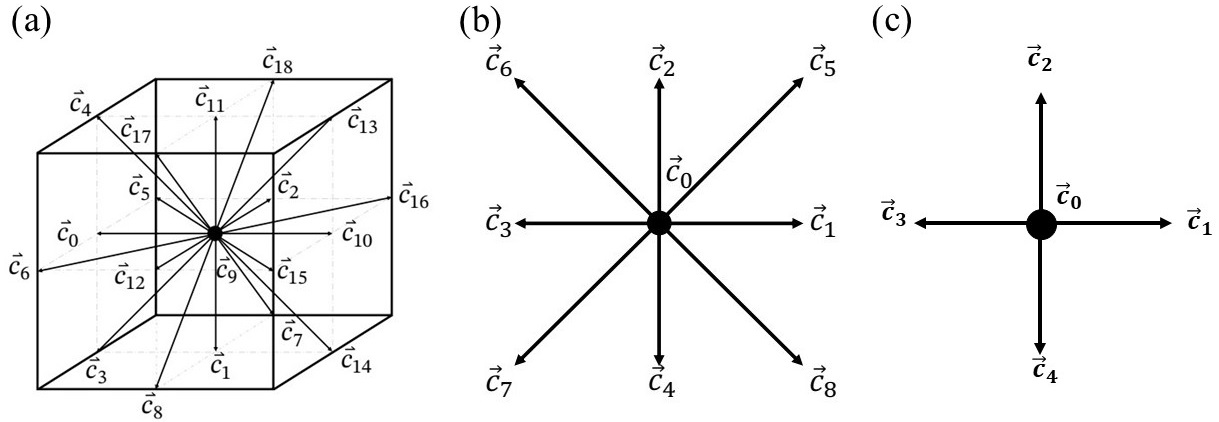}
    \caption{Different lattice velocity schemes: (a) D3Q19, (b) D2Q9, and (c) D2Q5.}
    \label{fig:lattice_components}
\end{figure}

\textcolor{black}{In the 3D version of CooLBM code, we adopt the D3Q19 scheme to solve fluid flow problems involving single-phase and multiphase, as well as multicomponent systems.  For 2D flow fields, the D2Q9 model is employed, whereas the D2Q5 model is used to obtain the 2D concentration field. The rationale for not using the D2Q9 model for concentration field calculations is explained in the following section. Note that the naming order of lattice velocity directions can be adjusted to facilitate implementation flexibility during programming. The arrangement of lattice velocity for these schemes is shown in the Figure \ref{fig:lattice_components}. The explicit forms of the velocity sets used in the CooLBM for D3Q19, D2Q9, and D2Q5 are provided in Tables \ref{tab:D3Q19}–\ref{tab:D2Q5}.}

\begin{table}[htbp]
\centering
\caption{The D3Q19 lattice velocities and weights}
    \label{tab:D3Q19}
\begin{tabularx}{\textwidth}
{ 
  | >{\raggedright\arraybackslash}X 
  | >{\centering\arraybackslash}X 
  | >{\raggedleft\arraybackslash}X 
  | >{\centering\arraybackslash}X
  | >{\centering\arraybackslash}X
  | >{\centering\arraybackslash}X
  | >{\centering\arraybackslash}X
  | >{\centering\arraybackslash}X
  | >{\centering\arraybackslash}X 
  | >{\centering\arraybackslash}X
   | >{\centering\arraybackslash}X
    | >{\centering\arraybackslash}X
     | >{\centering\arraybackslash}X
      | >{\centering\arraybackslash}X
       | >{\centering\arraybackslash}X
        | >{\centering\arraybackslash}X
         | >{\centering\arraybackslash}X
          | >{\centering\arraybackslash}X
           | >{\centering\arraybackslash}X
            | >{\centering\arraybackslash}X| }
 \hline
 $i$ & 0 & 1 & 2 & 3 & 4 & 5 & 6 & 7 & 8 & 9 & 10 & 11 & 12 & 13 & 14 & 15 & 16 & 17 & 18   \\
 \hline
  $w_i$ & $\frac{1}{3}$ & $\frac{1}{18}$ & $\frac{1}{18}$ & $\frac{1}{18}$ & $\frac{1}{18}$ & $\frac{1}{18}$ & $\frac{1}{18}$ & $\frac{1}{36}$ & $\frac{1}{36}$  & $\frac{1}{36}$  & $\frac{1}{36}$  & $\frac{1}{36}$  & $\frac{1}{36}$  & $\frac{1}{36}$  & $\frac{1}{36}$  & $\frac{1}{36}$  & $\frac{1}{36}$  & $\frac{1}{36}$  & $\frac{1}{36}$ \\
\hline
 $c_{ix}$ & 0 & +1 & -1 & 0 & 0 & 0 & 0 & +1 & -1  & +1 & -1 & 0 & 0 & +1 & -1 & +1 & -1 & 0 & 0 \\
 \hline
  $c_{iy}$ & 0 & 0 & 0 & +1 & -1 & 0 & 0 & +1 & -1 & 0 & 0 & +1 & -1 & -1 & +1 & 0 & 0 & +1 & -1 \\
 \hline
   $c_{iz}$ & 0 & 0 & 0 & 0 & 0 & +1 & -1 & 0 & 0 & +1 & -1 & +1 & -1 & 0 & 0 & -1 & +1 & -1 & +1  \\
 \hline
\end{tabularx}
\end{table}

\begin{table}[htbp]
\centering
\caption{The D2Q9 lattice velocities and weights}
    \label{tab:D2Q9}
\begin{tabularx}{0.8\textwidth}
{ 
  | >{\raggedright\arraybackslash}X 
  | >{\centering\arraybackslash}X 
  | >{\raggedleft\arraybackslash}X 
  | >{\centering\arraybackslash}X
  | >{\centering\arraybackslash}X
  | >{\centering\arraybackslash}X
  | >{\centering\arraybackslash}X
  | >{\centering\arraybackslash}X
  | >{\centering\arraybackslash}X 
   | >{\centering\arraybackslash}X| }
 \hline
 $i$ & 0 & 1 & 2 & 3 & 4 & 5 & 6 & 7 & 8  \\
 \hline
  $w_i$ & $\frac{4}{9}$ & $\frac{1}{9}$ & $\frac{1}{9}$ & $\frac{1}{9}$ & $\frac{1}{36}$ & $\frac{1}{36}$ & $\frac{1}{36}$ & $\frac{1}{36}$ & $\frac{1}{36}$ \\
\hline
 $c_{ix}$ & 0 & +1 & 0 & -1 & 0 & +1 & -1 & -1 & +1  \\
 \hline
  $c_{iy}$ & 0 & 0 & +1 & 0 & -1 & +1 & +1 & -1 & -1  \\
 \hline
\end{tabularx}
\end{table}

\begin{table}[htbp]
\centering
\caption{The D2Q5 lattice velocities and weights}
    \label{tab:D2Q5}
\begin{tabularx}{0.8\textwidth}
{ 
  | >{\raggedright\arraybackslash}X 
  | >{\centering\arraybackslash}X 
  | >{\raggedleft\arraybackslash}X 
  | >{\centering\arraybackslash}X
  | >{\centering\arraybackslash}X
   | >{\centering\arraybackslash}X| }
 \hline
 $i$ & 0 & 1 & 2 & 3 & 4  \\
 \hline
  $w_i$ & 0.4 & 0.15 & 0.15 & 0.15 & 0.15 \\
\hline
 $c_{ix}$ & 0 & +1 & 0 & -1 & 0   \\
 \hline
  $c_{iy}$ & 0 & 0 & +1 & 0 & -1   \\
 \hline
\end{tabularx}
\end{table}

%\begin{equation}
%w_i = \begin{cases}
%\frac{4}{9}, & i = 0 , \\
%\frac{1}{9},  & i = 1, 2, 3,4 , \\
%\frac{1}{36},  & i = 5, 6, 7, 8.
%\end{cases}
%\label{weight_factors}
%\end{equation}
%The nine discrete velocity vectors are 
%\begin{equation}
%\textbf{c}_i = \begin{cases}
%(0,0), & i = 0 , \\
%(\text{cos}[i-1]\frac{\pi}{2}, \text{sin}[i-1]\frac{\pi}{2})\frac{\Delta x}{\Delta t},  & i = 1, 2, 3,4 , \\
%(\text{cos}[2i-9]\frac{\pi}{2}, \text{sin}[2i-9]\frac{\pi}{2})\frac{\Delta x}{\Delta t},  & i = 5, 6, 7, 8.
%\end{cases}
%\label{velocity_vectors}
%\end{equation}
%%%%%%%%%%%%%%%%%%%%%%%%%%%%%%%%%%%%%%%%%%%%%%%%%%%%%%%%%%%%%%%%%%%%%%%%%%%%%%%
%%%%%%%%%%%%%%%%%%%%%%%%%%%%%%%% Reactive LBM %%%%%%%%%%%%%%%%%%%%%%%%%%%%%%%%%
%%%%%%%%%%%%%%%%%%%%%%%%%%%%%%%%%%%%%%%%%%%%%%%%%%%%%%%%%%%%%%%%%%%%%%%%%%%%%%%
\subsubsection{Reactive LBM}\label{sec:reactiveLBM}
%In this section, the Lattice Boltzmann Method is employed to model surface chemical reactions and conjugate heat transfer. 
This section focuses on the coupling flow, thermal, and species concentration fields to investigate the surface combustion accompanied by species mass transfer, solid consumption accompanied as well as fluid-solid conjugate heat transfer. The conjugate heat transfer is modeled using a thermal counter-slip method \cite{Alamian2024}, while surface chemical reactions and species transfer are modeled using the Kang et al. wet node scheme \cite{Kang2010}.

The governing equations for the bulk fluid domain, $\Omega_f$, define the convection-diffusion phenomena for the various fields, enabling the time evolution of the corresponding variables. Additionally, a separate set of equations is defined for the reactive interface at the solid surface $\partial\Omega$. These equations ensure continuity of temperature, heat flux, and species mass flux in the presence of heterogeneous chemical reactions.

The governing equations for the bulk fluid domain $\Omega_f$ are as follows:
\begin{subequations}\label{reactive1}
\begin{gather}
\nabla \cdot \mathbf{u} = 0
\label{eq:reactive1a}\\
\frac{\partial \mathbf{u}}{\partial t} + \mathbf{u}(\nabla \cdot \mathbf{u}) = -\frac{1}{\rho_f}\nabla P + \nu \Delta \mathbf{u}
\label{eq:reactive1b}\\
\frac{\partial C_{O_2}}{\partial t} + C_{O_2} (\nabla \cdot \mathbf{u}) = D_{O_2}\Delta C_{O_2}
\label{eq:reactive1c}\\
\frac{\partial C_{CO_2}}{\partial t} + C_{CO_2} (\nabla \cdot \mathbf{u}) = D_{CO_2}\Delta C_{CO_2}
\label{eq:reactive1d}\\
\frac{\partial (\rho_f \varepsilon_f)}{\partial t} + \nabla \cdot (\rho_f \varepsilon_f \mathbf{u}) = K_f \Delta T
\label{eq:reactive1e}\\
\rho_s c_{ps} \frac{\partial T}{\partial t} = K_s \Delta T
\label{eq:reactive1f}
\end{gather}
\end{subequations}	

In the above equations, $\mathbf{u}$ represents the fluid velocity, $\rho$ is density, $P$ is pressure, $\nu$ is kinematic viscosity, $C$ is concentration, $D$ is species diffusivity, $\varepsilon$ is internal energy, $K$ is thermal conductivity, and $c_p$ is specific heat capacity. Subscripts $f$ and $s$ refer to fluid and solid, respectively.

Within the bulk fluid domain $\Omega_f$, where gases flow, the incompressible continuity and momentum Eqs. \eqref{eq:reactive1a} and \eqref{eq:reactive1b}, also known as Navier-Stokes equations, are solved to determine \textcolor{black}{gas transport and the reaction at the solid interface.}
%pore-scale (XXX why pore-scale? why not meso-scale which is more generic? XXX) gas flow. 
The species concentrations, including O$_2$ and CO$_2$, evolve according to the unsteady convection-diffusion Eqs. \eqref{eq:reactive1c} and \eqref{eq:reactive1d}. \textcolor{black}{The heat transfer due to the chemical reactions within the bulk fluid domain $\Omega_f$, 
%(XXX we suddenly pass from fluid to gas!!XXX) 
is modeled by the convection-diffusion Eq. \eqref{eq:reactive1e},} while the heat conduction Eq. \eqref{eq:reactive1f} describes heat transfer in the solid phase (pure diffusion).

At the reactive interface $\partial\Omega$, carbon consumes O$_2$, and releases CO$_2$ and heat energy. The reactive boundary conditions ensure the continuity of temperature and thermal and mass fluxes as follows:
\begin{subequations}\label{domI}
\begin{gather}
C+O_2\rightarrow CO_2 \label{eq:react}\\
D_{O_2}\mathbf{n} \cdot \nabla C_{O_2}-A \exp\left(\frac{-E}{R_u T}\right)C_{b,O_2}=0 \label{eq:reactO2}\\
D_{CO_2}\mathbf{n} \cdot \nabla C_{CO_2}+A \exp\left(\frac{-E}{R_u T}\right)C_{b,O_2}=0 \label{eq:reactCO2}\\
T_s=T_f \label{eq:tempCon}\\
\mathbf{n} \cdot (-K_s \nabla _s T+K_f \nabla _f T)+\Dot H=0 \label{eq:fluxCon}\\
\Dot H=h_r A  \exp\left(\frac{-E}{R_u T}\right) C_{b,O_2} \label{eq:reacRate}
\end{gather}
\end{subequations}

In the above equations, $\mathbf{n}$ is the unit vector normal to the reactive interface pointing toward the fluid phase, $A$ is the pre-exponential factor, $E$ is the activation energy, $R_u$ is the ideal gas constant, $T$ is temperature, and $h_r$ is the standard enthalpy of reaction. \textcolor{black}{Additionally, $\Dot H$ is the heat flux released during the reaction between $C$ and $O_2$ at the interface, and  $C_{b,O_2}$ is the interface oxygen concentration.}

%XXX definition of some parameters such as $\Dot H$ and $C_{b}$ are missing !!XXX

For the modeling of fluid momentum and heat and mass transfer, five distinct sets of distribution functions are defined: $f_i$ for the flow field, $g_i$ for the fluid thermal energy, $g_{i,s}$ for solid thermal energy, $h_{i,O_2}$ for the oxygen species field, and $h_{i,CO_2}$ for the carbon dioxide species field. The index $i$ ranges from 0 to $Q-1$, where $Q$ represents the number of discrete velocities of the lattice. For reactive flows, the 9-velocity lattice D2Q9 is used for both flow and thermal fields for the reactive interface problem, while the 5-velocity lattice D2Q5 is used for the species field.

The thermal fields in the solid and fluid phases are modeled by introducing an energy density distribution function to the LBGK model. The heat transfer in the fluid and solid phases is accounted for by the Internal Energy Double Distribution Function (IEDDF) \cite{He1998}. The following IEDDF LBGK model applies to both the fluid and solid domains. For the fluid phase:

\begin{equation}
g_i(\mathbf{x} + \mathbf{c}_i \Delta t, t + \Delta t) = g_i(\mathbf{x}, t) - \frac{\Delta t}{\tau_{e,f}}\left[g_i(\mathbf{x}, t) - g_i^{\text{eq}}(\epsilon_f, \mathbf{u})\right]
\label{eq:therpop}
\end{equation}
where
%\begin{subequations}\label{macroreac}
%\end{gather}
%\end{subequations}
\begin{equation}
 \rho_f \mathbf{u} = \sum_{i=0}^{8} \mathbf{c}_i f_i \label{eq:macroreaca}   
\end{equation}
and macroscopic variable (internal energy) is:
\begin{equation}
 \epsilon_f = \frac{1}{\rho_f} \sum_{i=0}^{8} g_i. \label{eq:macroreacb}   
\end{equation}

Here, the constants $\mathbf{c}_i$ are the discrete lattice velocities, $\Delta t$ is the discrete time step, and \textcolor{black}{$\tau_{e,f}$ is the relaxation time to equilibrium, which is determined through the thermal diffusivity of the fluid i.e. $(3\alpha_f + 0.5)$. The thermal diffusivity of the fluid, $\alpha_f$, is defined as the ratio between the thermal conductivity of the fluid to its density and specific heat capacity i.e. $K_f/(\rho_f c_{pf})$.} The equilibrium distribution functions $g_i^{\text{eq}}$ are calculated in terms of the fluid internal energy $\epsilon_f$ and velocity $\mathbf{u}$:

\begin{equation}
g_i^{\text{eq}}(\epsilon_f, \mathbf{u}) = w_i \rho_f \epsilon_f \left[1 + \frac{\mathbf{u} \cdot \mathbf{c}_i}{c_s^2} + \frac{(\mathbf{u} \cdot \mathbf{c}_i)^2}{2c_s^4} - \frac{\mathbf{u} \cdot \mathbf{u}}{2c_s^2}\right]
\label{eq:therpopeq}
\end{equation}

Here, $c_s$ is the speed of sound in lattice units. In the adopted D2Q9 model, it adopts the value $\frac{1}{\sqrt{3}} \frac{\Delta x}{\Delta t}$, where $\Delta x$ is the discrete lattice spacing and $\Delta t$ is the temporal time step. \textcolor{black}{Similarly, for the solid phase: 
\begin{equation}
g_{s,i}(\mathbf{x} + \mathbf{c}_i \Delta t, t + \Delta t) = g_{s,i}(\mathbf{x}, t) - \frac{\Delta t}{\tau_{e,s}}\left[g_{s,i}(\mathbf{x}, t) - g_{s,i}^{\text{eq}}(\epsilon_s)\right]
\label{eq:therpop_solid}
\end{equation}
where the equilibrium distribution functions $g_{s,i}^{\text{eq}}$ are defined in terms of the solid internal energy $\epsilon_s$ as
\begin{equation}
g_{s,i}^{\text{eq}}(\epsilon_s) = w_i \rho_s \epsilon_s 
\label{eq:therpopeq_solid}
\end{equation}
and $\tau_{e,s}$ is the solid relaxation time to equilibrium, which is obtained from the thermal diffusivity of the solid i.e. $(3\alpha_s + 0.5)$. The thermal diffusivity of the solid $\alpha_s$ is defined as the ratio between the thermal conductivity of the solid to its density and specific heat capacity i.e. $K_s/(\rho_s c_{ps})$. The macroscopic internal energy for the solid is calculated as: 
\begin{equation}
 \epsilon_s = \frac{1}{\rho_s} \sum_{i=0}^{8} g_{s,i}. \label{eq:macroreacb_solid}   
\end{equation}}

\textcolor{black}{To simulate the convection and diffusion of species such as O$_2$ and CO$_2$, Eq. \eqref{eq:reactive1c} and \eqref{eq:reactive1d} are solved in the LBGK model using a D2Q5 lattice. The D2Q5 lattice model offers several advantages, including higher stability and accuracy for simulations with low Péclet numbers (diffusion-dominated flows) and reduced computational demand due to fewer velocity directions $\textbf{c}_i$  at each node in comparison to the D2Q9 lattice model.}

\textcolor{black}{The species in LBGK model i.e. $h_{i,O_2}$ and $h_{i,CO_2}$ is obtained as:
\begin{equation}
h_{i, \beta}(\mathbf{x} + \mathbf{c}_i \Delta t, t + \Delta t) = h_{i, \beta}(\mathbf{x}, t) - \frac{\Delta t}{\tau_{d,\beta}}\left[h_{i, \beta}(\mathbf{x}, t) - h_{i, \beta}^{\text{eq}}(C_\beta, \textbf{u})\right]
\label{eq:therpop_species}
\end{equation}
where $\tau_{d, \beta}$ is the relaxation time for species $\beta$, which is defined as $(10/3D_{\beta} + 0.5)$ and $\beta$ takes O$_2$ and CO$_2$. The equilibrium distribution functions $h_{i, \beta}^{\text{eq}}$ are defined in terms of the concentration field $C_\beta$ and velocity $\textbf{u}$ as:
\begin{equation}
h_{i, \beta}^{\text{eq}}(C_\beta, \textbf{u}) = w_i C_\beta \left(1 + \frac{\textbf{c}_i \cdot \textbf{u}} {c_s^2} \right)
\label{eq:therpopeq_species}
\end{equation}
where $w_0 = 0.4, w_1 = w_2 = w_3 = w_4 = 0.15$. The macroscopic species concentration is calculated as: 
\begin{equation}
C_\beta =  \sum_{i=0}^{4} h_{i, \beta} \label{eq:macroreacb_species }.   
\end{equation}}
The reactive boundary conditions reflect the fact that the interface between the solid and the fluid, where the chemical reaction is taking place, has both solid and fluid properties. To model the interface in CooLBM, wet-node boundary conditions are adopted at the reactive interface.

\begin{figure}[htbp]
\centering
\includegraphics[width=1\textwidth]{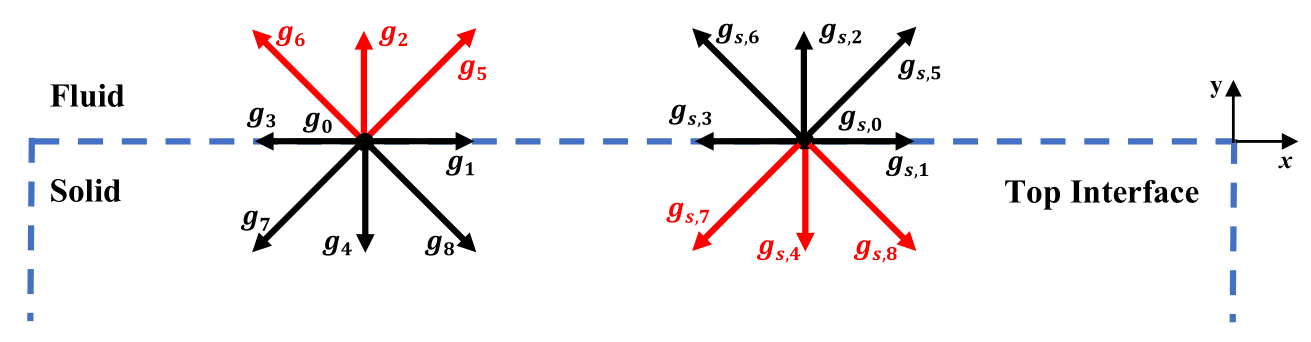}
\caption{Thermal counter slip method D2Q9 lattice at top surface reactive interface.}
\label{fig:wetnode}
\end{figure}

To illustrate the procedure, Figure \ref{fig:wetnode} shows interface nodes for the fluid and solid heat fields, where the interface normal is pointing to the $y^+$ direction from solid to fluid. The solid populations $g_{s,\alpha}$ have nine directions, three of which are coming from the fluid domain: $g_{s,7}$, $g_{s,4}$, $g_{s,8}$. These populations are unknown and must be given a suitable formula as they will stream into the solid domain in the next time step. Similarly, for the nine fluid node populations $g_{\alpha}$, three of them are coming from the solid domain: $g_2$, $g_5$, $g_6$, which are unknown.

The formulas for those unknown solid and fluid populations should reflect the temperature and heat flux continuity in the presence of the combustion process at the reactive interface.
\begin{subequations} \label{countSlipFluxTempCon}
\begin{align}
  \begin{split}
  \frac{\tau_{e,f}-0.5}{\tau_{e,f}} \left[(g_2+g_5+g_6)-(g_4+g_7+g_8)\right] 
  &= \frac{\tau_{e,s}-0.5}{\tau_{e,s}} \left[(g_{s,2}+g_{s,5}+g_{s,6})-(g_{s,4}+g_{s,7}+g_{s,8})\right] \\
  & \quad + h_r A \exp\left(\frac{-E}{R_u T}\right) C_{b,O_2},
  \end{split}
  \label{eq:countSlipFluxCon}
  \\
  \begin{split}
  \frac{1}{\rho_f c_{pf}} \left(g_0+g_1+g_2+g_3+g_4+g_5+g_6+g_7+g_8 \right) 
  &= \frac{1}{\rho_s c_{ps}} \left(g_{s,0}+g_{s,1}+g_{s,2}+g_{s,3}+g_{s,4}\right) \\
 & + \frac{1}{\rho_s c_{ps}} \left(g_{s,5}+g_{s,6}+g_{s,7}+g_{s,8} \right).
  \end{split}
  \label{eq:countSlipTempCon}
\end{align}
\end{subequations}

By defining new variables $c_{fs} = \frac{\rho_f c_{pf}}{\rho_s c_{ps}}$, $k_{sf} = \frac{\tau_{e,s} - 0.5}{\tau_{e,s}} \frac{\tau_{e,f}}{\tau_{e,f} - 0.5}$, and subtracting Eq. \eqref{eq:countSlipTempCon} from \eqref{eq:countSlipFluxCon}, the formula for the missing populations for the solid are obtained as:
\begin{equation}
\begin{split}
g_{s,\alpha}=\frac{\omega_\alpha}{(\omega_4+\omega_7+\omega_8)(k_{sf}+c_{fs})}[k_{sf}(g_{s,2}+g_{s,5}+g_{s,6})+(g_4+g_7+g_8)+& \\ (g_0+g_1+g_3+g_4+g_7+g_8)-c_{fs}(g_{s,0}+g_{s,1}+g_{s,2}+g_{s,3}+g_{s,5}+g_{s,6})+& \\ \frac{\tau_{e,f}}{\tau_{e,f}-0.5} h_r A \exp\left(\frac{-E}{R_u T}\right) C_{b,O_2}]
\end{split}
\label{eq:unknownpopTempS}
\end{equation}
where $g_{s,\alpha} = g_{s,4}$, $g_{s,7}$, and $g_{s,8}$ are the missing populations for the solid-heat domain. The coupling term of $\dot H$ contains the concentration of oxygen at the reactive boundary $C_{b,O_2}$. This term prevents explicitly solving the unknown populations without addressing the field species concentration. After solving $g_{s,\alpha}$, the missing fluid populations can be easily found as:
\begin{equation}
g_\alpha=\frac{\omega_\alpha}{(\omega_2+\omega_5+\omega_6)}[c_{fs}(g_{s,0}+g_{s,1}+g_{s,2}+g_{s,3}+g_{s,4}+g_{s,5}+g_{s,6}+g_{s,7}+g_{s,8})-(g_0+g_1+g_3+g_4+g_7+g_8)]
\label{eq:unknownpopTempF}
\end{equation}
where $g_{\alpha} = g_2$, $g_5$, and $g_6$ are the missing populations for the fluid heat domain. The thermal counter slip method shows that the unknown populations for the nodes on the reactive boundary are coupled with the concentration of oxygen at the boundary $C_{b,O_2}$. It cannot be solved without addressing the reactive boundary conditions for the species concentration fields for O$_2$ and CO$_2$.

Kang et al. \cite{Kang2010} developed several schemes for modeling heterogeneous chemical reactions using the wet-node approach. They directly calculated the species concentrations on the reactive boundary $C_{b,O_2}$ and $C_{b,CO_2}$ from the unknown species populations. Afterwards, they used these concentrations to calculate the missing species populations streaming to the fluid domain as a result of the chemical reaction. They used Non-Equilibrium Bounce Back method to determine the missing population. Based on their calculation, the concentration of oxygen at the interface $C_{b,O_2}$ can be calculated as \cite{Xu2018}:
\begin{equation}
C_{b, O_2}=\frac{2h_{4,O_2}}{\frac{1-J_{0,O_2}}{2}+\frac{\tau_{d,O_2}}{\tau_{d,O_2}-0.5}A \exp\left(\frac{-E}{R_u T}\right)}
\label{eq:wetnode9}
\end{equation}
\textcolor{black}{where $J_{0,O_2} = 0.4$. }

Equations \eqref{eq:unknownpopTempS} to \eqref{eq:wetnode9} are coupled by the transient heat term, which is a function of temperature and oxygen concentration at the reactive boundary. Therefore, these set of equations must be solved  
%(\textcolor{blue}{XXX as? XXX})
 until we obtain a converged value for both the oxygen concentration at the boundary $C_{b,O_2}$ and the temperature $T$, before advancing to the next time step. The implementation for this iterative procedure is shown in Figure \ref{fig:flowchart}. Additional information on implementing initial and boundary conditions, along with detailed formulations, can be found in ref. \cite{Alamian2024}.
\begin{figure}[htbp]
\centering
\includegraphics[width=0.8\textwidth]{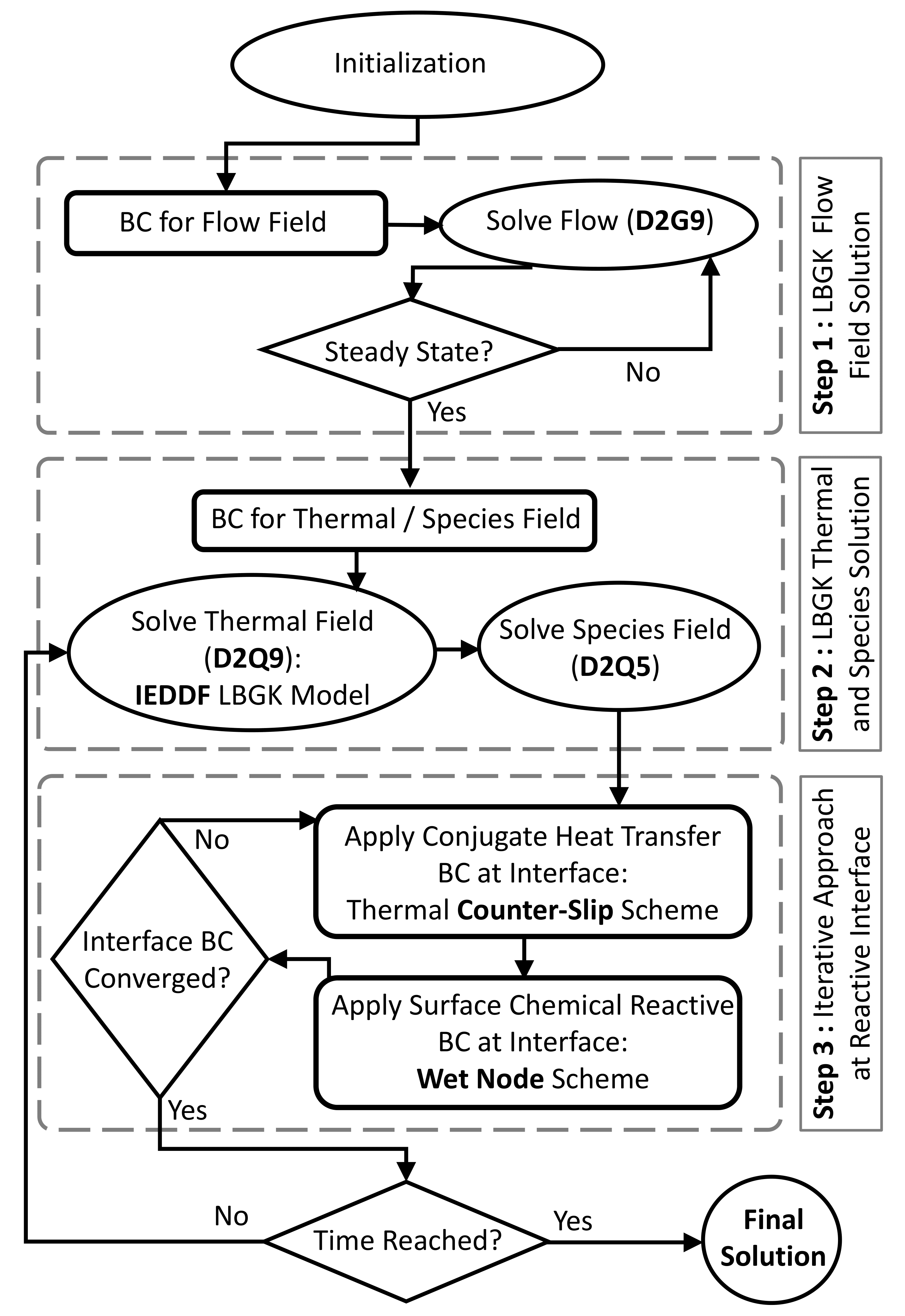}
\caption{Flowchart for solving the heterogeneous reaction, conjugate heat transfer simulation.}
\label{fig:flowchart}
\end{figure}

%%%%%%%%%%%%%%%%%%%%%%%%%%%%%%%%%%%%%%%%%%%%%%%%%%%%%%%%%%%%%%%%%%%%%%%%%%%%%%%
%%%%%%%%%%%%%%%%%%%%%%%% Parallel Computing Strategy %%%%%%%%%%%%%%%%%%%%%%%%%%
%%%%%%%%%%%%%%%%%%%%%%%%%%%%%%%%%%%%%%%%%%%%%%%%%%%%%%%%%%%%%%%%%%%%%%%%%%%%%%%
\subsection{Parallel Computing Strategy}\label{sec:parallel}
The numerical framework employed in CooLBM harnesses the power of parallel computing through industry-standard tools and libraries. It employs Threading Building Blocks (TBB) of Intel, the Standard Template Library (STL) implementation of Visual C++, and the NVIDIA HPC SDK to facilitate hardware-independent program development across many-core platforms. These tools serve as backends that optimize the execution of parallel code on both CPUs and GPUs. By leveraging TBB and the Visual C++ STL backend, CooLBM efficiently implements parallel algorithms introduced in the C++17 standard, enabling effective parallelization of computational tasks on homogeneous many-core systems. Additionally, the recent integration of the NVIDIA HPC SDK provides a GPU backend, enhancing opportunities for accelerated computations on NVIDIA GPUs.

%In our approach to parallel computing, we employ the C++ Standard Template Library (STL) and its comprehensive set of parallel algorithms. 
The functional programming style of STL, augmented with element access functions, allows for the expression of complex computations as a series of high-level algorithms operating on data containers. By adopting parallel algorithms of STL, such as \enquote{for\_each}, we seamlessly distribute computational tasks across multiple cores, thereby transforming explicit loops into concise and expressive parallelized routines. This simplifies the code structure and enhances its readability and maintainability. \textcolor{black}{A simple example of C++ code demonstrating the implementation of \enquote{for\_each} using the STL, along with its equivalent without \enquote{for\_each}, is given in \ref{appendixA}.} %\textcolor{blue}{XXX it is good to put a simplified example algorithm here! XXX}

Within CooLBM, parallel algorithms are utilized to handle computationally intensive tasks, including the computation of macroscopic variables and the evolution of fluid flow fields. The \enquote{for\_each} algorithm, when applied to data containers representing fluid elements, enables the simultaneous processing of large sets of fluid cells using multiple threads. This strategy efficiently distributes the computational workload, reducing the overall execution time and improving the responsiveness of code to changes in system configurations.

The use of TBB, the STL implementation of Visual C++, and the NVIDIA HPC SDK within the CooLBM framework provides several benefits to our numerical simulations. Firstly, TBB and the Visual C++ STL backend offer high-performance parallel algorithms and data structures that efficiently utilize the multiple cores available on CPUs. This leads to significant speedups in executing computationally intensive operations, such as the collision and streaming steps in the LBM. Furthermore, the seamless integration of the NVIDIA HPC SDK allows us to exploit the processing power of NVIDIA GPUs by offloading selected tasks to the GPU backend, thereby accelerating our simulations. These tools also enable hardware abstraction, allowing CooLBM to run efficiently on various many-core platforms and leverage available resources for improved performance and scalability.

Lattice Boltzmann method, known for its suitability for efficient parallel computing, inherently supports such parallelism. Its structure conveniently separates into a local collision step and a streaming step, which involves limited non-local memory accesses. This separation is ideal for parallel execution, $i.e.$ the local collision step can be independently processed on assigned lattice sites by multiple computing units, such as cores or processors, while the streaming step facilitates efficient communication and data sharing among these units. Capitalizing on this inherent parallelizability, LBM has demonstrated significant computational acceleration and scalability on parallel computing platforms \cite{Obrecht2013, Xian2011}.

%%%%%%%%%%%%%%%%%%%%%%%%%%%%%%%%%%%%%%%%%%%%%%%%%%%%%%%%%%%%%%%%%%%%%%%%%%%%%%%
%%%%%%%%%%%%%%%%%% Hardware Specifications and Setup %%%%%%%%%%%%%%%%%%%%%%%%%%
%%%%%%%%%%%%%%%%%%%%%%%%%%%%%%%%%%%%%%%%%%%%%%%%%%%%%%%%%%%%%%%%%%%%%%%%%%%%%%%
\subsection{Hardware Specifications and Setup}\label{sec:hardware}
In this section, we provide a detailed comparison of the technical specifications of the processors used in our performance evaluation. The processors encompass both CPUs and GPUs, each designed for specific computational tasks. \textcolor{black}{For instance, the CPU is used to manage the primary control and sequential processes, while the GPU handles highly parallel, computationally intensive tasks}. %(XXX what do you mean? XXX). 
The CPUs examined in this CooLBM code development study include the Intel® Core™ i7-10700K and the Intel® Core™ i9-10900K, featuring 8 and 10 cores, respectively, and supporting up to 16 and 20 threads for parallel processing. Additionally, the 3rd Gen Intel Xeon Scalable processors are also utilized for applications requiring more CPU cores. The GPUs evaluated are the NVIDIA RTX A2000, RTX A4000, NVIDIA P100, and NVIDIA V100, each equipped with varying numbers of CUDA and Tensor cores and different memory sizes.

We conducted performance evaluations for the CooLBM code on these processors. The Intel® Core™ CPUs demonstrated their prowess in general-purpose computation, while the 3rd Gen Intel Xeon scalable processors offered scalable configurations ranging from 2 to 128 cores, ideal for data center applications. Among the GPUs, the NVIDIA RTX A2000 and RTX A4000 impressed with their CUDA and Tensor cores, offering 4GB and 16GB of GDDR6 memory, respectively. The NVIDIA P100 GPU showcased the CUDA processing capability
%with 24GB of GDDR5 and 
 with 16GB of HBM2 memory, respectively. The high-performance NVIDIA V100 GPU emerged as a powerful contender, featuring 5120 CUDA cores and 640 Tensor cores, and either 16GB or 32GB of HBM2 memory. A summary of these information are presented in Table \ref{tab:processor_specs}.

Throughout our evaluations, we ensure that the CooLBM code was meticulously tested on each processor and GPU. This is done by carefully selecting a range of test cases to assess their computational efficiency and parallelizability. The comprehensive comparison of these hardware configurations offers valuable insights into the performance characteristics of the CooLBM code, assisting us in identifying the most suitable processors for efficient LBM simulations. By understanding the specific strengths and capabilities of each processor, we aim to optimize our simulations and achieve more accurate and faster results for various scientific applications.

\begin{table}[htbp]
\centering
\caption{Comparison of Processor Technical Specifications}
\label{tab:processor_specs}
\begin{adjustbox}{width=\textwidth}
\begin{tabular}{|l|l|c|c|c|c|c|}
\hline
\multirow{2}{*}{Processor Name} & \multirow{2}{*}{Type} & \multirow{2}{*}{Cores/Threads} & \multirow{2}{*}{CUDA Cores} & \multirow{2}{*}{Tensor Cores} & \multirow{2}{*}{Memory Type} & \multirow{2}{*}{Memory Size (GB)} \\
 & & & & & & \\
\hline
Intel® Core™ i7-10700K CPU @ 3.80GHz & CPU & 8 / 16 & N/A & N/A & DDR4 & 32 \\
\hline
Intel® Core™ i9-10900K CPU @ 3.70GHz & CPU & 10 / 20 & N/A & N/A & DDR4 & 64 \\
\hline
3rd Gen Intel Xeon Scalable (Ice Lake 8375C) & CPU & 2 / 4 / 8 / 16 / 32 / 64/ 128 & N/A & N/A & DDR4 & Up to 64 \\
\hline
NVIDIA RTX A2000 & GPU & N/A & 3328 & 104 & GDDR6 & 4 \\
\hline
NVIDIA RTX A4000 & GPU & N/A & 6144 & 192 & GDDR6 & 16 \\
\hline
NVIDIA P100 GPU & GPU & N/A & 3584 & N/A & HBM2 & 16 \\
\hline
NVIDIA V100 GPU & GPU & N/A & 5120 & 640 & HBM2 & 16 / 32 \\
\hline
\end{tabular}
\end{adjustbox}
\end{table}

%%%%%%%%%%%%%%%%%%%%%%%%%%%%%%%%%%%%%%%%%%%%%%%%%%%%%%%%%%%%%%%%%%%%%%%%%%%%%%%
%%%%%%%%%%%%%%%%%%%%%%% Validation and Test Cases %%%%%%%%%%%%%%%%%%%%%%%%%%%%%
%%%%%%%%%%%%%%%%%%%%%%%%%%%%%%%%%%%%%%%%%%%%%%%%%%%%%%%%%%%%%%%%%%%%%%%%%%%%%%%

\section{Validation and Test Cases}\label{sec:validation}
This section presents the assessment of the CooLBM performance of code and accuracy by employing well-known benchmark test cases. These test cases have been widely recognized and utilized in the scientific community to validate various aspects of fluid dynamics simulations. Our validation efforts encompass a range of scenarios, including single-phase and multi-phase simulations, as well as simulations involving multiple fluid components. Additionally, we extend our validation to include reactive interfaces, where the behavior of chemical reactions at the fluid-solid interface is a key focus. By subjecting the CooLBM code to these established test cases, we aim to assess its ability to reliably reproduce the expected behavior and validate its applicability across different fluid systems and complex phenomena.

%%%%%%%%%%%%%%%%%%%%%%%%%%%%%%%%%%%%%%%%%%%%%%%%%%%%%%%%%%%%%%%%%%%%%%%%%%%%%%%
%%%%%%%%%%%%%%%%%%%%%%%%% Single-Phase Test Cases %%%%%%%%%%%%%%%%%%%%%%%%%%%%%
%%%%%%%%%%%%%%%%%%%%%%%%%%%%%%%%%%%%%%%%%%%%%%%%%%%%%%%%%%%%%%%%%%%%%%%%%%%%%%%
\subsection{Single-Phase Test Cases}\label{sec:singlevalid}
This subsection focuses on the validation of the CooLBM code using three well-established test cases, namely, Couette flow, Poiseuille flow and Taylor-Green vortex. These test cases have been extensively studied and widely adopted as standard benchmarks for evaluating the accuracy and performance of computational fluid dynamics (CFD) codes. By applying these single-phase test cases, we aim to assess the ability of the CooLBM code to accurately model the boundary conditions and capture fundamental flow phenomena, such as shear-driven flows, pressure-driven flows, and complex vortex dynamics. Through detailed comparisons between simulation results and analytical or numerical solutions, we can quantitatively evaluate the ability of code to reproduce key flow features and validate its robustness and accuracy in single-phase scenarios. 

%%%%%%%%%%%%%%%%%%%%%%%%%%%%%%%%%%%%%%%%%%%%%%%%%%%%%%%%%%%%%%%%%%%%%%%%%%%%%%%
%%%%%%%%%%%%%%%%%%%%%%%%%%%%%%% Couette Flow %%%%%%%%%%%%%%%%%%%%%%%%%%%%%%%%%%
%%%%%%%%%%%%%%%%%%%%%%%%%%%%%%%%%%%%%%%%%%%%%%%%%%%%%%%%%%%%%%%%%%%%%%%%%%%%%%%
\subsubsection{Couette Flow}\label{sec:Couette}
Couette flow refers to the viscous flow between two parallel plates, where the upper plate moves horizontally. This causes a shear stress on the fluid and results in a flow with a linear velocity profile. This problem offers an analytical solution that serves as an ideal validation case for flow simulation codes. In Couette flow, the velocity profile exhibits a linear variation ranging from zero at the fixed bottom boundary to a maximum velocity at the moving top boundary, as illustrated in Figure \ref{fig:couette_flow}.

\begin{figure}[htbp]
\begin{center}
\includegraphics[trim=7.5cm 3.7cm 12.0cm 2.4cm, clip=true, width=0.6\textwidth]{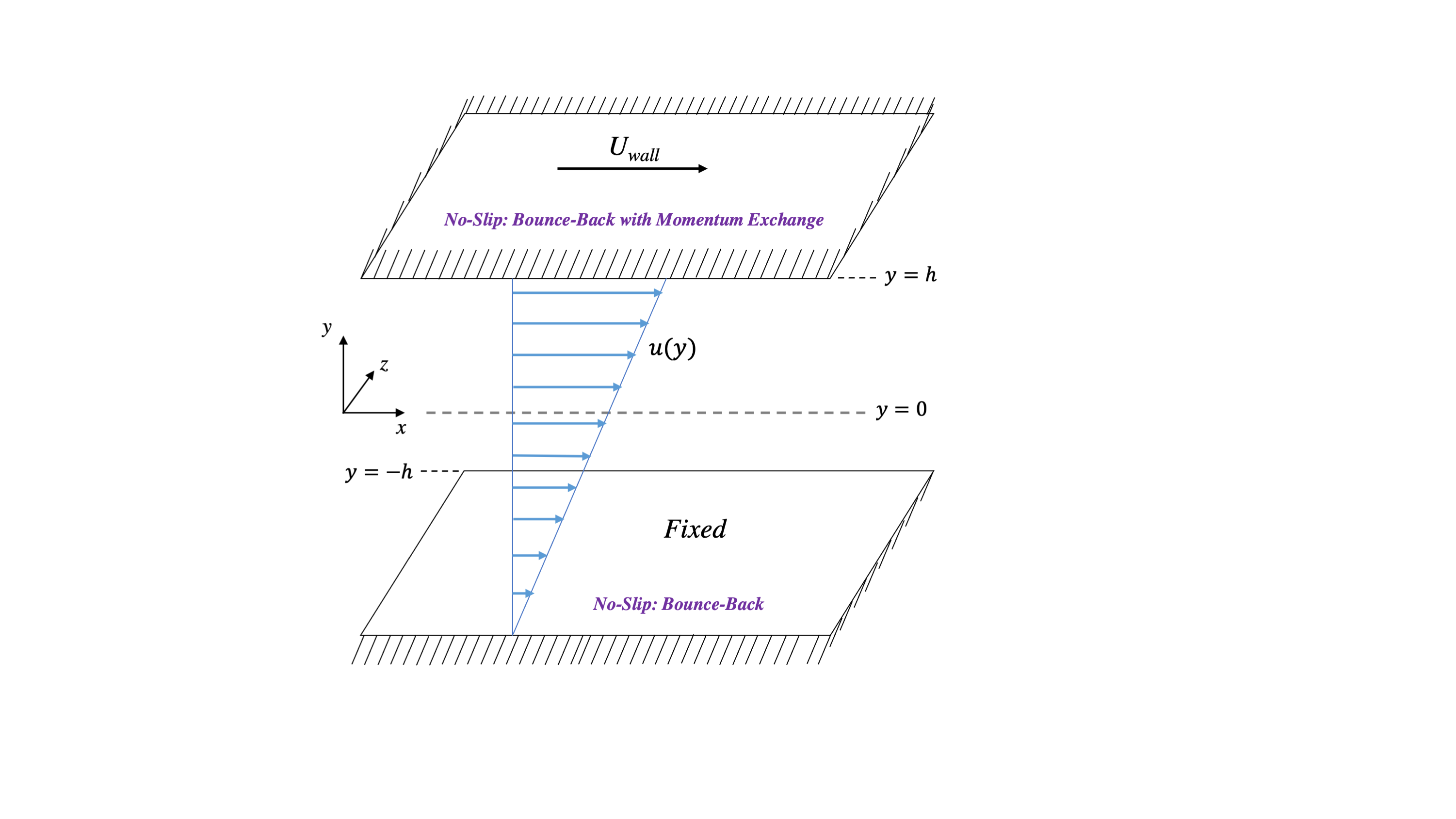}
\caption{\label{fig:couette_flow} {\small Geometry and boundary condition definition for the Couette flow problem.}}
\end{center}
\end{figure}

The analytical solution for the velocity profile in Couette flow is given by \cite{Kreith1999}:

\begin{equation}
u = \frac{U_{\text{wall}}}{2h}(y+h) \quad -h \leq y \leq +h \,,
\label{eq:couette_velocity_profile}
\end{equation}
\textcolor{black}{where $U_{\text{wall}}$ is the velocity of the top wall, and $h$ is the half-width of the channel.}
This test case is simulated in three-dimension using CooLBM. The computational domain is set up with periodic boundaries for the side walls, and inlet and outlet. The top boundary is considered as a no-slip boundary, where a horizontal lattice velocity of $U_{\text{wall}} = 2\times10^{-2}$ is imposed. In the LBM framework, this boundary is treated as a Bounce Back boundary, incorporating an additional term to capture momentum exchange between the moving top plate and the adjacent fluid. Conversely, the bottom boundary is treated as a no-slip boundary without any momentum exchange. Therefore, a simple Bounce Back boundary condition is applied without any extra terms. \textcolor{black}{The specified flow parameters for the simulation include the Reynolds  number ($Re$), the ratio between inertial and viscous forces, of 10:}

\begin{equation}
\textcolor{black}{Re = \frac{U_{\text{wall}} N_y}{\nu}}
\label{eq:reynolods_couette}
\end{equation}
\textcolor{black}{where $\nu$ is the kinematic viscosity of the fluid in lattice units.} Flow simulation is conducted using a lattice resolution of $N_x\times N_y \times N_z = 15 \times 15 \times 15$ nodes for the validation study. Interested readers are refereed to our recent work for details of all conversion from the lattice unit to the physical space \cite{stockinger2024lattice}.
\begin{lstlisting}[frame=none, label={lst:Couette_flow}]
//CooLBM 3D Couette flow canonical C++ code
#include "couette3D.h"
using namespace std;

string problem = "couette3D";

int main() {   
    if (problem == "couette3D") {
        couette3D();  //Call couette3D function 
    }   
    return 0;
}
//Definition of function "couette3D"
//Execute an iteration (collision and streaming) on a cell called through "for_each"
void operator() () {
    macro();  //Calculate macroscopic variables
    collideBgk(); //Perform BGK collision step
    stream();     //Perform streaming step
}
void couette3D()
{
    std::ifstream coufile("../apps/Config_Files/config_couette3D.txt"); //Input file
    //Read data from the input file
    ......................................
    //Variable initialization for the LBM operations
    ......................................
    //Define simulation geometry
    inigeom_couette3D();
    //Start main time step loop
    for (int time_iter = 0; time_iter < max_time_iter; ++time_iter) {
        //Save data for post-processing
        ..............................
        //With the double-population scheme, collision and streaming are executed in the following loop.
        for_each(execution::par_unseq, lattice, lattice + dim.nelem, lbm);
    }
}

\end{lstlisting}

The canonical form of the code is shown in the Listing~\ref{lst:Couette_flow}. Post-processing of the simulation results included generating three-dimensional velocity contours for Couette flow, as depicted in Figure \ref{fig:velocity_contour_couette}, providing visual representations of the flow field. The velocity profiles obtained from CooLBM simulations are compared with the analytical solution. As shown in Figure \ref{fig:couette_analytic}, the simulations accurately replicated the analytical solution using a lattice resolution of $N_x\times N_y \times N_z = 15 \times 15 \times 15$, highlighting the efficiency of CooLBM and the accuracy of the implemented governing equations and boundary conditions.

\begin{figure}[htbp]
\begin{center}

\begin{subfigure}[b]{0.49\textwidth}
	\includegraphics[width=1\textwidth]{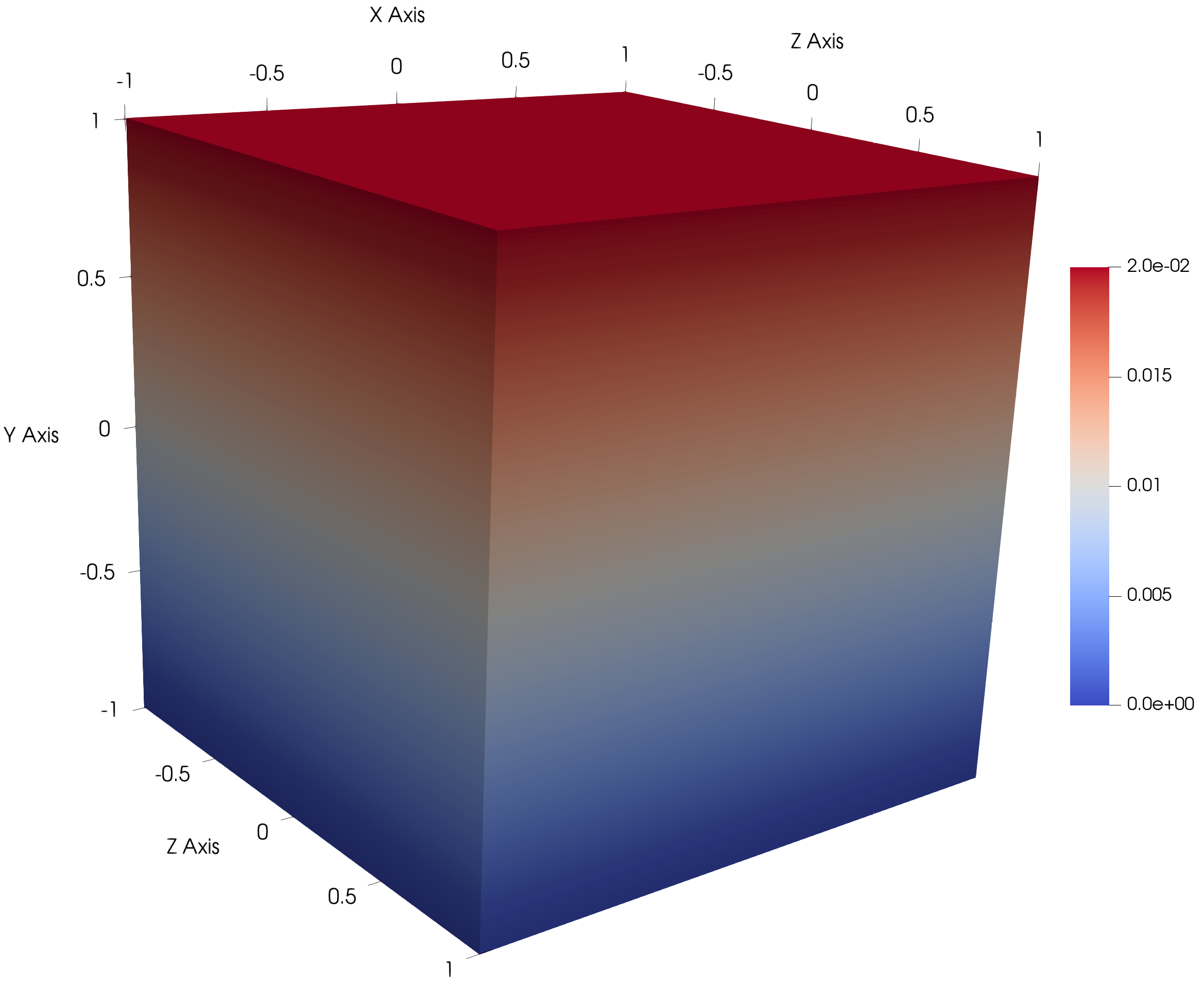}
	\caption{}
	\label{fig:velocity_contour_couette}
\end{subfigure}
%\hspace{5mm}
\begin{subfigure}[b]{0.49\textwidth}
	\includegraphics[width=1\textwidth]{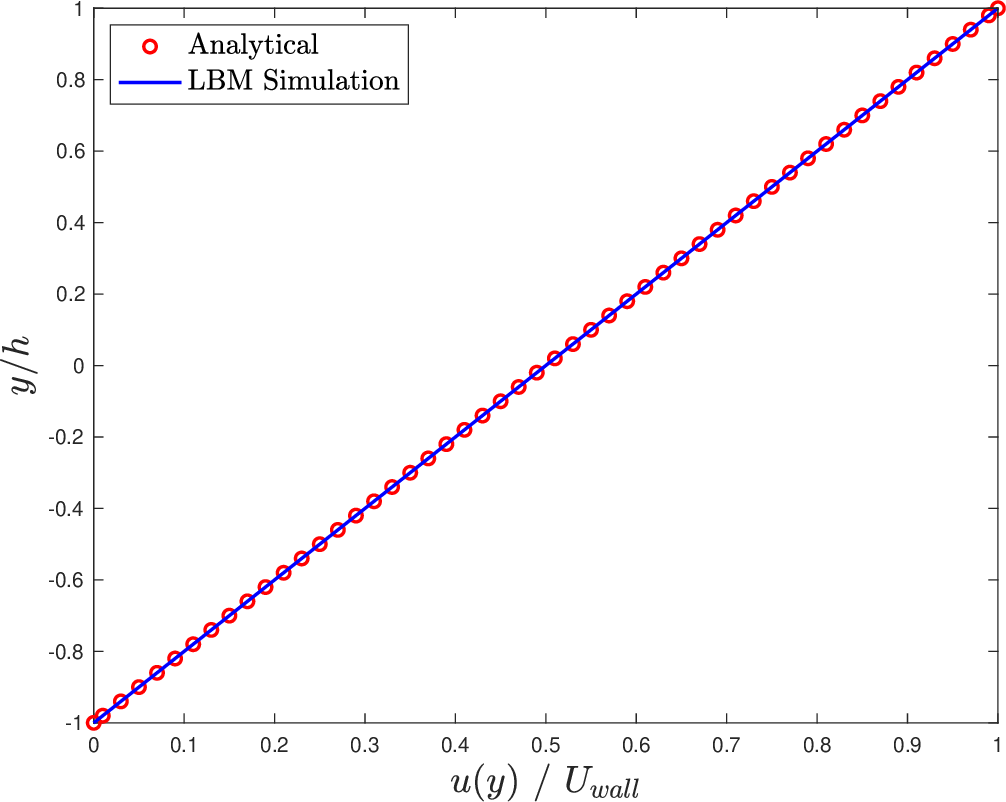}
	\caption{}
	\label{fig:couette_analytic}
\end{subfigure}

\caption{\label{fig:couette_valid} {\small Results for $x$-direction velocity in three-dimensional Couette flow: (a) velocity contour, and (b) validation against analytical solution.}}

\end{center}
\end{figure}

%%%%%%%%%%%%%%%%%%%%%%%%%%%%%%%%%%%%%%%%%%%%%%%%%%%%%%%%%%%%%%%%%%%%%%%%%%%%%%%
%%%%%%%%%%%%%%%%%%%%%%%%%%%%% Poiseuille Flow %%%%%%%%%%%%%%%%%%%%%%%%%%%%%%%%%
%%%%%%%%%%%%%%%%%%%%%%%%%%%%%%%%%%%%%%%%%%%%%%%%%%%%%%%%%%%%%%%%%%%%%%%%%%%%%%%
\subsubsection{Poiseuille Flow}\label{Poiseuille}
The flow between two parallel plates could also involve the Poiseuille problem, which represents a pressure-induced flow unlike the shear-induced flow observed in the Couette problem. Poiseuille flow is characterized by a hyperbolic velocity profile, with the maximum velocity developing along the centerline between the plates, as illustrated in Figure \ref{fig:poiseuille_flow}.

\begin{figure}[htbp]
\begin{center}
\includegraphics[width=0.6\textwidth]{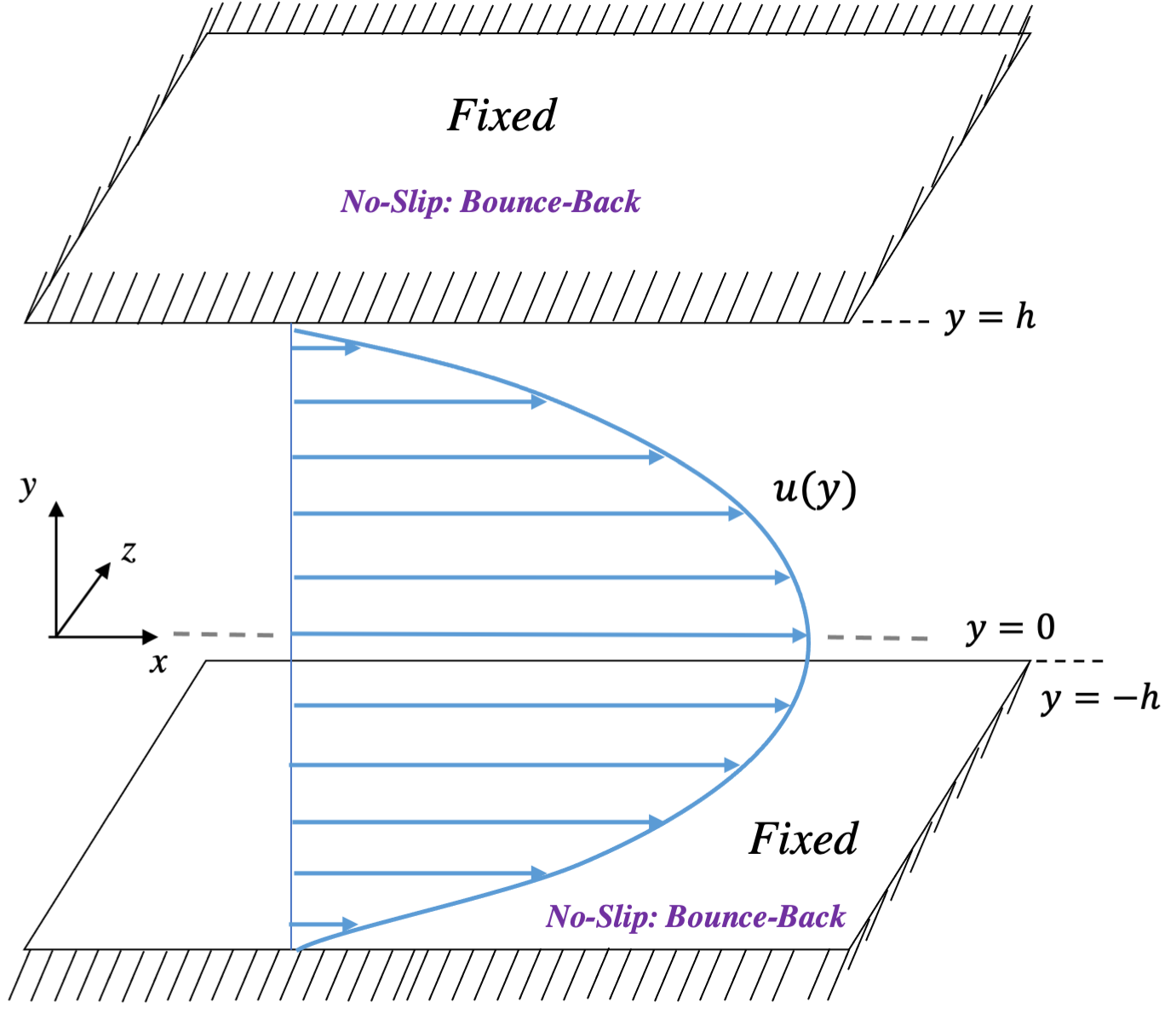}
\caption{\label{fig:poiseuille_flow} {\small Geometry and boundary condition definition for the Poiseuille flow problem.}}
\end{center}
\end{figure}

Analytical solutions are available to validate the results obtained by CooLBM. The analytical solution for the velocity profile and value of the maximum velocity in the Poiseuille flow are, respectively, as follows \cite{Kreith1999}:

\begin{equation}
u (y)= \frac{{h^2}}{{\mu}} \frac{dp}{dx}\left(1 - \frac{{y^2}}{{h^2}}\right)
\end{equation}

\begin{equation}
u_{\text{max}} = \frac{{h^2}}{{\mu}}\frac{dp}{dx}.
\end{equation}

In the given equations, the variable $u (y)$ represents the $x$-component of the velocity field at a specific $y$-position in the flow. The term ${dp}/{dx}$ denotes the pressure gradient, which indicates the rate of change of pressure with respect to the distance $x$, and $\mu$ refers to the dynamic viscosity of the fluid. 
The flow parameters in the lattice units are specified as follows: Reynolds number $Re = 10$, lattice velocity $u_{\text{max}} = 2\times10^{-2}$, and pressure gradient $dp/dx = 5\times10^{-5}$. The boundary conditions used in the simulation are similar to those employed in the Couette problem. Except, the no-slip condition at the top boundary is modeled as a simple Bounce Back boundary without any momentum exchange terms.

\begin{figure}[htbp]
\begin{center}
\begin{subfigure}[b]{0.49\textwidth}
	\includegraphics[width=1\textwidth]{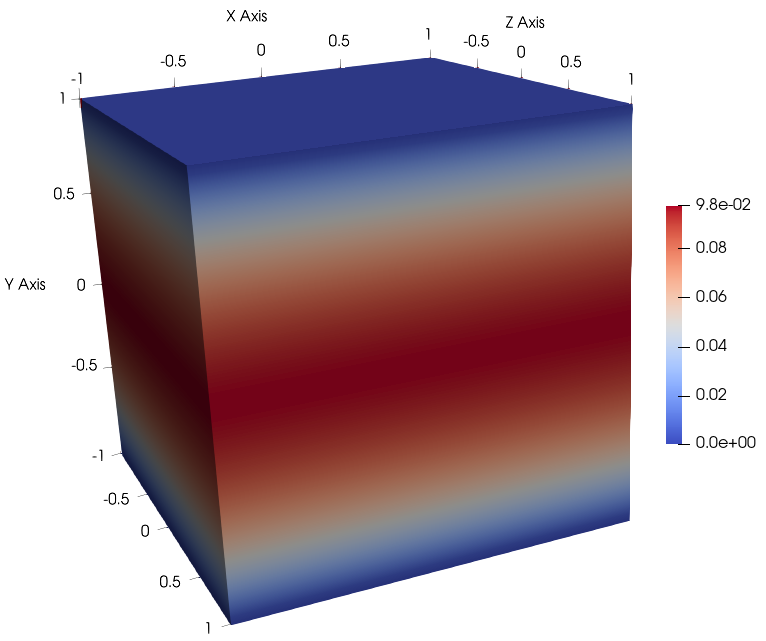}
	\caption{}
	\label{fig:velocity_contour_poiseuille}
\end{subfigure}
%\hspace{5mm}
\begin{subfigure}[b]{0.49\textwidth}
	\includegraphics[width=1\textwidth]{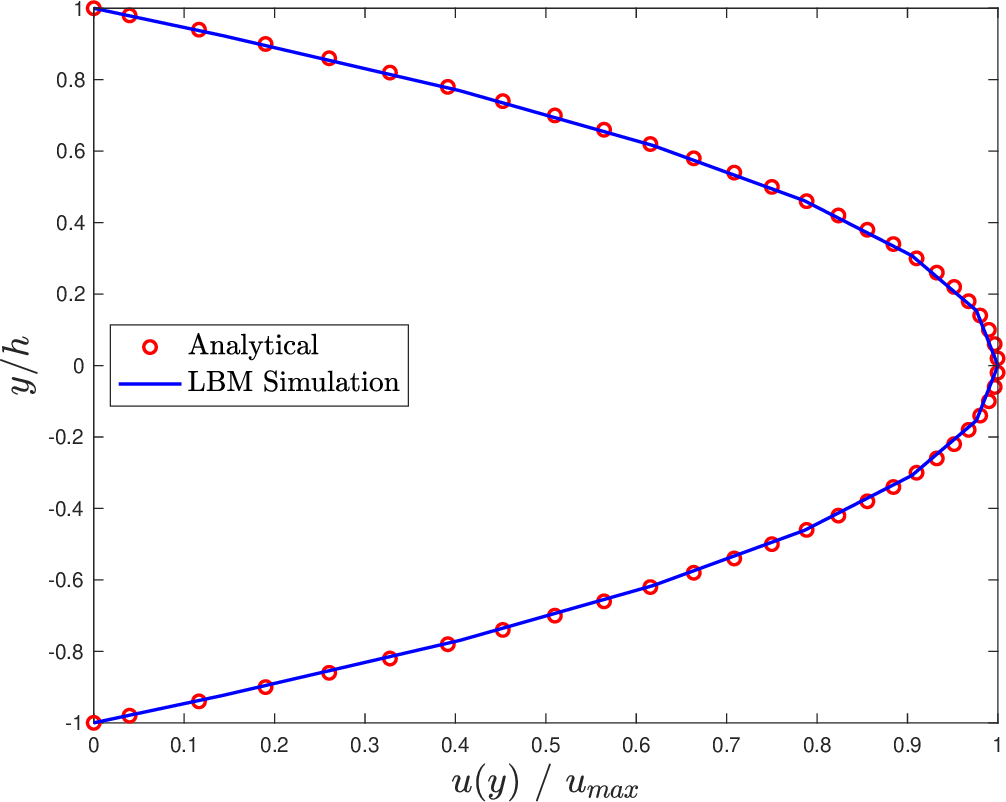}
	\caption{}
	\label{fig:poiseuille_analytic}
\end{subfigure}

\caption{\label{fig:poiseuille_valid} {\small Results for $x$-direction velocity in three-dimensional Poiseuille flow: (a) velocity contour, and (b) validation against analytical solution.}}

\end{center}
\end{figure}

Figure \ref{fig:velocity_contour_poiseuille} illustrates the contour for the $x$-component of the velocity field in the three-dimensional Poiseuille flow. Figure \ref{fig:poiseuille_analytic} demonstrates the validation of the CooLBM simulation results against the analytical solution. Remarkably, CooLBM is able to accurately replicate the hyperbolic velocity profile predicted by the analytical solution, even with the utilization of a coarse lattice resolution of $N_x\times N_y \times N_z = 15 \times 15 \times 15$.

%%%%%%%%%%%%%%%%%%%%%%%%%%%%%%%%%%%%%%%%%%%%%%%%%%%%%%%%%%%%%%%%%%%%%%%%%%%%%%%
%%%%%%%%%%%%%%%%%%%%%%%%%% Taylor-Green Vortex %%%%%%%%%%%%%%%%%%%%%%%%%%%%%%%%
%%%%%%%%%%%%%%%%%%%%%%%%%%%%%%%%%%%%%%%%%%%%%%%%%%%%%%%%%%%%%%%%%%%%%%%%%%%%%%%

\subsubsection{Taylor-Green Vortex}\label{taylor}
 The Taylor-Green Vortex flow offers a test case to investigate the generation of various turbulence scales through three-dimensional vortex stretching, leading to turbulence development. To ensure the robustness of the obtained results, an extensive examination is performed, comparing them with the findings of Abdelsamie et al. \cite{Abdelsamie2021}, which encompasses the direct numerical simulation (DNS) of Navier-Stokes equations.

In this problem, it is assumed that all faces are periodic. The initial conditions for velocity and pressure distributions throughout the computational domain are defined as:

\begin{subequations}\label{taylorgreen}
\begin{gather}
u = U_0 \cos\left(\frac{2\pi x}{L}\right) \sin\left(\frac{2\pi y}{L}\right) \sin\left(\frac{2\pi z}{L}\right)
\label{eq:taylorgreenU}\\
v = -U_0 \sin\left(\frac{2\pi x}{L}\right) \cos\left(\frac{2\pi y}{L}\right) \sin\left(\frac{2\pi z}{L}\right)
\label{eq:taylorgreenV}\\
w = 0
\label{eq:taylorgreenW}\\
P = P_0\frac{\rho_0 U_0^2}{16} \left[\cos\left(\frac{2\pi z}{L}\right)\right]\left[\cos\left(\frac{2\pi x}{L}\right)+\cos\left(\frac{2\pi y}{L}\right)\right]
\label{eq:taylorgreenP}
\end{gather}
\end{subequations}	
where $u$, $v$ and $w$, respectively, represent the velocity component in the $x-$, $y-$ and $z-$ directions, $L$ is the domain size, and $P$ represents the pressure. 
%(XXX define all initial values with $X_0$ XXX). 
\textcolor{black}{A moderately high Reynolds number of $Re = \frac{U_0}{\nu k} = 1600$, where $k$ is the wave number and is defined as $\frac{2 \pi}{L}$. The simulation time of $20 t_{\text{ref}}$ are selected for the study.} %This corresponds to the lattice velocity of $u = 0.05$. 
%XXX also length and mu XXX.
Here, $t_{\text{ref}}$ is defined as $t_{\text{ref}} = \frac{L}{U_0}$, where $L$ represents a characteristic length with a value of 128, and $U_0$ denotes the reference velocity. \textcolor{black}{The reference velocity $U_0$, density $\rho_0$, and pressure $P_0$ are 0.05, 1.0, and 1.0, respectively}. The validation study involves performing the flow simulation with a lattice resolution of $N_x\times N_y \times N_z = 128 \times 128 \times 128$. 

Figure \ref{fig:vortex_evolution} presents the three-dimensional velocity contours obtained by CooLBM, illustrating the temporal evolution of the Taylor-Green vortex. This figure consists of four frames, each depicts the progressive development of small-scale instabilities driven by viscosity, which leads to the emergence of a highly turbulent and chaotic flow. %Notably, regions of intense vorticity can be observed away from the walls (XXX but there is no wall!! XXX), as highlighted by the post-processing analysis conducted on the CooLBM results. 

\begin{figure*}[t!]
    \subfloat[$t=0$]{%
        \includegraphics[width=.48\linewidth]{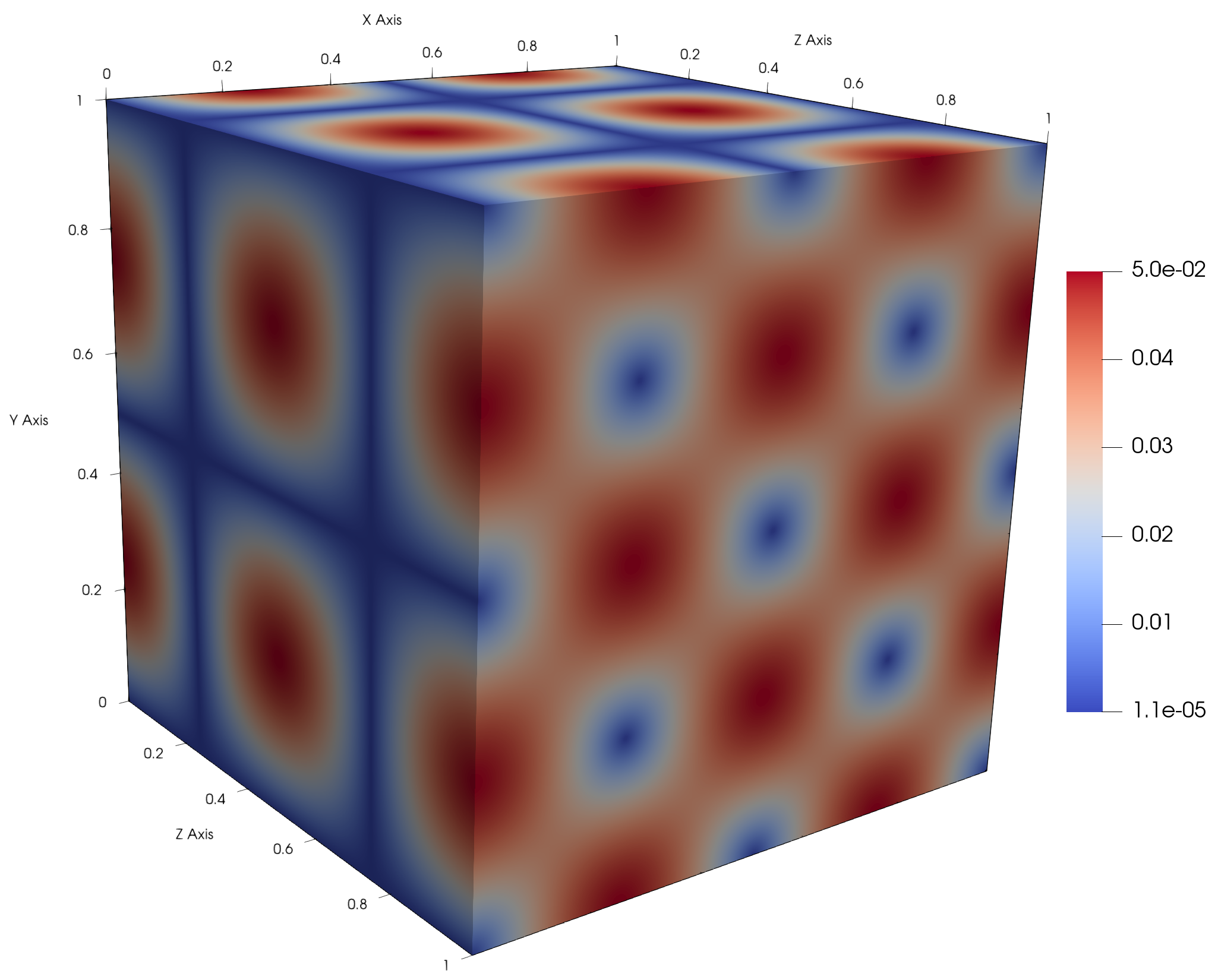}%
        \label{vortex_evolution:a}%
    }\hfill
    \subfloat[$t=7.5 t_\text{ref}$]{%
        \includegraphics[width=.48\linewidth]{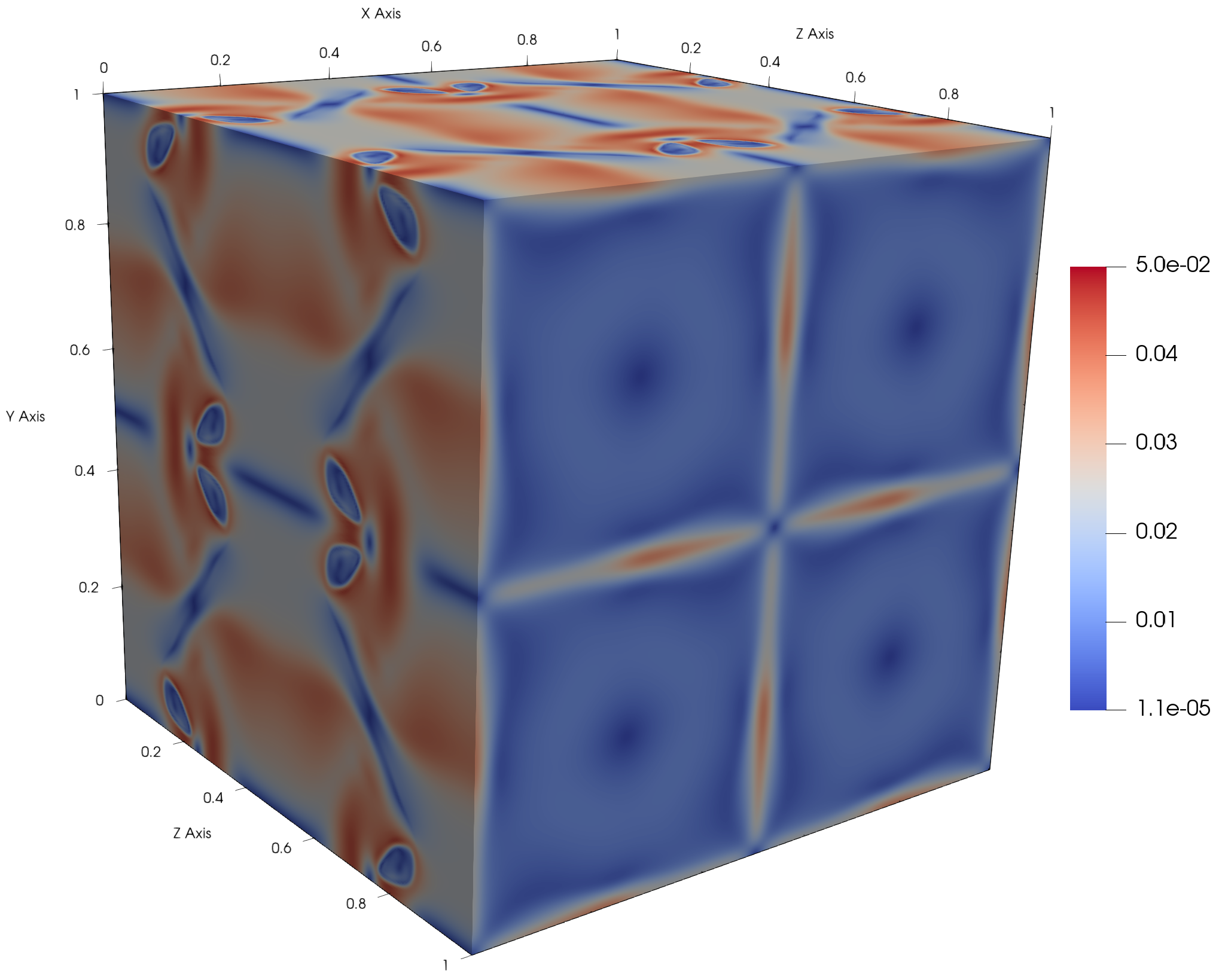}%
        \label{vortex_evolution:b}%
    }\\
    \subfloat[$t=15 t_\text{ref}$]{%
        \includegraphics[width=.48\linewidth]{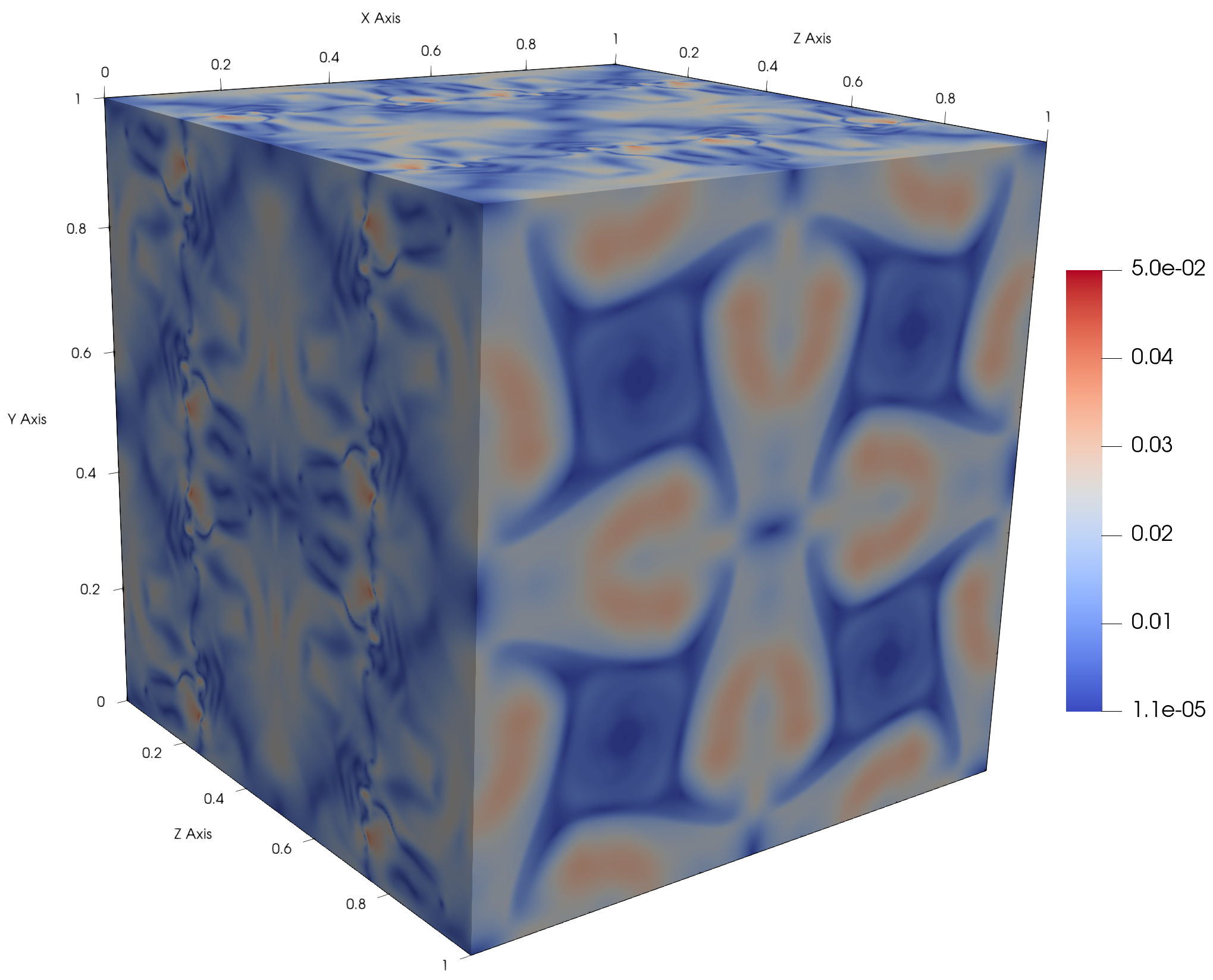}%
        \label{vortex_evolution:c}%
    }\hfill
    \subfloat[$t=20 t_\text{ref}$]{%
        \includegraphics[width=.48\linewidth]{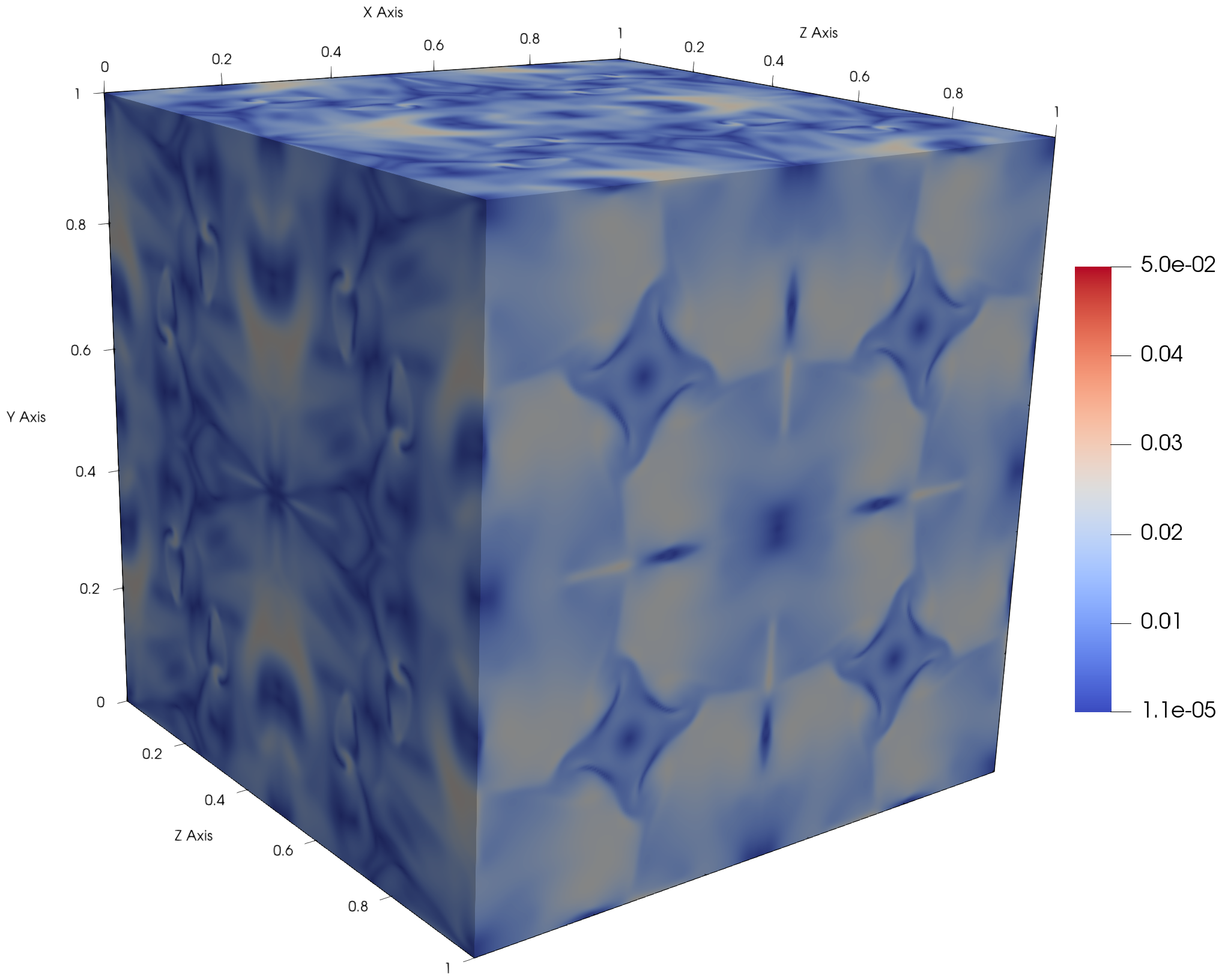}%
        \label{vortex_evolution:d}%
    }
    \caption{\label{fig:vortex_evolution} {\small Four frames (a) to (d) captured from different time steps showing the evolution of small scales of instabilities driven by viscosity.}}
\end{figure*}

\textcolor{black}{Figure \ref{fig:taylorgreen_valid} presents a comparison between the current simulation results with the direct numerical simulation (DNS) reported  by Abdelsamie et al. \cite{Abdelsamie2021}.} 
%(XXX using which method?XXX).
Specifically, it examines the temporal evolution of normalized turbulent kinetic energy, as presented in Figure \ref{fig:kineticenergy} and the normalized dissipation rate of turbulent kinetic energy, as shown in Figure \ref{fig:DissipationRate}. The results depicted in Figure \ref{fig:taylorgreen_valid} highlight a very good agreement between the current LBM simulations and the available DNS results in literature. 

As can be seen, CooLBM exhibits a level of accuracy akin to DNS, effectively capturing the dynamic changes in turbulent kinetic energy and the dissipation rate within the Taylor-Green Vortex. Notably, CooLBM accurately represents the peak dissipation of turbulent energy and successfully reproduces the small-scale instabilities induced by viscosity. This leads to the emergence of a highly chaotic and turbulent flow, featuring prominent vorticity patches located away from the boundaries, as revealed through post-processing analysis.

\begin{figure}[htbp]
\begin{center}

\begin{subfigure}[b]{0.49\textwidth}
	\includegraphics[width=1\textwidth]{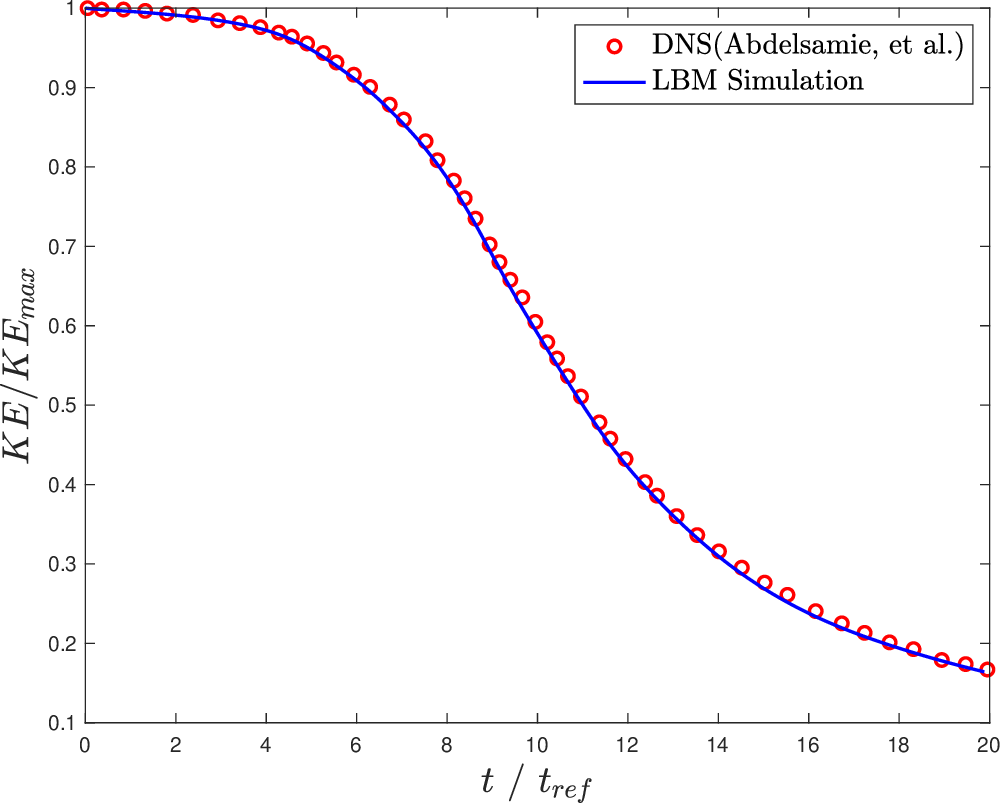}
	\caption{}
	\label{fig:kineticenergy}
\end{subfigure}
%\hspace{5mm}
\begin{subfigure}[b]{0.49\textwidth}
	\includegraphics[width=1\textwidth]{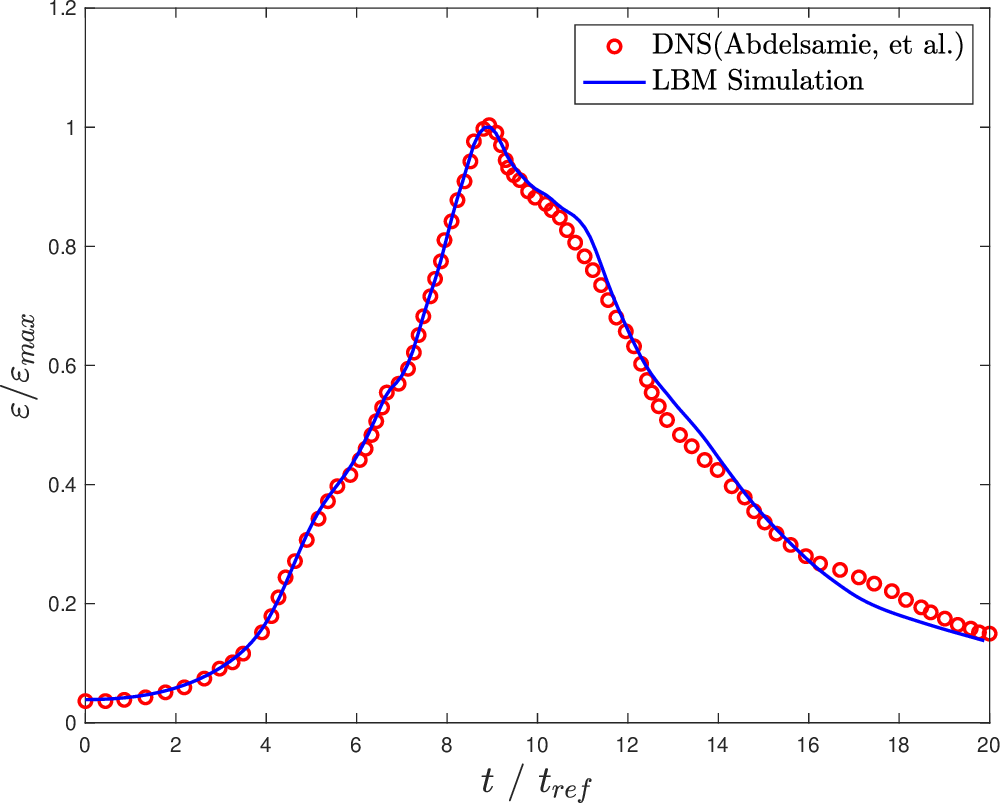}
	\caption{}
	\label{fig:DissipationRate}
\end{subfigure}

\caption{\label{fig:taylorgreen_valid} {\small Comparison of CooLBM $128 \times 128 \times 128$ Lattice results with Abdelsamie et al. \cite{Abdelsamie2021}: (a) Temporal evolution of normalized turbulent kinetic energy, and (b) Temporal evolution of dissipation rate of turbulent kinetic energy.}}

\end{center}
\end{figure}

%%%%%%%%%%%%%%%%%%%%%%%%%%%%%%%%%%%%%%%%%%%%%%%%%%%%%%%%%%%%%%%%%%%%%%%%%%%%%%%
%%%%%%%%%%%%%%%%%% Multi-Phase Multi-Component Test Cases %%%%%%%%%%%%%%%%%%%%%
%%%%%%%%%%%%%%%%%%%%%%%%%%%%%%%%%%%%%%%%%%%%%%%%%%%%%%%%%%%%%%%%%%%%%%%%%%%%%%%
\subsection{Multi-Phase and Multi-Component Test Cases}\label{sec:multivalid}
This section focuses on the validation of the CooLBM code in the context of multi-phase multi-component test cases. The presnted test cases are phase separation, cavitation, Rayleigh-Taylor instability (RTI), and contact angle problems. These test cases have been selected to assess the CooLBM capabilities in accurately capturing complex phenomena related to phase transitions, interface dynamics, and interfacial phenomena. By successfully validating CooLBM with these multi-phase multi-component test cases, we will demonstrate its suitability for a wider range of practical applications.

%\textcolor{blue}{XXX explain why from here are results are 2D! XXX you can use argument for comparison and potential lack of memory in 3D for these cases! XXX}

%%%%%%%%%%%%%%%%%%%%%%%%%%%%%%%%%%%%%%%%%%%%%%%%%%%%%%%%%%%%%%%%%%%%%%%%%%%%%%%
%%%%%%%%%%%%%%%%%%%%%%%%%%%%% Phase Separation %%%%%%%%%%%%%%%%%%%%%%%%%%%%%%%%
%%%%%%%%%%%%%%%%%%%%%%%%%%%%%%%%%%%%%%%%%%%%%%%%%%%%%%%%%%%%%%%%%%%%%%%%%%%%%%%
\subsubsection{Phase Separation}\label{sec:phaseseparation}
%Fluid mixing (\textcolor{blue}{XXX why we talk about mixing in phase separation? why not separation? XXX}) is a fundamental unit operation in industrial process engineering, which involves handling heterogeneous physical systems, such as gases or liquids, with the aim of achieving a more homogeneous composition. The process of fluid mixing is commonly employed to enhance heat and/or mass transfer between different components within a single system. It can encompass various scenarios, including the incorporation of one immiscible liquid into another immiscible liquid, such as in the case of emulsions ($e.g.$ water in oil or oil in water), or the introduction of gas into a liquid phase.
\textcolor{black}{Phase separation occurs when a homogeneous mixture spontaneously separates into distinct phases with different physical properties, such as density, under certain thermodynamic conditions. This process is driven by the minimization of free energy, resulting in the formation of interfaces between phases. Phase separation in single-component multiphase systems is indispensable in engineering, playing a crucial role in material synthesis, separation technologies, and multiphase flow systems. Understanding and controlling it is vital for optimizing the performance and stability of processes involving emulsions, foams, and colloidal dispersions, with direct implications for designing systems in energy storage, catalysis, and fluid transport.}

CooLBM's multi-phase solver incorporates the theoretical framework described in Eq. \eqref{eq:1.12} to simulate the mixing of two immiscible fluid phases within a randomly distributed mixture in a container. Over time, the dispersed fluid forms larger droplets, indicating the progress of the mixing process. In this simulation, the interaction coefficient $G$ is set to $-5.0$, while both $\psi_0$ and $\rho_0$ are set to $1.0$. The simulations are conducted for the Reynolds number of $Re=25$, indicating a lattice velocity of $u_{lb}=0.05$. The lattice resolution for this two-dimensional simulation is set to be $N_x \times N_y=128 \times 128$.

\begin{figure*}[t!]
    \subfloat[$t=0s$]{%
        \includegraphics[trim=1.8cm 0.5cm 2.7cm 0.5cm, clip=true, width=0.48\linewidth]{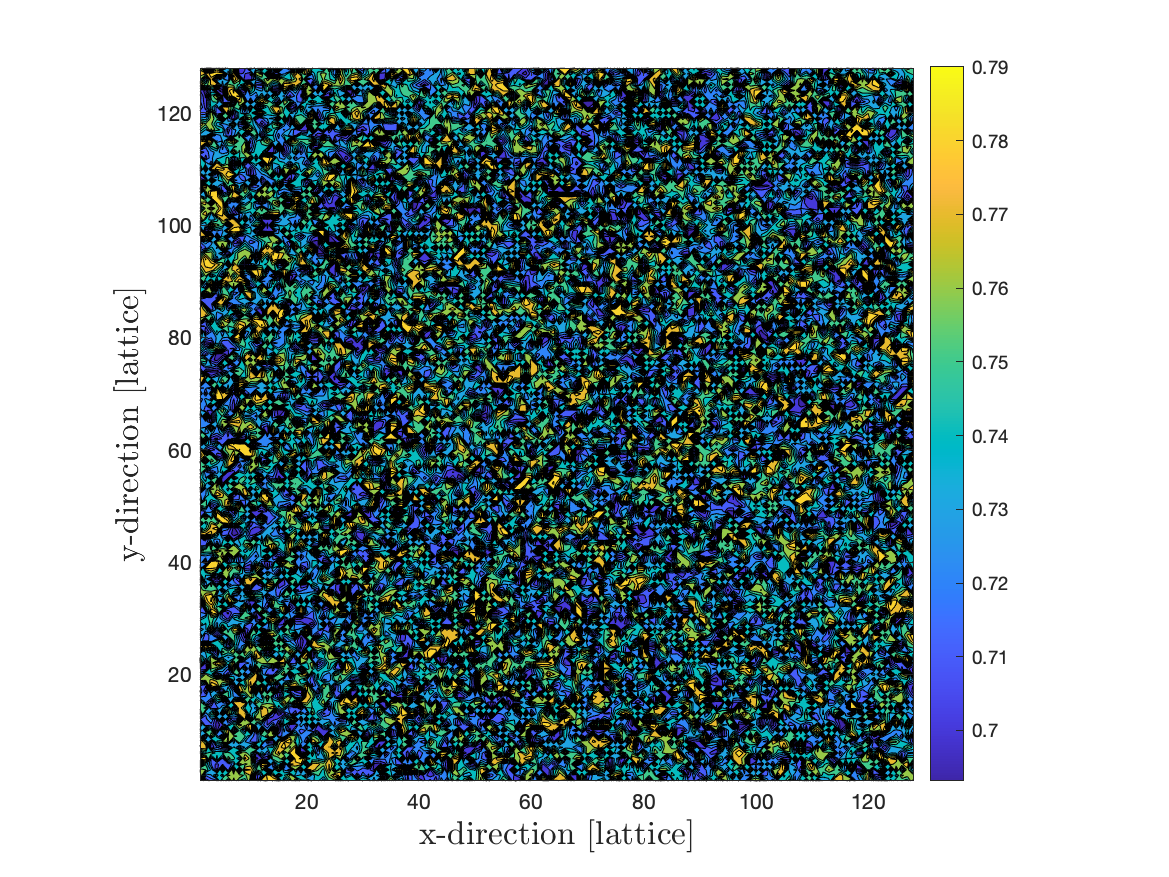}%
        \label{phase_evolution:a}%
    }\hfill
    \subfloat[$t=0.04s$]{%
        \includegraphics[trim=1.8cm 0.5cm 2.7cm 0.5cm, clip=true, width=0.48\linewidth]{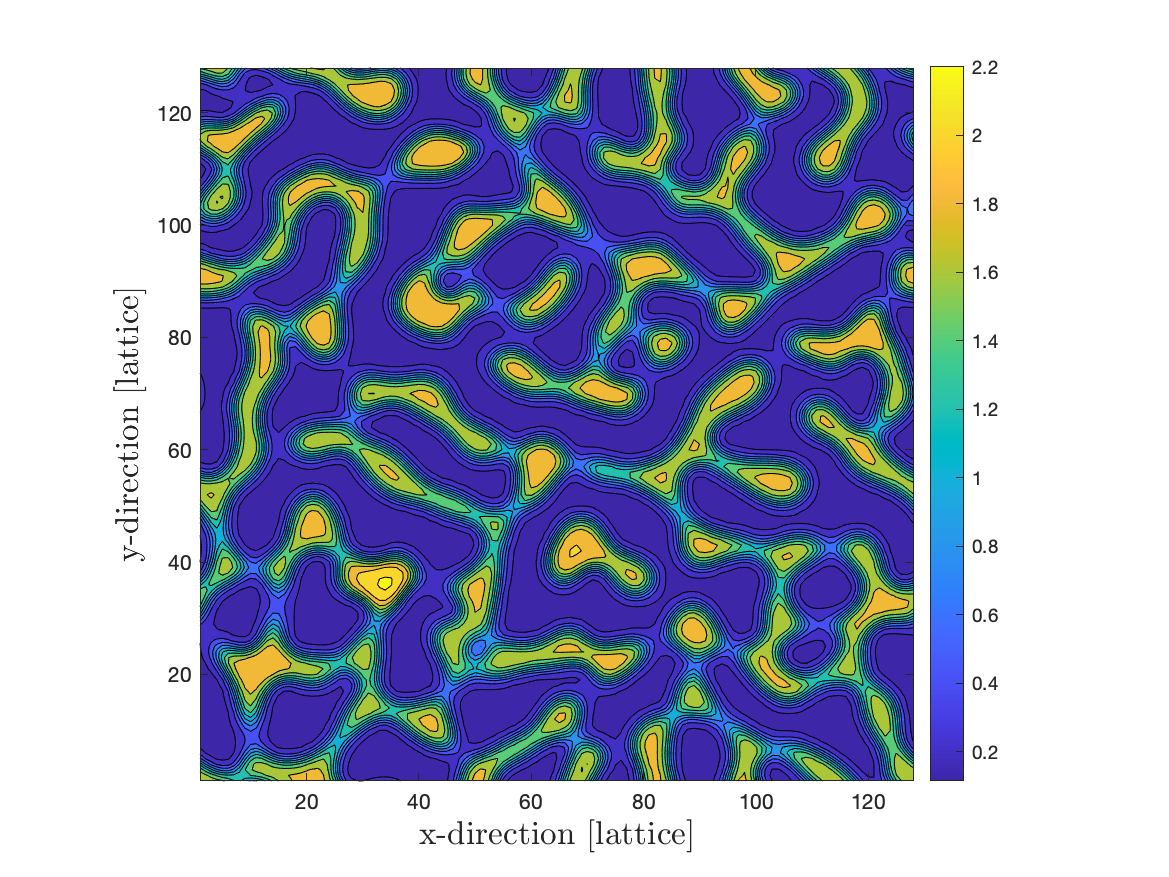}%
        \label{phase_evolution:b}%
    }\\
    \subfloat[$t=0.4s$]{%
        \includegraphics[width=0.48\linewidth]{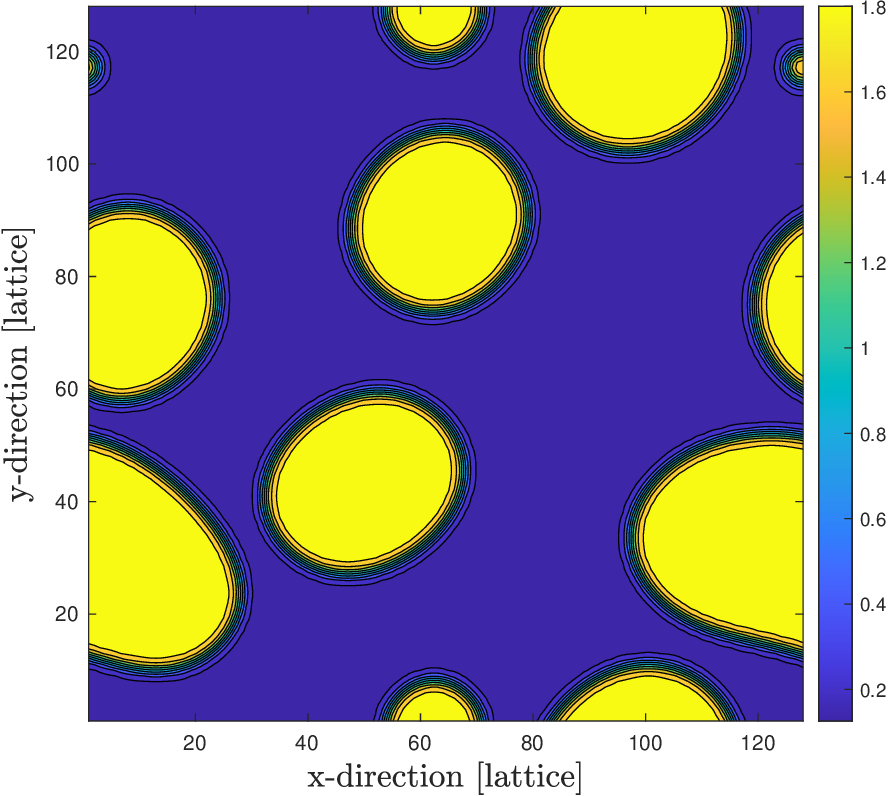}%
        \label{phase_evolution:c}%
    }\hfill
    \subfloat[$t=4s$]{%
        \includegraphics[width=0.48\linewidth]{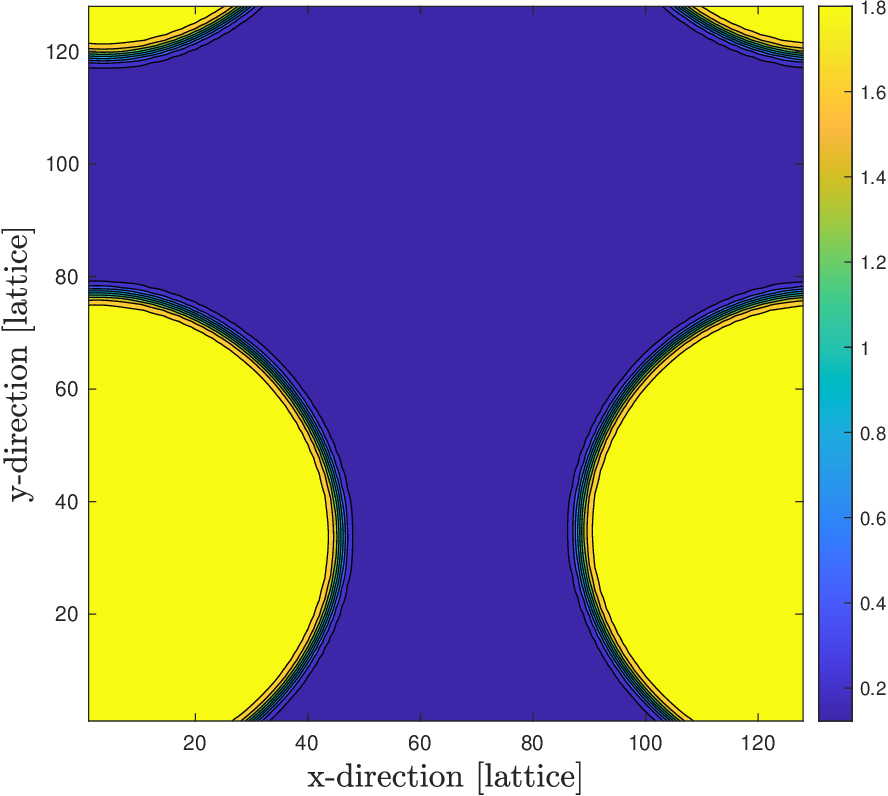}%
        \label{phase_evolution:d}%
    }
    \caption{\label{fig:phase_evolution} {\small Four frames (a) to (d) captured from different time steps showing the evolution of initial perturbations in density for the multi-phase single-component case and the formation of droplets.}}
\end{figure*}

The evolution of the mixing process is illustrated in Figure \ref{fig:phase_evolution}, where a sequence of 4 frames (a) to (d) captured at different time steps, demonstrating the development of initial perturbations in density and the subsequent formation of droplets. These visualizations provide insight into the progression of the mixing phenomenon, as the initially perturbed density field undergoes significant changes, resulting in the formation of distinct droplets within the fluid mixture. \textcolor{black}{These results show a qualitative consistency with Basit et al. \cite{Basit2010}, although our findings differ due to the use of a different set of parameters. Nevertheless, the objective of both studies was to investigate phase separation using the SC multiphase model. In our study, phase separation was simulated with perturbed random density fluctuations ranging from 0.001 to 0.1 relative to the average density.}
%\textcolor{blue}{XXX the presented results are quantitatively/qualitatively compatable with those of \cite{} in terms if XXX and YYY, showing ZZZ and WWW !}

%%%%%%%%%%%%%%%%%%%%%%%%%%%%%%%%%%%%%%%%%%%%%%%%%%%%%%%%%%%%%%%%%%%%%%%%%%%%%%%
%%%%%%%%%%%%%%%%%%%%%%%%%%%%%%%% Cavitation %%%%%%%%%%%%%%%%%%%%%%%%%%%%%%%%%%%
%%%%%%%%%%%%%%%%%%%%%%%%%%%%%%%%%%%%%%%%%%%%%%%%%%%%%%%%%%%%%%%%%%%%%%%%%%%%%%%
\subsubsection{Cavitation}\label{sec:cavitation}

Cavitation is a phenomenon in which the static pressure of a liquid reduces below its vapor pressure, leading to the formation of small vapor-filled cavities within the liquid. It serves as a classic case for studying multi-phase (or multi-component) systems, allowing us to further validate our code.

%Phase separation 
\textcolor{black}{Cavitation plays an important role in understanding the formation of bubbles/drops.} %(XXX again the word cavitation is missing XXX). 
The Shan-Chen model relies on the cohesion parameter $G_c$, which controls fluid-fluid interactions. \textcolor{black}{This $G_c$ is similar to the term without subscript $c$ i.e. $G$ used in Eq. \eqref{eq:1.12}, maintaining consistency with the literature \cite{Huang2007}}. When $G_c$ surpasses the threshold value of $1/(\rho_1+\rho_2)$, phase separation occurs. In this case, Fluid 1 occupies the bubble with a main density of $\rho_1$, while Fluid 2 has a lower dissolved density of $\rho_2$. Huang et al. \cite{Huang2007} conducted a series of simulations with two fluids, each having an initial density of $\rho_i=\rho_1+\rho_2$. The chosen density was 2, while varying $G_c$ from 0 to 3.6. It should be noted that exceeding certain limits of $G_c$ can induce instabilities in the numerical simulation. The canonical CooLBM code for cavitation, an example of a multiphase simulation, is provided in Listing~\ref{lst:Cavitation}. 
\begin{lstlisting}[frame=none, label={lst:Cavitation}]
//CooLBM 2D cavitation simulation canonical C++ code
//Example of a multiphase system
#include "cavitation2D.h"
using namespace std;

string problem = "cavitation2D";

int main() {  
    if (problem == "cavitation2D") {
        cavitation2D();  //cavitation2D function called 
    }   
    return 0;
}
//Definition "cavitation2D" function
//Execute an iteration (collision and streaming) on a cell called through "for_each"
void operator() () {
    density(); //Obtain density of the fluid
    force ();  //Calculate force
    macro();  //Calculate macroscopic variables
    collideBgk(); //Perform BGK collision step
    stream();     //Perform streaming step
}
void cavitation2D()                                
{                                                                                                            
    std::ifstream cavfile("../apps/Config_Files/config_cavitation2D.txt"); //Input file
    //Read data from the input file
    ......................................
    //Variable initialization for the LBM operations
    ......................................
    //Define simulation geometry
    inigeom_cavitation2D();
    // Start main time step loop
    for (int time_iter = 0; time_iter < max_time_iter; ++time_iter) { 
        //Save data for post-processing
        ..............................
        //Calculate surface tension
        ..............................
        //With the double-population scheme, collision and streaming are executed in the following loop.
        for_each(execution::par_unseq, lattice, lattice + dim.nelem, lbm); 
    }
}
\end{lstlisting}
Considering the pressure distribution within the system, the pressure at point $\mathbf{x}$ can be determined using the following equation:
\begin{equation}
P(x) = \frac{\rho_1(\mathbf{x}) + \rho_2(\mathbf{x})}{3} + G_c \frac{\rho_1(\mathbf{x}) + \rho_2(\mathbf{x})}{3}
\end{equation}

Furthermore, the interfacial tension, the key parameter in characterizing the bubble/drop interface, can be calculated by measuring the component densities inside and outside the bubble/drop, according to the Laplace law:
\begin{equation}
P(\mathbf{x}_{\text{in}}) - P(\mathbf{x}_{\text{out}}) = \frac{\gamma}{R}
\end{equation}

\textcolor{black}{Here, $\gamma$ 
%(XXX you used $\sigma$ before! replace all with $\gamma$ XXX) 
represents the surface tension}, $P$ denotes the pressure, and $R$ represents the radius of the bubble. Figure \ref{fig:cavitationvalidation} presents a comparison between CooLBM simulation results and the results reported by Huang et al. \cite{Huang2007} \textcolor{black}{for a domain of size $N_x \times N_y = 100 \times 100$.} Figure \ref{fig:scaleddensity} illustrate the scaled component densities as a function of the scaled cohesion parameter, while Figure \ref{fig:scaledsurfacetension} depicts the scaled lattice surface tension as a function of the same parameter. Both the scaled component densities and scaled surface tension exhibit excellent agreement with the findings of Huang \cite{Huang2007}, providing further validation and confidence in the accuracy and reliability of the developed code.

\begin{figure}[htbp]
\begin{center}

\begin{subfigure}[b]{0.49\textwidth}
	\includegraphics[width=1\textwidth]{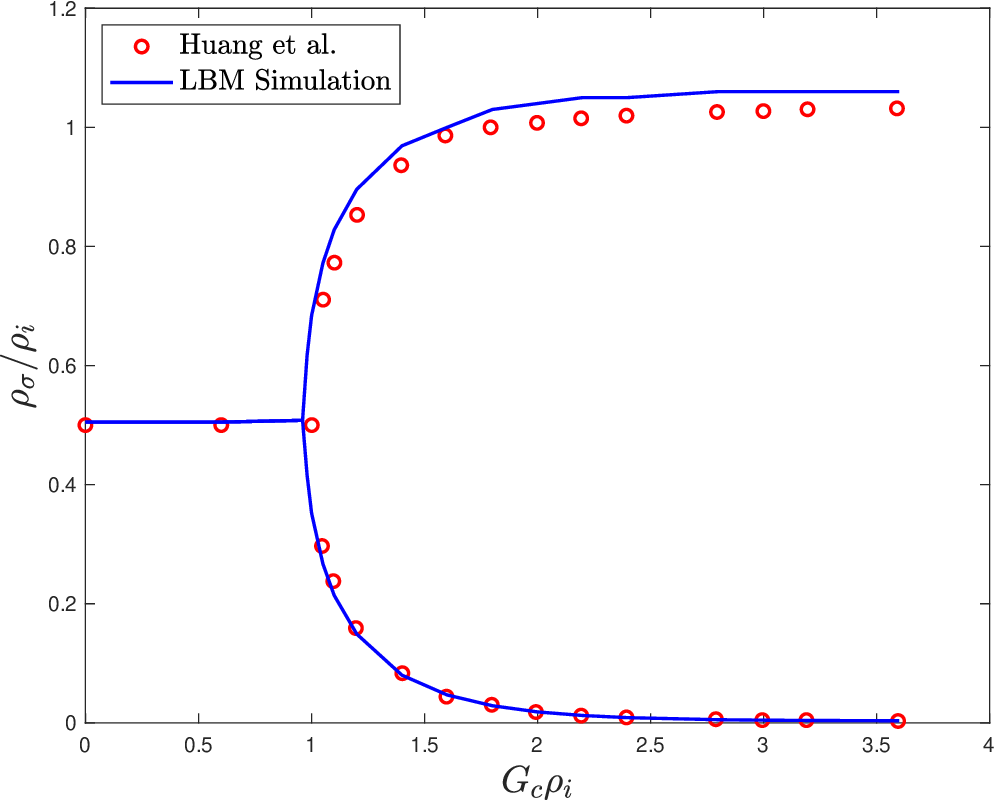}
	\caption{}
	\label{fig:scaleddensity}
\end{subfigure}
%\hspace{5mm}
\begin{subfigure}[b]{0.49\textwidth}
	\includegraphics[width=1\textwidth]{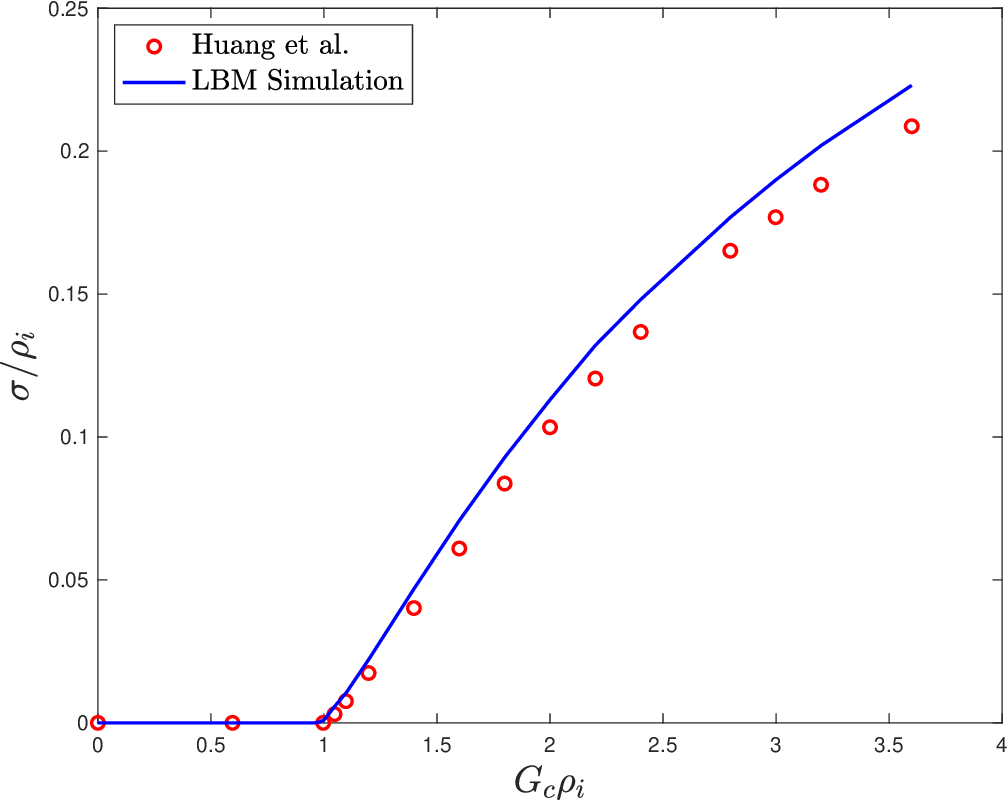}
	\caption{}
	\label{fig:scaledsurfacetension}
\end{subfigure}

\caption{\label{fig:cavitationvalidation} {\small Comparison of CooLBM results with the findings of Huang et al. \cite{Huang2007}  for (a) scaled component densities as a function of the scaled cohesion parameter, and (b) scaled lattice surface tension as a function of the scaled cohesion parameter.}}
\end{center}
\end{figure}

Figure \ref{fig:scaleddensity} illustrates the density of Fluid 1 and the small density of dissolved Fluid 2 inside the bubble, scaled to the initial density $\rho_\sigma/\rho_i$, plotted against the scaled cohesion parameter $G_c\rho_i$. \textcolor{black}{When $G_c\rho_i$ exceeds the critical value of 1, the bubble initially filled with Fluid 1 begins to mix with Fluid 2. However, Fluid 1 remains the dominant fluid inside the bubble. With a further increase in $G_c\rho_i$, the scaled density of Fluid 1 exceeds its initial scaled density as shown in the upper bifurcation branch of Figure \ref{fig:scaleddensity}.} 
%(\textcolor{blue}{XXX I didnt understand this sentence!! XXX}). 

\begin{figure}[htbp]
\centering
\includegraphics[width=\textwidth]{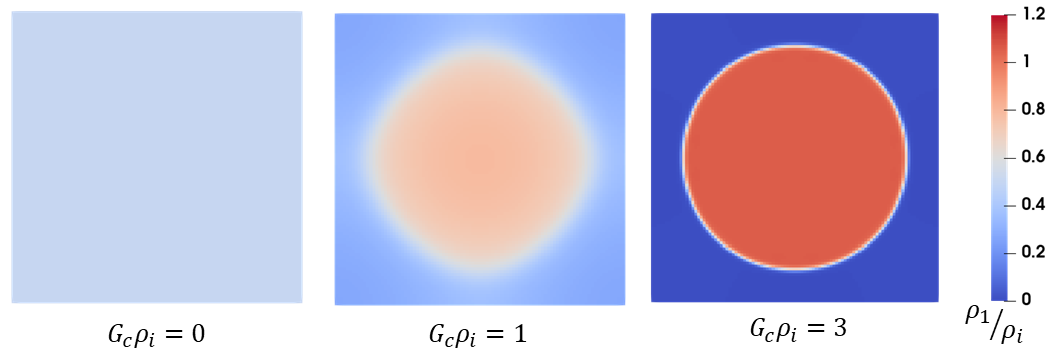}
\caption{\label{fig:cavitation_different_Gc_rhoi} \small Scaled density of Fluid 1 represented sequentially for $G_c\rho_i = 0.0, 1.0$ and $3.0$.}
\end{figure}
This suggests that larger values of $G_c$ are favorable for stable phase separation in multi-phase multi-component simulations. Conversely, when $G_c\rho_i$ is below $1$, no phase separation occurs, as shown in Figure \ref{fig:scaleddensity}, where the scaled densities for both components are $0.5$. This outcome arises from diffusion, which drives the two-fluid solution to become homogeneous. \textcolor{black}{Figure \ref{fig:cavitation_different_Gc_rhoi} shows the observations of scaled density of Fluid 1 within the computational domain for three different values of $G_c\rho_i$: 0.0, 1.0, and 3.0.}

%\textcolor{blue}{XXX dont we have a illustration figure for this test case like other cases?? XXX}

%%%%%%%%%%%%%%%%%%%%%%%%%%%%%%%%%%%%%%%%%%%%%%%%%%%%%%%%%%%%%%%%%%%%%%%%%%%%%%%
%%%%%%%%%%%%%%%%%%%%%%% Rayleigh-Taylor Instability %%%%%%%%%%%%%%%%%%%%%%%%%%%
%%%%%%%%%%%%%%%%%%%%%%%%%%%%%%%%%%%%%%%%%%%%%%%%%%%%%%%%%%%%%%%%%%%%%%%%%%%%%%%
\subsubsection{Rayleigh-Taylor Instability}\label{sec:RayleighTaylor}

The Rayleigh-Taylor instability (RTI) is an instability that occurs at the interface between two fluids of different densities when the heavier fluid displaces the lighter one and vice versa. Here, we employ the multi-component solver of CooLBM to investigate the evolution of the RTI through its four main stages, namely: Stage 1: At the initial stage, the perturbation amplitudes are small compared to the wavelength but exhibit exponential growth. Stage 2: Nonlinear effects become prominent, leading to the formation of mushroom-shaped spikes where the heavier fluid penetrates into the lighter fluid, and the creation of bubbles where the lighter fluid displaces the heavier fluid. Stage 3: The spikes and bubbles interact with each other, resulting in the formation of larger spikes and bubbles. Stage 4: Regions of turbulent mixing start to form as a consequence of the interaction between spikes and bubbles.

Figure \ref{fig:rayleightaylor} displays the initial setup of the components and the boundary conditions in the computational domain. \textcolor{black}{The simulation parameters were set as follows: the density of Fluid 1 is equal to the initial density, $i.e.$ 1, plus density fluctuations of 0.01 multiplied by an arbitrary random number in the range of 0 to 99. Similarly, the density of Fluid 2 is set to the initial density of 1, plus density fluctuations of 0.01 multiplied by an arbitrary random number in the range of 0 to 99.} The numerical grid size is set to $N_x \times N_y = 1200 \times 400$ lattices. The external forces such as gravity and electric forces act on the fluid, they can be combined with Shan-Chen force $\mathbf{F}^{\text{SC}(\sigma)}$. These forces have to be added in Eqs.\eqref{eq:1.17} and \eqref{Shan_Chen_forcing_actual_velocity} in the Shan-Chen forcing scheme, or Eqs.\eqref{eq:1.19}, \eqref{eq:1.21}, and \eqref{Guo_forcing_macroscopic} in the Guo forcing scheme. Here, we employ the Guo forcing scheme to add gravitational force in this simulation.
\begin{figure}[htbp]
\centering
\includegraphics[width=0.55\textwidth]{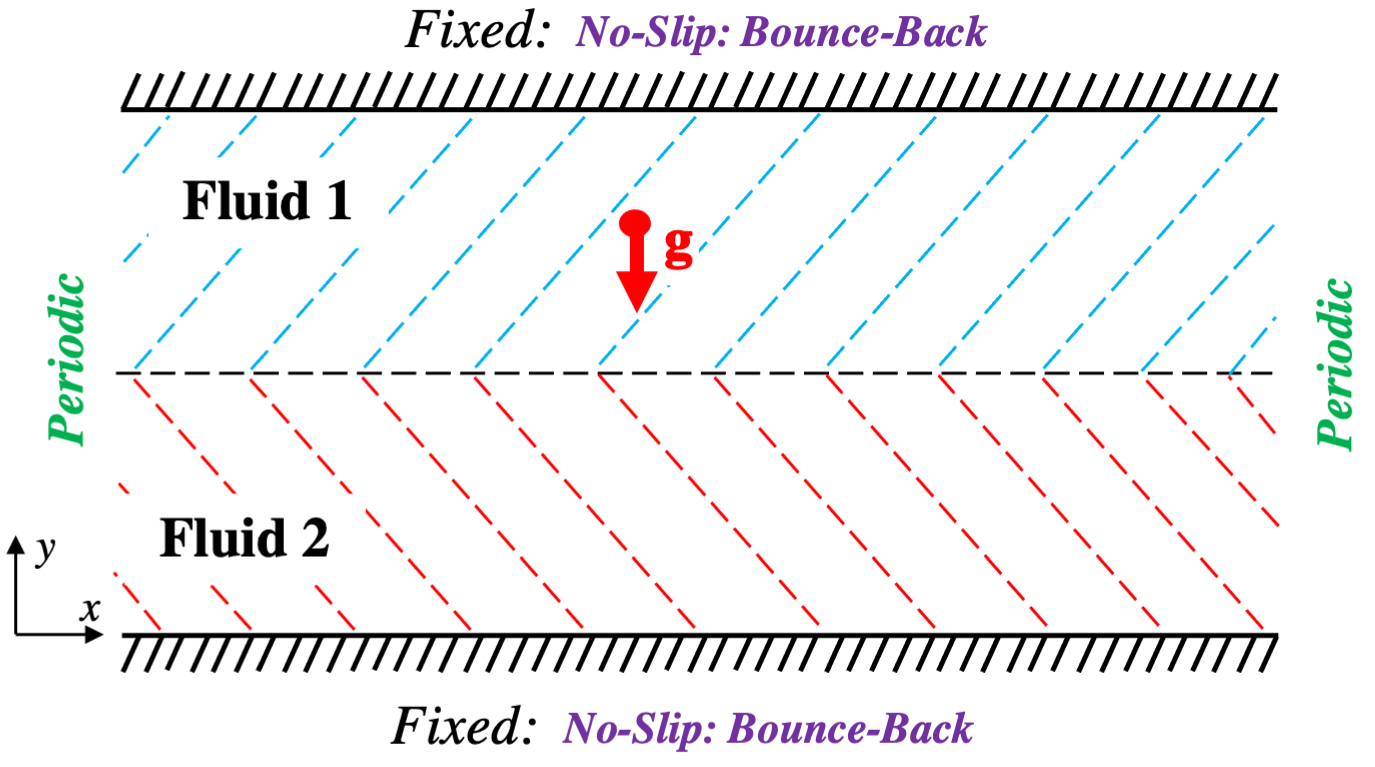}
\caption{\label{fig:rayleightaylor} \small Initial setup of components and boundary conditions in the computational domain [7].}
\end{figure}

Figure \ref{fig:RT_evolution} depicts the time evolution of the Rayleigh-Taylor instability using four frames (a) to (d). These frames correspond to the accurate representation of all four stages by the CooLBM simulation: (i) small perturbation amplitudes in Figure \ref{fig:RT_evolution:a}; (ii) formation of mushroom-shaped spikes in Figure \ref{fig:RT_evolution:b}; (iii) interaction of spikes and bubbles resulting in larger structures in Figure \ref{fig:RT_evolution:c}; and (iv) formation of turbulent mixing regions in Figure \ref{fig:RT_evolution:d}. Therefore, it can be concluded that the CooLBM multi-component solver accurately captures the four stages of RTI, starting from small perturbations to the formation of turbulent mixing, as demonstrated in this figure. 

\begin{figure*}[t!]
    \subfloat[$t=0.08s$]{%
        \includegraphics[trim=0.8cm 4.1cm 1.7cm 4.5cm, clip=true, width=0.48\linewidth]{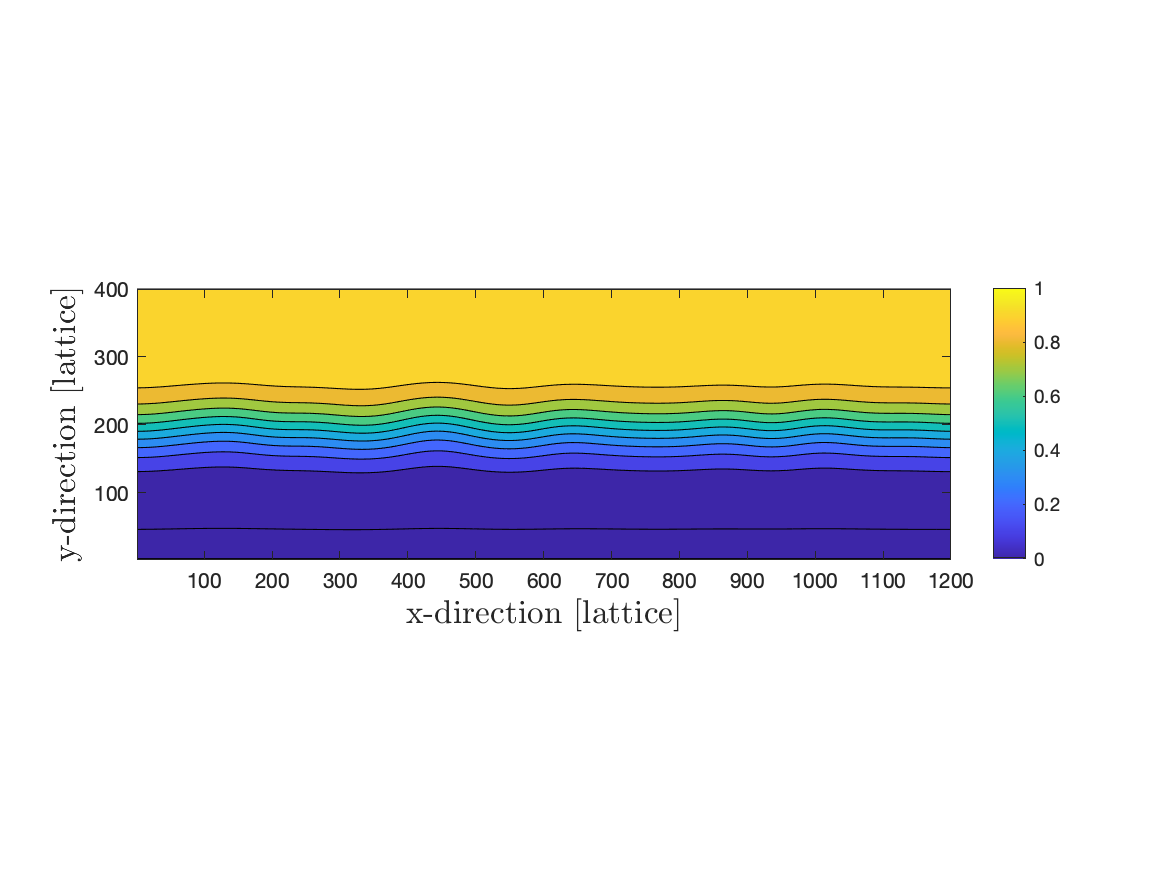}%
        \label{fig:RT_evolution:a}%
    }\hfill
    \subfloat[$t=0.13s$]{%
        \includegraphics[trim=0.8cm 4.1cm 1.7cm 4.5cm, clip=true, width=0.48\linewidth]{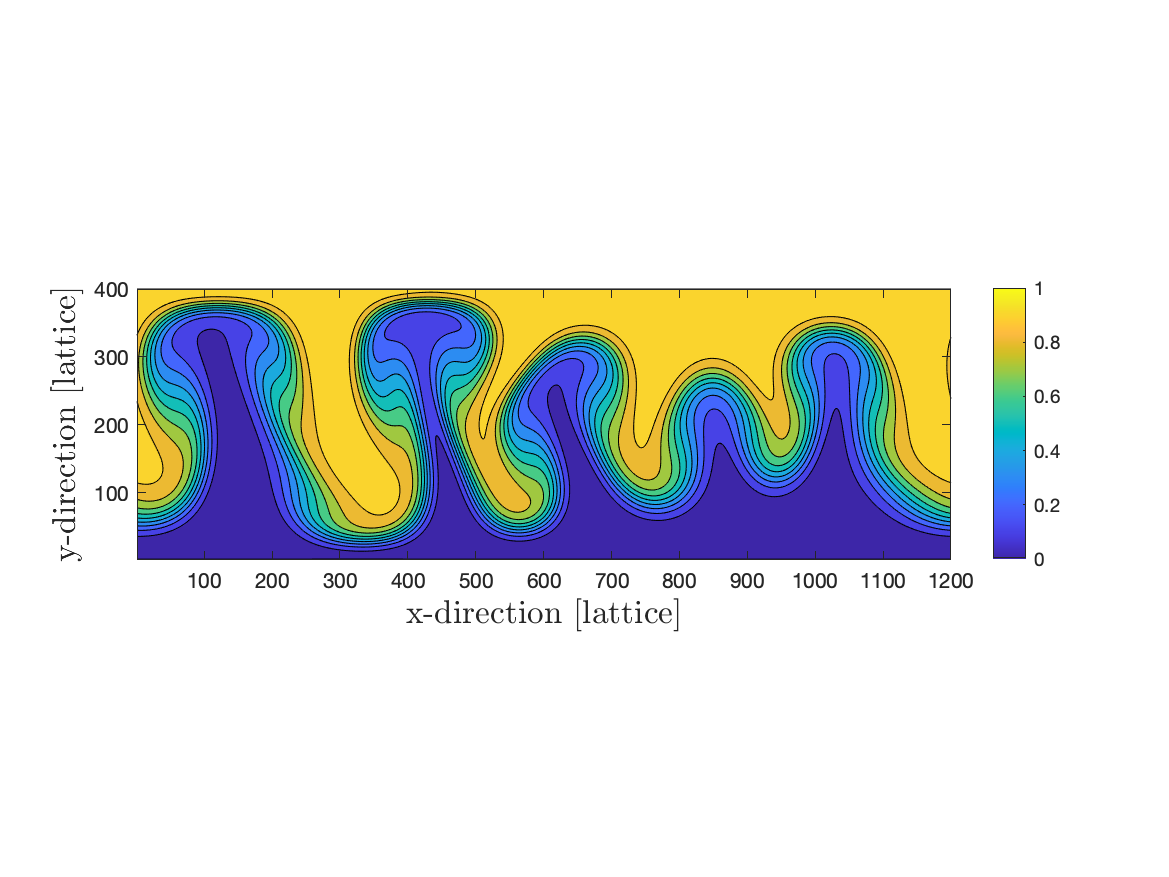}%
        \label{fig:RT_evolution:b}%
    }\\
    \subfloat[$t=0.17s$]{%
        \includegraphics[trim=0.8cm 4.1cm 1.7cm 4.5cm, clip=true, width=0.48\linewidth]{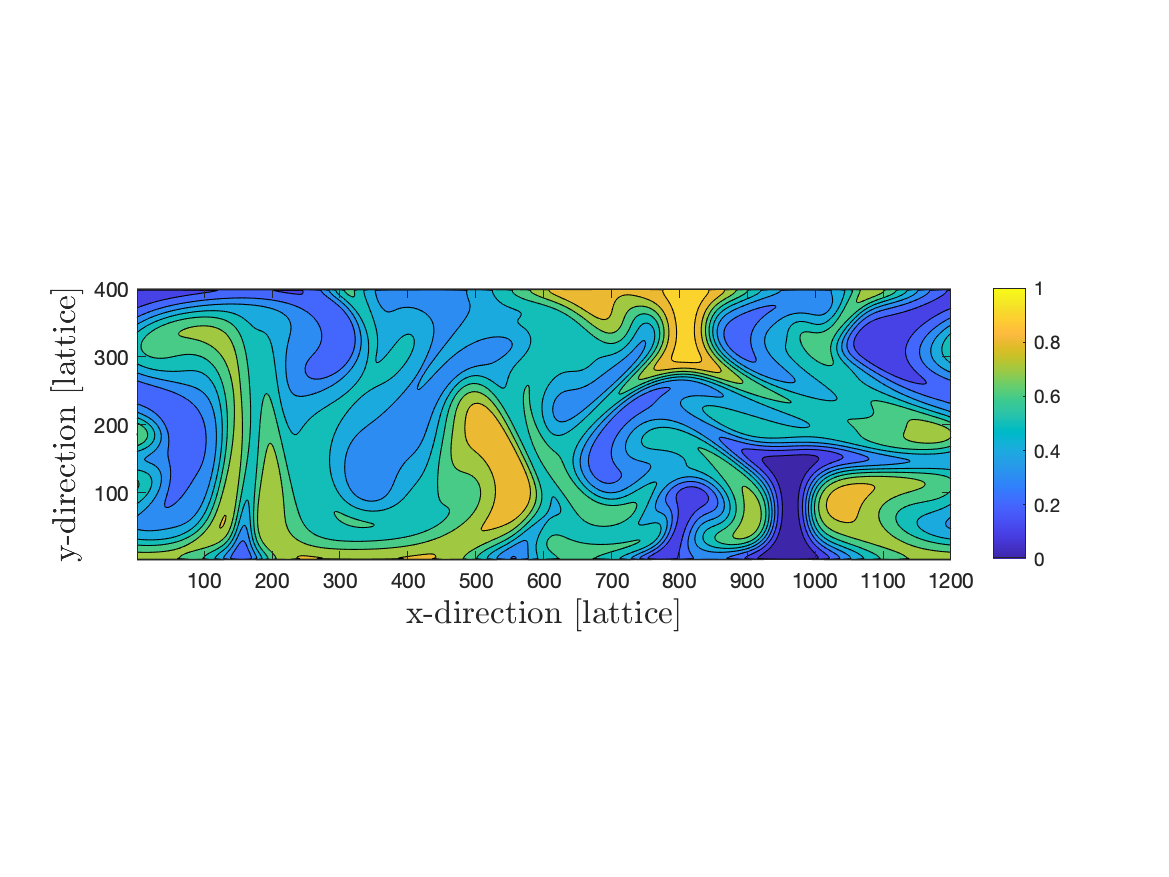}%
        \label{fig:RT_evolution:c}%
    }\hfill
    \subfloat[$t=0.21s$]{%
        \includegraphics[trim=0.8cm 4.1cm 1.7cm 4.5cm, clip=true, width=0.48\linewidth]{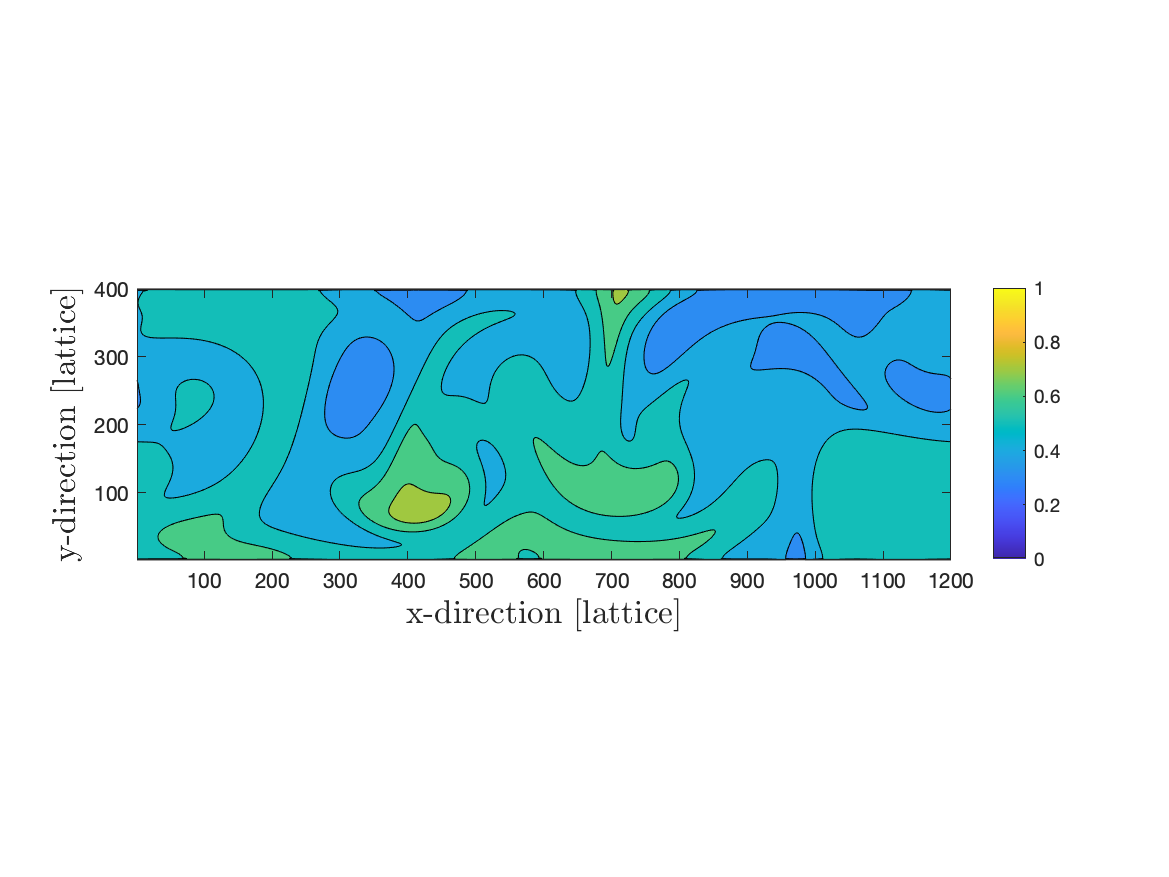}%
        \label{fig:RT_evolution:d}%
    }
    \caption{\label{fig:RT_evolution} {\small Time evolution of the Rayleigh-Taylor instability as captured by the CooLBM multi-component simulation.}}
\end{figure*}

%%%%%%%%%%%%%%%%%%%%%%%%%%%%%%%%%%%%%%%%%%%%%%%%%%%%%%%%%%%%%%%%%%%%%%%%%%%%%%%
%%%%%%%%%%%%%%%%%%%%%%%%%%%%%%% Contact Angle %%%%%%%%%%%%%%%%%%%%%%%%%%%%%%%%%
%%%%%%%%%%%%%%%%%%%%%%%%%%%%%%%%%%%%%%%%%%%%%%%%%%%%%%%%%%%%%%%%%%%%%%%%%%%%%%%
\subsubsection{Contact Angle}\label{ContactAngle}
\textcolor{black}{The phenomenon of contact angle in the fluid domain is simulated using the same way of implementation of velocity and pressure boundary conditions for multi-phase and multi-component SC models as in the standard LBM.}
%(\textcolor{blue}{XXX what do you mean by standard? you mean single phase ? XXX}) LBM.
However, near to the boundary, the multi-phase effect needs to be taken into account. When fluid interfaces come into contact with a solid boundary, they assume a specific contact angle. \textcolor{black}{To accurately simulate wettability}, we introduce an interaction similar to the SC force, but this time it describes the interaction between solid nodes and boundary nodes. The strength of interaction between each fluid and a wall can be adjusted using the adhesion parameters, denoted as $G_{\text{ads}}$.

The theoretical contact angle, in terms of the adhesion and cohesion coefficients ($G_{\text{ads}}$ and $G_c$), can be expressed as \cite{Huang2007}:
\begin{equation}
\cos\theta = \frac{{G_{\text{ads},2} - G_{\text{ads},1}}}{{G_c (\rho_1 - \rho_2)/2}}.
\end{equation}

\begin{figure}[htbp]
\centering
\includegraphics[width=0.5\textwidth]{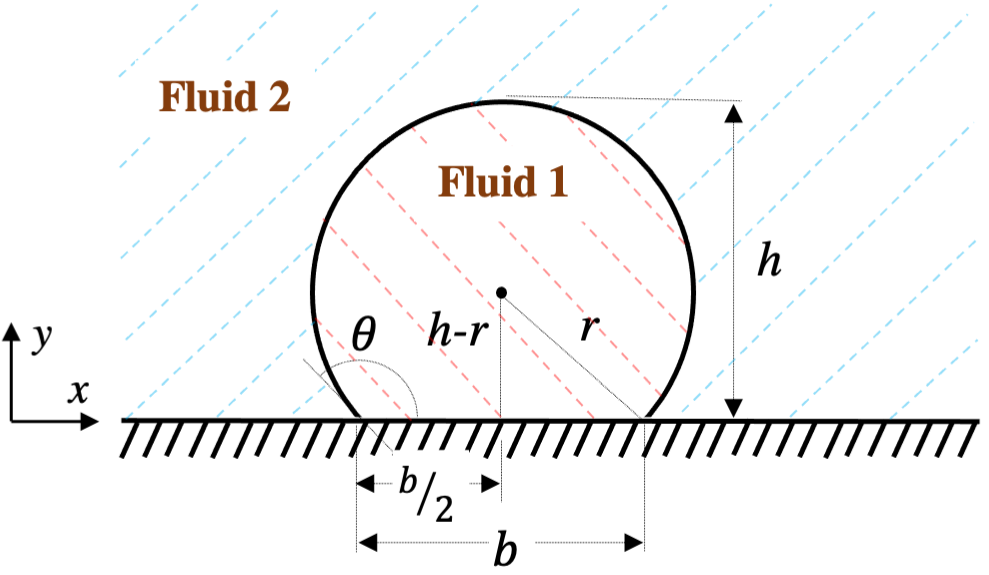}
\caption{\label{fig:contactanglegeom} \small Calculation of contact angle in terms of radius r, height h, and contact length b.}
\end{figure}

The calculation of the contact angle in CooLBM is based on the geometric configuration depicted in Figure \ref{fig:contactanglegeom}. CooLBM calculates the height, $h$, and contact length, $b$, of the droplet based on the post-processing data of regions with certain densities. Then, a relationship is implemented to calculate the radius, $r$, and the contact angle, $\theta$, using the following equations:
\begin{equation}
r = \frac{{4h^2 + b^2}}{{8h}}
\end{equation}
\begin{equation}
\theta = \pi - \arctan\left[\frac{{b}}{{2(h-r)}}\right].
\end{equation}

Figure \ref{fig:contactanglevalid} displays the contact angle results for different coefficients of fluid-solid interaction, $G_{\text{ads}}$, obtained by CooLBM. These results are in very good agreement with the theoretical values reported in the work of Huang et al. \cite{Huang2007}. The comparison demonstrates a high level of concurrence between these data, indicating that CooLBM accurately predicts the contact angle behavior for varying $G_{\text{ads}}$ values. 
\begin{figure}[htbp]
\centering
\includegraphics[width=0.6\textwidth]{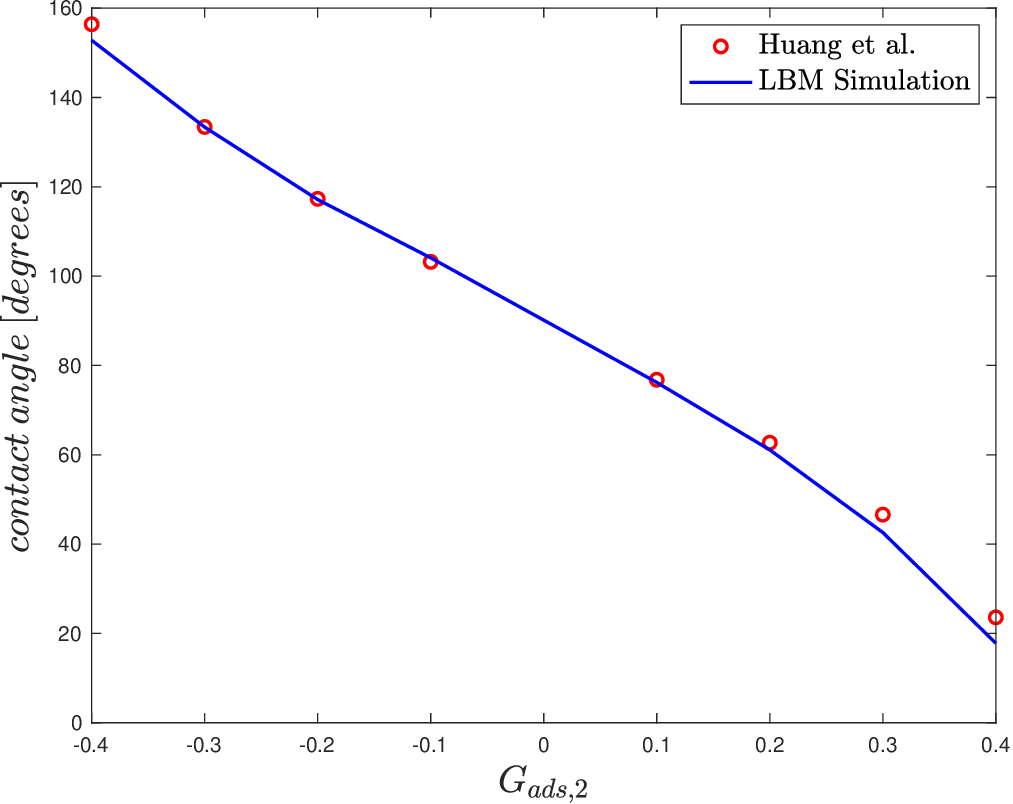}
\caption{\label{fig:contactanglevalid} \small Contact angle obtained by the code were plotted and compared to theoretical values and previous numerical work of Huang et al. \cite{Huang2007}.}
\end{figure}
%%%%%%%%%%%%%%%%%%%%%%%%%%%%%%%%%%%%%%%%%%%%%%%%%%%%%%%%%%%%%%%%%%%%%%%%%%%%%%%
%%%%%%%%%%%%%%%%%%%%%%%%%%%%%%%%%%%%%%%%%%%%%%%%%%%%%%%%%%%%%%%%%%%%%%%%%%%%%%%
%%%%%%%%%%%Particle deposition%%%%%%%%%%%%%%%%%%%%%%%%%%%%%%%%%%%%%%%%%%%%%%%%%
\subsection{\textcolor{black}{Particle Deposition}}\label{sec:particle_deposition}
\textcolor{black}{The deposition of particles is a ubiquitous process in industrial applications \cite{Slinn1980, Elimelech2013}. One example is the capture of particulate matter (PM) produced during fuel combustion in automobile engines and its subsequent treatment, which is required to prevent interference with engine operations as well as to control environmental pollution \cite{Cavallo2023}. To accomplish these goals, porous particulate filters, which have complex interconnected-porous structures, are employed to capture PM. These captured particles are then treated using various prescribed procedures. In this section, we will use a naive deposition model to simulate the deposition of particles in a 2-D geometry obtained by Focused Ion Beam Scanning Electron Microscopy (FIB-SEM). Although this will not completely elucidate the deposition behavior at the particle scale, it will provide a qualitative explanation of particle deposition and its implications on the operation of a system. In this model, instead of explicitly considering nano-sized particles, which requires enormous computational power, we use the concentration field for particles, and its transport is obtained through the advection-diffusion equation, as in Eq. \eqref{eq:reactive1c} or \eqref{eq:reactive1d}. The deposition of particles on the solid substrates is obtained through the simplest model, as proposed by Yamamoto et al. \cite{yamamoto2009simulation}, which is given as follows,
\begin{equation}
    C_{p,s}(\mathbf{x},t+\Delta t) =  C_{p,s}(\mathbf{x},t) + DP \sum_{i = 0}^{8} h_{p,i}(\mathbf{x},t + \Delta t) \,,
    \label{surface_deposition}
\end{equation}
where $C_{p,s}$ and $DP$ denote the particle mass fraction and the particle deposition probability at lattice nodes near the substrate, respectively, and $h_{p,i}$ is the distribution function for particles.
Once the particle mass fraction, $C_{p,s}$ at the lattice nodes near the substrate exceeds unity, these fluid lattice nodes are converted into solid nodes. Then solid nodes are treated as no-slip boundary conditions, resulting in changes to the boundary conditions for both the fluid and the particle mass fraction each time this occurs.} 

\textcolor{black}{We simulate the particle deposition using CooLBM in a domain of size $N_x \times N_y =$ 419 $\times$ 415, with other parameters set as follows: the diffusion coefficient, $D_p = 0.156$, the particle deposition probability, $PD = 0.002$, the inlet maximum parabolic velocity, $u_\text{max} = 5.65 \times 10^{-5}$, the relaxation time, $\tau_{d,p} = 0.9687$, and the inlet mass fraction of particles, $C_{p, inlet} = 0.01$. The physical lattice spacing ($\delta x$) is 10 nm, and the lattice model used for both fluid and particle mass fraction is D2Q9.}
\begin{figure}[htbp]
    \centering
    \includegraphics[width=\linewidth]{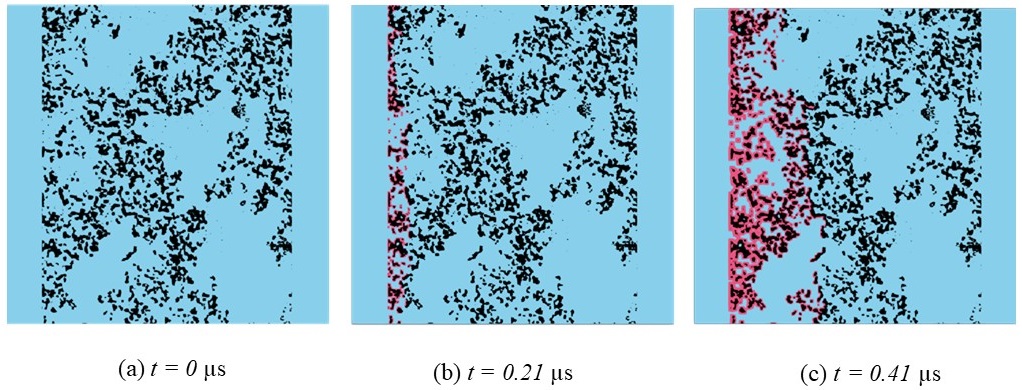}
    \caption{Particle deposition, represented in red, captured at different time intervals: (a) 0  \textmu s, (b) 0.21 \textmu s, and (c) 0.41  \textmu s.}
    \label{fig:deposition_plot}
\end{figure}
\textcolor{black}{Figure \ref{fig:deposition_plot} illustrates particle deposition over porous substrates at various time intervals. Initially, as shown in Figure \ref{fig:deposition_plot}(a), the FIB-SEM image reveals a highly complex porous structure. As the simulation progresses, particles begin to deposit near the channel inlet, as depicted in Figure \ref{fig:deposition_plot}(b). The deposition of particles reduces the gaps between substrates, leading to an increase in flow velocity to maintain mass conservation in that region. This increased flow velocity causes particles to be transported further downstream, resulting in deposition away from the inlet. Finally, Figure \ref{fig:deposition_plot}(c) demonstrates significant particle accumulation at the inlet, eventually obstructing the flow of particles downstream. This case study of CooLBM qualitatively aligns with the reported results of Yamamoto et al. \cite{yamamoto2009simulation}, despite our focus on 2D images compared to their study on 3D porous medium}
%%%%%%%%%%%%%%%%%%%%%%%%%%%%%%%%%%%%%%%%%%%%%%%%%%%%%%%%%%%%%%%%%%%%%%%%%%%%%%%
%%%%%%%%%%%%%%%%%%%%%%%%%%%% Reactive Interface %%%%%%%%%%%%%%%%%%%%%%%%%%%%%%%
%%%%%%%%%%%%%%%%%%%%%%%%%%%%%%%%%%%%%%%%%%%%%%%%%%%%%%%%%%%%%%%%%%%%%%%%%%%%%%%
\subsection{Reactive Interface}\label{sec:reactive}
To validate the reactive interface problem, we employ a simplified two-dimensional channel as the computational domain, containing four square-shaped obstacles, as depicted in Figure \ref{fig:reactivegeometry}. The geometric parameters in the figure are defined as follows: $H=80$, $h=w=s=20$, $l_{\text{in}}=100$, and $l_{\text{out}}=180$ in lattice units.

In this problem, specific inlet conditions are set for temperature, species mass fractions, and flow fields. The temperature at the inlet is $T_{f,\text{inlet}}=$ 773 K, the inlet mass fraction of oxygen is $C_{O_2,\text{inlet}}=0.22$, and the inlet mass fraction of carbon dioxide is $C_{CO_2,\text{inlet}}=0$. The flow velocity profile is defined as $u=6y(1-y)$, where $y$ represents the spatial coordinate. The top and bottom surfaces of the domain follow the no-slip condition for the flow field, while the outlet has a zero-gradient condition for all fields. The temperature and species fields at the top and bottom surfaces also have zero-gradient conditions. Additionally, the inlet air pressure is set to 10 bars, and all physical properties of the inlet air, such as thermal diffusivity, species diffusivity, density, and specific heat, are extrapolated to correspond to the pressure of 10 bars and the temperature of 773 K. Further details regarding the chosen parameters for the numerical simulation can be found in reference \cite{Alamian2024}.
\begin{figure}[htbp]
\begin{center}
\includegraphics[width=1\textwidth]{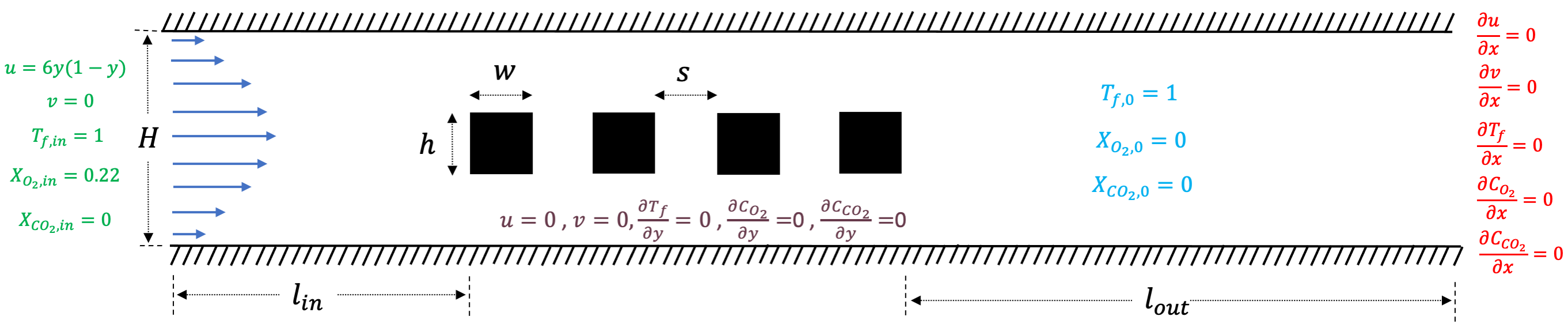}
\caption{\label{fig:reactivegeometry} {\small The reactive gas transfer simulation shows the fluid domain, the reactive obstacle positioning within the channel, the physical boundary conditions for inlet, outlet, top and bottom surfaces and the initial condition inside the domain.}}
\end{center}
\end{figure}

We choose a Péclet number of $Pe=10^{-2}$ and a Damköhler number of $Da=3.525\times10^{-2}$ to model the complex phenomena of heterogeneous combustion and conjugate heat transfer simultaneously, utilizing the presented models. To assess the reliability of CooLBM, we compare our obtained results with their counterparts reported by Xu et al. \cite{Xu2018}, which are presented in Figure \ref{fig:reactivevalidationgraph}. The canonical CooLBM code for reactive interface is provided in Listing~\ref{lst:reactive_flow}.

\begin{lstlisting}[frame=none, label={lst:reactive_flow}]
//CooLBM 3D Couette flow canonical C++ code
#include "combustion2D.h"
using namespace std;

string problem = "combustion2D";

int main() {   
    if (problem == "combustion2D") {
        combustion2D();  //Call combustion2D function 
    }   
    return 0;
}
//Definition of function "combustion2D"
//Execute an iteration (collision and streaming) on a cell called through "for_each"
void operator() () {
    macro();  //Calculate macroscopic variables
    collideBgk(); //Perform BGK collision step
    stream();     //Perform streaming step
}
void couette3D()
{
    std::ifstream coufile("../apps/Config_Files/config_couette3D.txt"); //Input file
    //Read data from the input file
    ......................................
    //Variable initialization for the LBM operations
    ......................................
    //Define simulation geometry
    inigeom_couette3D();
    //Start main time step loop
    for (int time_iter = 0; time_iter < max_time_iter; ++time_iter) {
        //Save data for post-processing
        ..............................
        //With the double-population scheme, collision and streaming are executed in the following loop.
        for_each(execution::par_unseq, lattice, lattice + dim.nelem, lbm);
    }
}

\end{lstlisting}

Analyzing the temperature profile along the mid-path, we find a commendable level of agreement with the temperature data published in \cite{Xu2018} (Figure \ref{fig:validationTemp}). Notably, the temperature plot exhibits distinct flattening, indicating the temperature distribution within the solid obstacles. This distribution gradually decreases from the first obstacle near the inlet to the last obstacle near the outlet. The remaining section of the temperature profile represents the temperature field within the fluid domain. Moreover, the validation of the oxygen and carbon dioxide mass fractions further supports the reliability of presented model, as the trend observed in Figures \ref{fig:validationConcO2} and \ref{fig:validationConcCO2} aligns well with those presented in the study of Xu et al. \cite{Xu2018}.

\begin{figure}[htbp]
     \centering
     \begin{subfigure}[b]{0.45\textwidth}
         \centering
         \includegraphics[width=\textwidth]{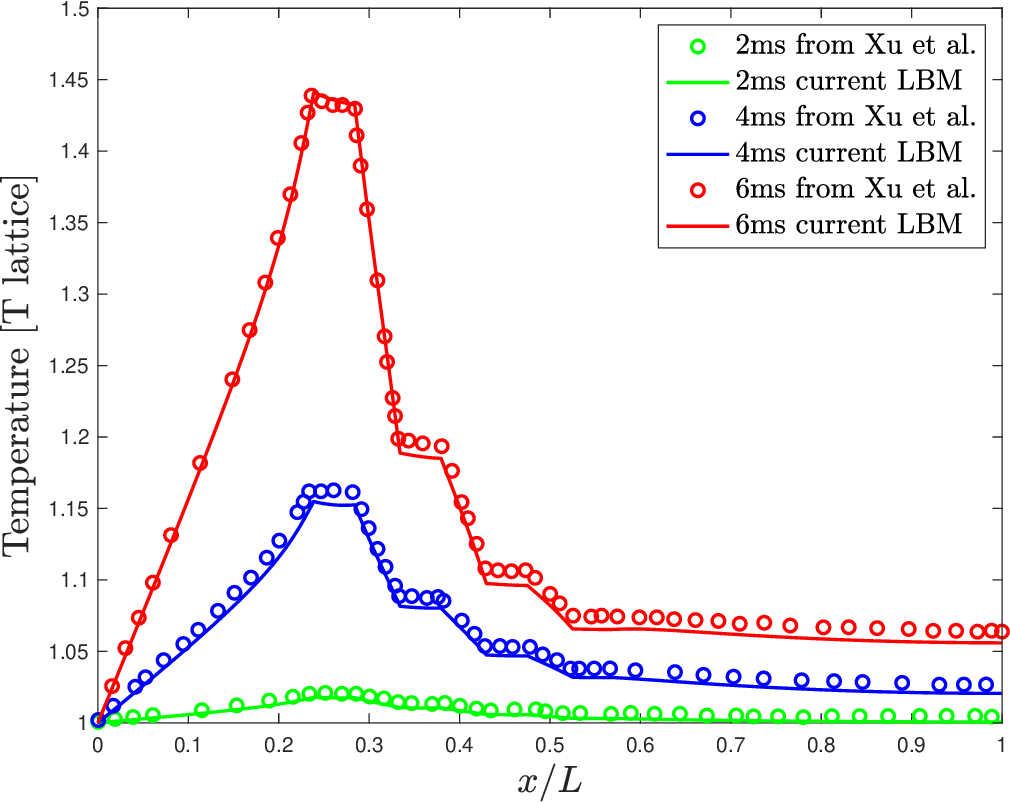}
         \caption{}
         \label{fig:validationTemp}
     \end{subfigure}
     \hfill
     \begin{subfigure}[b]{0.45\textwidth}
         \centering
         \includegraphics[width=\textwidth]{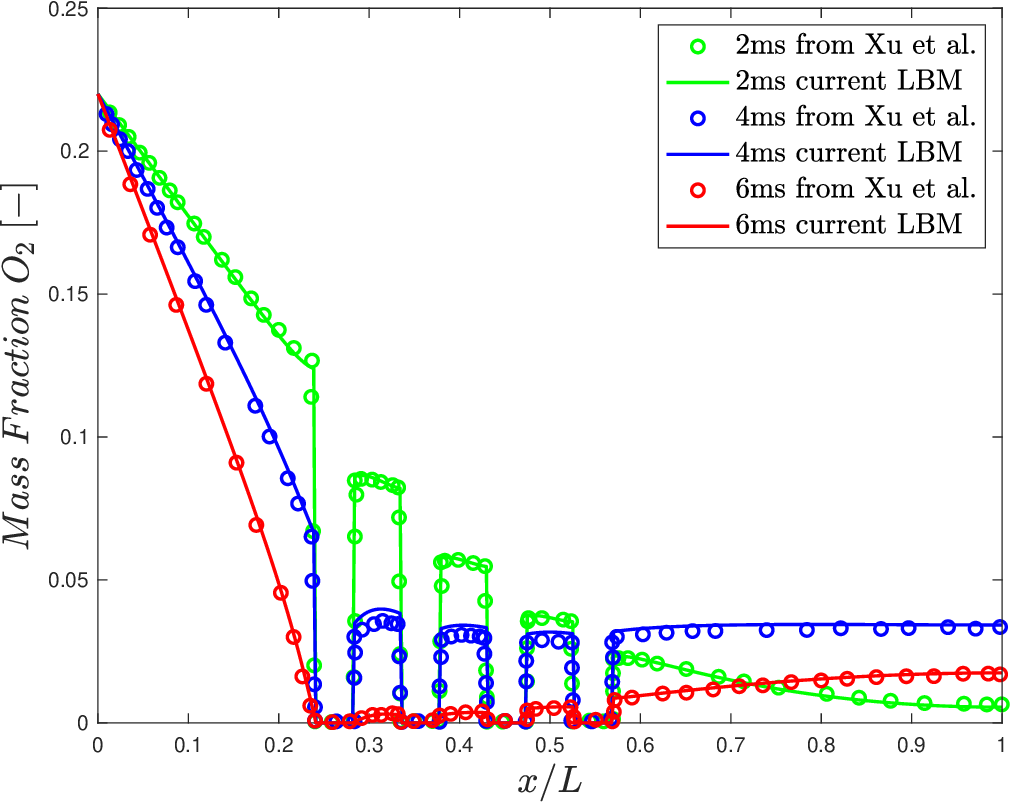}
         \caption{}
         \label{fig:validationConcO2}
     \end{subfigure}
     \hfill
     \begin{subfigure}[b]{0.45\textwidth}
         \centering
         \includegraphics[width=\textwidth]{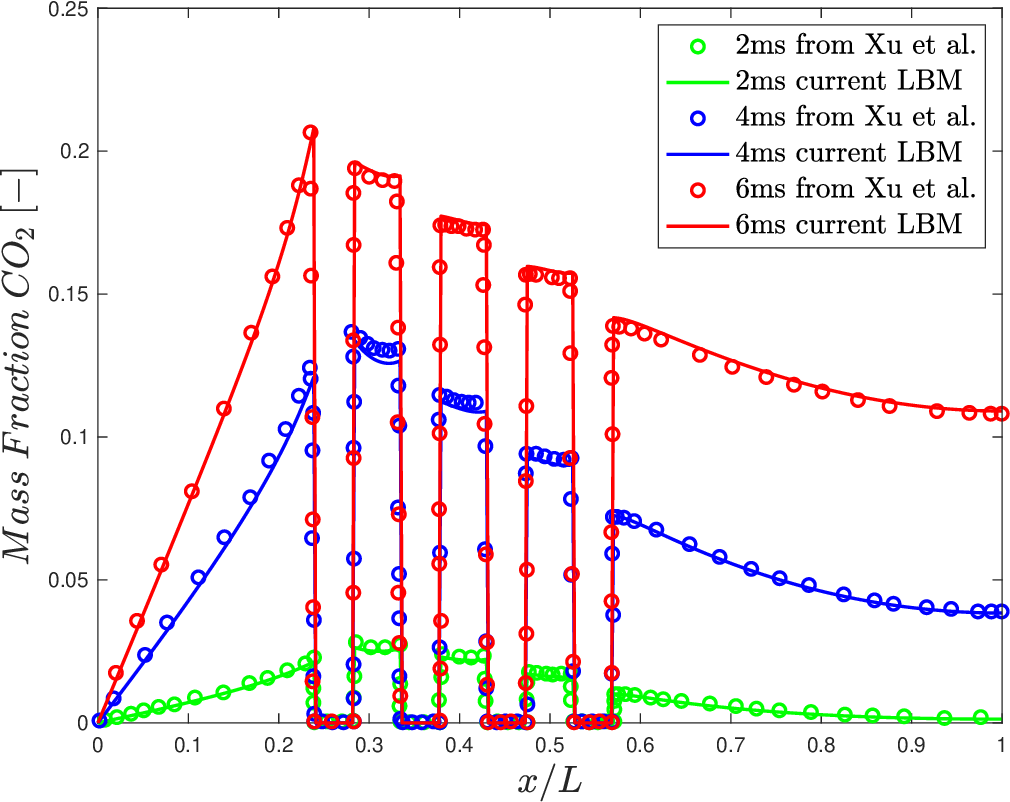}
         \caption{}
         \label{fig:validationConcCO2}
     \end{subfigure}
        \caption{\small Validation of the results with ref \cite{Xu2018} at 2, 6, and 10 ms for (a) temperature profile, (b) Oxygen mass fraction, and (c) Carbon-dioxide mass fraction}
        \label{fig:reactivevalidationgraph}
\end{figure}

%%%%%%%%%%%%%%%%%%%%%%%%%%%%%%%%%%%%%%%%%%%%%%%%%%%%%%%%%%%%%%%%%%%%%%%%%%%%%%%
%%%%%%%%%%%%%%%%%%%%%%% Results and Discussions  %%%%%%%%%%%%%%%%%%%%%%%%%%%%%%
%%%%%%%%%%%%%%%%%%%%%%%%%%%%%%%%%%%%%%%%%%%%%%%%%%%%%%%%%%%%%%%%%%%%%%%%%%%%%%%

\section{Discussions on Computational Costs}\label{sec:results}

This section aims to present a detailed analysis and interpretation of the findings obtained from three significant numerical experiments conducted to evaluate the performance of different computing units in lattice-based simulations. The central focus of this section is Mega Lattice Updates Per Second (MLUPS), an important metric for assessing the efficiency of lattice computations. The experiments include comparisons between processors, core configurations, and an examination of various hardware options. The primary goal is to gain insights into the strengths and limitations of each computing unit to optimize lattice simulations and inform researchers' hardware choices for their future computational studies.

\subsection{Parallel Lattices MLUPS Comparison}
Figure \ref{fig:paralattices} presents a direct comparison between the Intel Core i7-10700K CPU and the NVIDIA RTX A2000 GPU in terms of MLUPS performance with varying numbers of lattices. \textcolor{black}{These simulations are conducted for the three-dimensional Couette flow.}  Figure \ref{fig:paralattices} highlights the consistent superiority of the NVIDIA RTX A2000 over the Intel CPU, demonstrating the benefits of GPU-based computing in parallel lattice simulations. Initially, both configurations exhibit significant MLUPS gains as the number of lattices increases, showcasing the advantages of parallelization. However, beyond a certain point, both performance of hardware improvements start to plateau, indicating the presence of diminishing returns. Despite this observation, the NVIDIA RTX A2000 continues to maintain a performance advantage even at larger numbers of lattices, reaffirming its efficiency in handling the increased workload.

%\textcolor{blue}{XXX which test case is this? Or this does not matter? please specify in the text! XXX}

\begin{figure}[htbp]
\begin{center}
\includegraphics[width=0.7\textwidth]{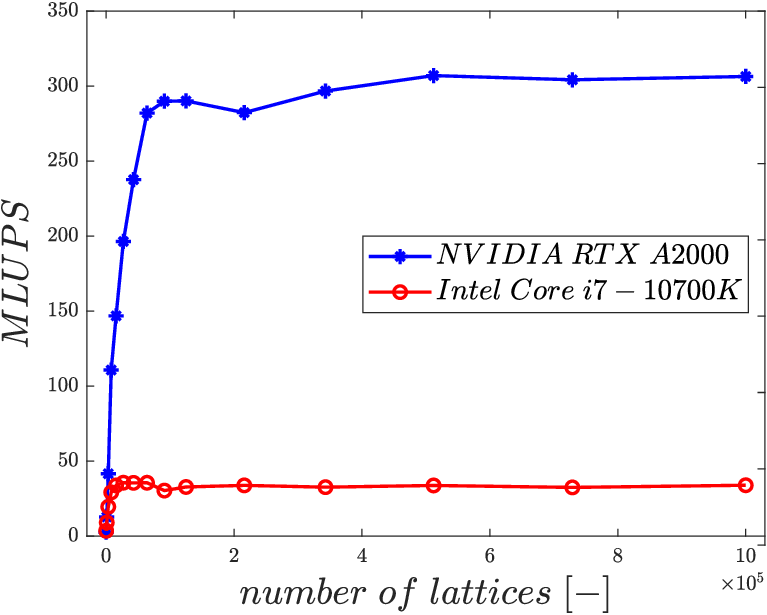}
\caption{\label{fig:paralattices} {\small Performance comparison on CPU and GPU for different number of lattices.}}
\end{center}
\end{figure}

\subsection{Parallel Cores Performance}
The MLUPS performance of the $3^{rd}$ Gen Intel Xeon Scalable hardware is evaluated for various numbers of cores using the LBM implementation in CooLBM. The simulation case focused on three-dimensional Couette flow with a fixed number of 125,000 lattices. The achieved results were analyzed in terms of MLUPS to understand the parallel processing capabilities of hardware.

Figure \ref{fig:paracore} presents a comparison between the MLUPS performance of this hardware while using CooLBM for simulation against the ideal linear increment case. The red curve represents the actual MLUPS performance obtained from CooLBM, while the blue dashed line indicates the ideal linear increment in MLUPS as the number of cores increases.

\begin{figure}[htbp]
\begin{center}
\includegraphics[width=0.7\textwidth]{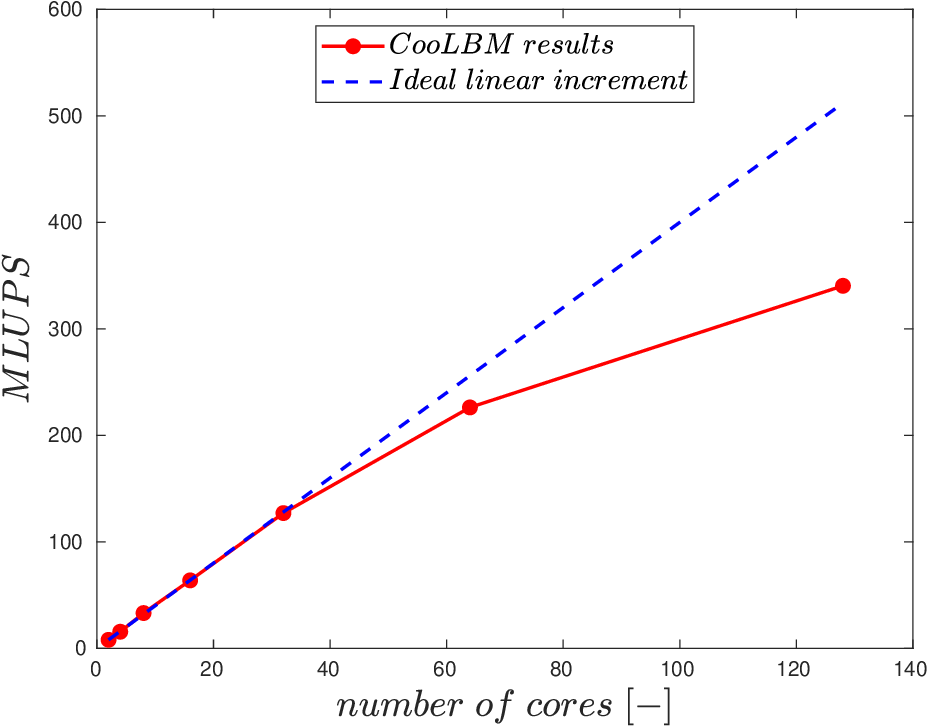}
\caption{\label{fig:paracore} {\small Performance comparison on CPU for different number of cores.}}
\end{center}
\end{figure}

As shown in the graph, the CooLBM results demonstrate a significant improvement in MLUPS with an increasing number of cores, showcasing the benefits of parallelization. The performance scales almost linearly up to 32 cores. Beyond 32 cores, the performance gains start to diminish, suggesting that additional cores may not lead to proportional speedup due to potential bottlenecks or limitations in the hardware architecture.

For 64 cores, the CooLBM performance achieves around 88.36\% of the ideal linear increment, while with 128 cores, it reaches approximately to 46.95\%. These results highlight the strong parallel processing capabilities of hardware, demonstrating efficient scaling with increasing core count and enhancing MLUPS performance in lattice-based simulations. 

However, it is important to note that the ideal linear increment in MLUPS is represented by the blue dashed line, serving as a reference for the expected performance of hardware with perfect parallel scaling. While the CooLBM results approach the ideal linear increment up to moderate numbers of cores, they deviate slightly as the number of cores increases. This deviation can be attributed to various factors, including communication overhead, contention for shared resources, and load imbalances in the parallel execution.

%
%\textcolor{blue}{XXX to show if it is coming from communication, we need to do another sets of simulations with larger number of lattices, for instance 100*100*100 to put on this figure XXX}

\subsection{MLUPS Performance Comparison Across Processors and Simulation Cases}
Figure \ref{fig:paraall} presents a comparison of MLUPS performance for different processors and three simulation cases: three-dimensional Couette flow with 1,000,000 lattices, three-dimensional Taylor-Green flow with 2,097,152 lattices, and two-dimensional reactive interface with 33,600 lattices. \textcolor{black}{The higher number of lattice nodes in the three-dimensional study compared to the two-dimensional reactive study is due to the inclusion of the third dimension, which is absent in the two-dimensional reactive study}. The bar graph displays the achieved MLUPS values for each processor and simulation case, providing insights into the efficiency of hardware in lattice-based simulations.

%XXX justify the reason why the 2D reactive flow has much less latices!! XXX

\begin{figure}[htbp]
\begin{center}
\includegraphics[width=0.7\textwidth]{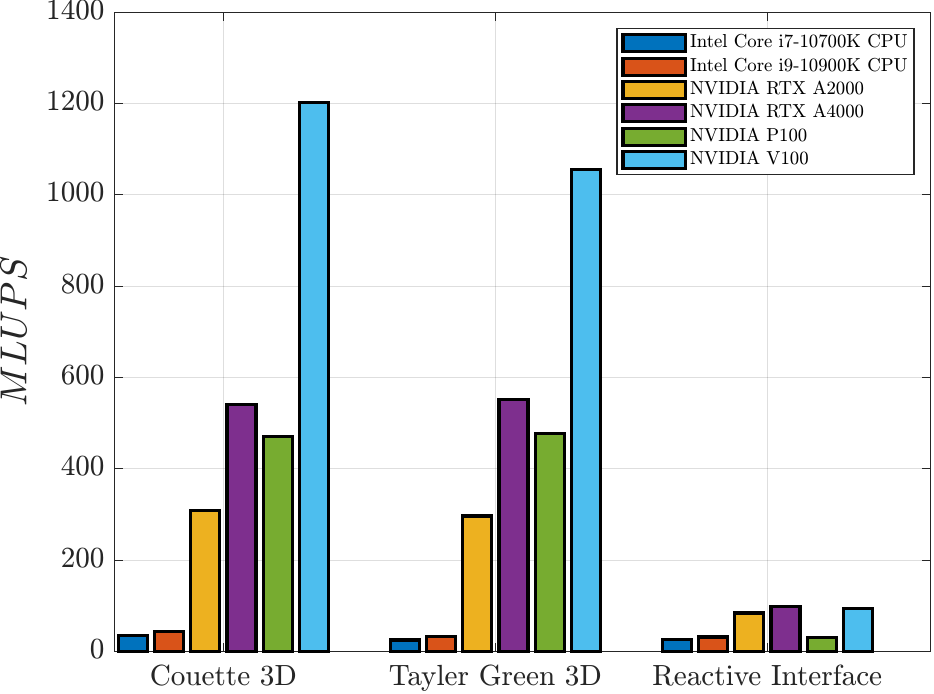}
\caption{\label{fig:paraall} {\small Comparison of MLUPS performance for different processors and three simulation cases.}}
%\textcolor{blue}{XXX figure legend can be on right side instead of top sideXXX}}
\end{center}
\end{figure}

Among the processors tested, the NVIDIA V100 GPU consistently outperforms others across all simulation cases. In the three-dimensional Couette flow test case with one million lattices, the NVIDIA V100 GPU achieves an impressive MLUPS of 1202, showcasing its powerful parallel processing capabilities and suitability for computationally demanding simulations. Similarly, in the three-dimensional Taylor-Green test case, the V100 GPU excels, achieving an MLUPS of 1056, further reinforcing its superiority for lattice simulations.

While the NVIDIA V100 demonstrates superior performance, the NVIDIA A4000 and RTX A2000 GPUs also deliver impressive MLUPS results. The A4000 achieves approximately 541 and 552 MLUPS, respectively, in the Couette flow and the Tayler Green simulations. The RTX A2000 and A4000 in the reactive interface case perform well, highlighting the substantial improvements achieved in MLUPS performance with the latest NVIDIA GPUs.

In contrast, the two Intel CPUs, i9-10900K and i7-10700K, demonstrate lower MLUPS values compared to the NVIDIA GPUs. However, it is essential to note that CPUs are general-purpose processors, while GPUs are specialized for parallel processing tasks. The i9-10900K CPU achieves around 44 MLUPS in the Couette flow problem and 33 MLUPS in the Tayler Green test case, while the i7-10700K CPU reaches, respectively, 34 and 25 MLUPS for the same test cases.

Regarding the reactive interface case, the lower MLUPS value can be attributed to the nature of the simulation. The reactive Interface involves iterative methods in the inner loop to handle reactions at the interface, which can increase the computational complexity. Additionally, this simulation case requires a higher number of populations, leading to a larger workload compared to other cases. Consequently, this test case presents unique challenges for parallelization, resulting in relatively lower MLUPS values compared to other simulation cases.

%\textcolor{blue}{XXX we need to add one more test case here. For instance RTI with 2 populations so that it is in between the 3D case with one population and 2D reactive case with 5 popilation!! XXX In this way, we also have at least 1 test case from each subsection of the results. I mean: 1- single phase, 2- 2phase, and 3- reactive interface.XXX}

Overall, the MLUPS performance comparison across different processors and simulation cases provides valuable insights into the capabilities of hardware for lattice-based simulations. The NVIDIA V100 GPU stands out as the most efficient processor, while the latest NVIDIA A4000 and RTX A2000 GPUs also deliver impressive performance. These findings offer researchers and engineers valuable guidance in selecting the most suitable hardware for their specific simulation requirements, ultimately leading to optimized and efficient lattice-based simulations.

%%%%%%%%%%%%%%%%%%%%%%%%%%%%%%%%%%%%%%%%%%%%%%%%%%%%%%%%%%%%%%%%%%%%%%%%%%%%%%%
%%%%%%%%%%%%%%%%%%%%%%%%%%%% Conclusions  %%%%%%%%%%%%%%%%%%%%%%%%%%%%%%%%%%%%%
%%%%%%%%%%%%%%%%%%%%%%%%%%%%%%%%%%%%%%%%%%%%%%%%%%%%%%%%%%%%%%%%%%%%%%%%%%%%%%%

\section{Conclusions }\label{sec:Conclusions}
In this work, we presented CooLBM, a novel computational framework developed for the simulation of single and multi-component multi-phase problems, along with a reactive interface simulation and conjugate fluid-solid heat transfer using the lattice Boltzmann method. CooLBM incorporates effective numerical methods and algorithms to accurately capture complex fluid dynamics and phase interactions. It is implemented in C++ and utilizes the Standard Template Library (STL) for improved code modularity, flexibility, and re-usability.

The performance and accuracy of CooLBM were demonstrated through various benchmark simulations, illustrating its ability to capture intricate fluid behaviors and interface dynamics. The code was tested on different hardware configurations, including CPUs and GPUs, to assess its efficiency and scalability. The parallelization capabilities of CooLBM using both CPUs and GPUs allowed for efficient and parallelized simulations, making it suitable for computationally demanding problems.

One of the significant contributions of CooLBM is its reactive interface simulation module, enabling the study of chemical reactions occurring at the fluid-\textcolor{black}{solid} 
%(XXX fluid-solid?XX)
interface. This expands the applicability of the framework to reactive multi-phase systems, which are of great importance in various scientific and engineering fields.

The modular structure of CooLBM allows for easy customization and extension, facilitating the implementation of additional components, models and boundary conditions. Visualization tools are also provided for the analysis and interpretation of simulation results.

Through extensive validation against analytical solutions and existing numerical models, CooLBM proved to be a reliable and accurate tool for simulating single and multi-component multi-phase problems, including reactive interfaces. The code demonstrated significant performance gains on GPU architectures, especially on the NVIDIA V100 GPU consistently outperforming other processors across all simulation cases.

Overall, CooLBM offers a robust and efficient computational framework for studying complex multi-phase systems and reactive interfaces, making it a valuable tool for researchers and engineers in fields such as chemical engineering, materials science, and environmental engineering. With its ability to handle large-scale simulations and its parallelization capabilities, CooLBM has the potential to accelerate scientific discovery and engineering advancements in a wide range of applications. Future work includes expanding the capabilities of code and optimizing its performance for even more demanding simulations, further enhancing its usability and impact in various scientific and engineering domains.

%%%%%%%%%%%%%%%%%%%%%%%%%%%%%%%%%%%%%%%%%%%%%%%%%%%%%%%%%%%%%%%%%%%%%%%%%%%%%%%
%%%%%%%%%%%%%%%%%%%%%%%%%%% Acknowledgment %%%%%%%%%%%%%%%%%%%%%%%%%%%%%%%%%%%%
%%%%%%%%%%%%%%%%%%%%%%%%%%%%%%%%%%%%%%%%%%%%%%%%%%%%%%%%%%%%%%%%%%%%%%%%%%%%%%%

\section*{Acknowledgment}\label{sec:Acknowledgment}
The authors would like to express their gratitude for the financial support provided by ANR under the grant number of ANR-20-CE92-0007, and the “CARNOT ESP” program, through the project “HYBRID”. The authors wish to acknowledge the French regional computing center of Normandy CRIANN (2022006) for providing access to the HPC resources.

%%%%%%%%%%%%%%%%%%%%%%%%%%%%%%%%%%%%%%%%%%%%%%%%%%%%%%%%%%%%%%%%%%%%%%%%%%%%%%%
%%%%%%%%%%%%%%%%%%%%%%%%%%%%%%%% Appendices %%%%%%%%%%%%%%%%%%%%%%%%%%%%%%%%%%%
%%%%%%%%%%%%%%%%%%%%%%%%%%%%%%%%%%%%%%%%%%%%%%%%%%%%%%%%%%%%%%%%%%%%%%%%%%%%%%%

%\appendix
%\section{Non-Dimensional System and Units Conversion}\label{unitconversion}
%In lattice Boltzmann simulations, the unit system differs from the physical unit system, whether it is a metric or imperial unit system. The lattice unit system is a system where the spatial and temporal discretization are 


\begin{thebibliography}{00}

\bibliographystyle{model1-num-names}
\bibitem{Guo2016}
Guo K, Li H,  Yu Z. In-situ heavy and extra-heavy oil recovery: A review. \textit{Fuel}. 2016; 185, 886-902. \href{https://doi.org/10.1016/j.fuel.2016.08.047}{DOI: 10.1016/j.fuel.2016.08.047}

\bibitem{Zhao2023}
Zhao X, Jiang J, Zuo H, Jia G. Soot combustion characteristics of oxygen concentration and regeneration temperature effect on continuous pulsation regeneration in diesel particulate filter for heavy-duty truck. \textit{Energy}. 2023; 264, 126265. \href{https://doi.org/10.1016/j.energy.2022.126265}{DOI: 10.1016/j.energy.2022.126265}

\bibitem{Chong2016}
Chong ZR, Yang SHB, Babu P, Linga P,  Li XS. Review of natural gas hydrates as an energy resource: Prospects and challenges. \textit{Applied energy}. 2016; 162, 1633-1652. \href{https://doi.org/10.1016/j.apenergy.2014.12.061}{DOI: 10.1016/j.apenergy.2014.12.061}

\bibitem{Pintelon2012}
Pintelon TR, Picioreanu C, van Loosdrecht MC,  Johns ML. The effect of biofilm permeability on bio‐clogging of porous media. \textit{Biotechnology and bioengineering}. 2012; 109(4), 1031-1042. \href{ https://doi.org/10.1002/bit.24381}{DOI: 10.1002/bit.24381}

\bibitem{Prieto2003}
Prieto M, Cubillas P,  Fernández-Gonzalez Á. Uptake of dissolved Cd by biogenic and abiogenic aragonite: a comparison with sorption onto calcite. \textit{Geochimica et Cosmochimica Acta}. 2003; 67(20), 3859-3869. \href{https://doi.org/10.1016/S0016-7037(03)00309-0}{DOI: 10.1016/S0016-7037(03)00309-0}

\bibitem{Stockmann2011}
Stockmann GJ, Wolff-Boenisch D, Gislason SR, Oelkers EH. Do carbonate precipitates affect dissolution kinetics? 1: Basaltic glass. \textit{Chemical Geology}. 2011; 284(3-4), 306-316. \href{https://doi.org/10.1016/j.chemgeo.2011.03.010}{10.1016/j.chemgeo.2011.03.010}

\bibitem{Daval2009}
Daval D, Martinez I, Corvisier J, Findling N, Goffé B, Guyot F. Carbonation of Ca-bearing silicates, the case of wollastonite: Experimental investigations and kinetic modeling. \textit{Chemical Geology}. 2009; 265(1-2), 63-78. \href{https://doi.org/10.1016/j.chemgeo.2009.01.022}{DOI: 10.1016/j.chemgeo.2009.01.022}

\bibitem{Gu2018}
Gu Q, Liu H, Zhang Y. Lattice Boltzmann simulation of immiscible two-phase displacement in two-dimensional Berea sandstone. \textit{Appl. Sci}. 2018; 8, 1497. \href{https://doi.org/10.3390/app8091497}{DOI: 10.3390/app8091497}

\bibitem{Liu2011}
Liu H, Zhang Y. Lattice Boltzmann simulation of droplet generation in a microfluidic cross-junction. \textit{Communications in Computational Physics}. 2011 May; 9(5):1235-5. \href{https://doi.org/10.4208/cicp.231009.101110s}{DOI}

\bibitem{Lintermann2020}
Lintermann A, Schröder W. Lattice–Boltzmann simulations for complex geometries on high-performance computers. \textit{CEAS Aeronautical Journal}. 2020; 11:745-66. \href{https://doi.org/10.1007/s13272-020-00450-1}{DOI: 10.1007/s13272-020-00450-1}

\bibitem{Simonis2023}
Simonis S, Nguyen J, Avis SJ, Dörfler W, Krause MJ. Binary fluid flow simulations with free energy lattice Boltzmann methods. \textit{Discrete and Continuous Dynamical Systems - S}. 2023; \href{https://doi.org/10.3934/dcdss.2023069}{DOI: 10.3934/dcdss.2023069}

\bibitem{Sukop2004}
Sukop MC, Or D. Lattice Boltzmann method for modeling liquid‐vapor interface configurations in porous media. \textit{Water Resources Research}. 2004; 40(1):1-11. \href{https://doi.org/10.1029/2003WR002333}{DOI: 10.1029/2003WR002333}

\bibitem{Verhaeghe2006}
Verhaeghe F, Arnout S, Blanpain B, Wollants P. Lattice-Boltzmann modeling of dissolution phenomena. \textit{Physical Review E}. 2006; 73(3):036316. \href{https://doi.org/10.1103/PhysRevE.73.036316}{DOI: 10.1103/PhysRevE.73.036316}

\bibitem{He2000}
He X, Li N. Lattice Boltzmann simulation of electrochemical systems. \textit{Computer Physics Communications}. 2000; 129(1-3):158-66. \href{https://doi.org/10.1016/S0010-4655(00)00103-X}{DOI: 10.1016/S0010-4655(00)00103-X}

\bibitem{Li2021}
Li Q, Niu X, Lu Z, Li Y, Khan A, Yu Z. An improved single-relaxation-time multiphase lattice Boltzmann model for multiphase flows with large density ratios and high Reynolds numbers. \textit{Adv. Appl. Math. Mech.} 2021; 13:426-54. \href{https://doi.org/10.4208/aamm.OA-2019-0232}{DOI: 10.4208/aamm.OA-2019-0232}

\bibitem{Leclaire2017}
Leclaire S, Parmigiani A, Malaspinas O, Chopard B, Latt J. Generalized three-dimensional lattice Boltzmann color-gradient method for immiscible two-phase pore-scale imbibition and drainage in porous media. \textit{Physical Review E}. 2017; 95(3):033306. \href{https://doi.org/10.1103/PhysRevE.95.033306}{DOI: 10.1103/PhysRevE.95.033306}

\bibitem{Su2017}
Su Y, Ng T, Davidson JH. A parallel non-dimensional lattice Boltzmann method for fluid flow and heat transfer with solid–liquid phase change. \textit{International Journal of Heat and Mass Transfer}. 2017; 106:503-17. \href{https://doi.org/10.1016/j.ijheatmasstransfer.2016.08.109}{DOI: 10.1016/j.ijheatmasstransfer.2016.08.109}

\bibitem{Pan2004}
Pan C, Prins JF, Miller CT. A high-performance lattice Boltzmann implementation to model flow in porous media. \textit{Computer Physics Communications}. 2004; 158(2):89-105. \href{https://doi.org/10.1016/j.cpc.2003.12.003}{DOI: 10.1016/j.cpc.2003.12.003}

\bibitem{Chen2016}
Chen C, Wang Z, Majeti D, Vrvilo N, Warburton T, Sarkar V, Li G. Optimization of lattice Boltzmann simulation with graphics-processing-unit parallel computing and the application in reservoir characterization. \textit{SPE Journal}. 2016; 21(04):1425-35. \href{https://doi.org/10.2118/179733-PA}{DOI: 10.2118/179733-PA}

\bibitem{Krafczyk2014}
Krafczyk M, Kucher K, Wang Y, Geier M. DNS/LES studies of turbulent flows based on the cumulant lattice Boltzmann approach. \textit{In High Performance Computing in Science and Engineering ‘14: Transactions of the High Performance Computing Center}, Stuttgart (HLRS) 2014 (pp. 519-531). Springer International Publishing. \href{https://doi.org/10.1007/978-3-319-10810-0_34}{DOI: 10.1007/978-3-319-10810-0\_34}

\bibitem{Latt2021}
Latt J, Coreixas C, Beny J. Cross-platform programming model for many-core lattice Boltzmann simulations. \textit{Plos one}. 2021; 16(4):e0250306. \href{https://doi.org/10.1371/journal.pone.0250306}{DOI: 10.1371/journal.pone.0250306}

\bibitem{Mohamad2019}
Mohamad AA. Lattice Boltzmann Method: Fundamentals and Engineering Applications with Computer Codes. 2nd ed. \textit{Springer London}; 2019. \href{https://doi.org/10.1007/978-1-4471-7423-3}{DOI: 10.1007/978-1-4471-7423-3}

\bibitem{Kruger2017}
Krüger T, Kusumaatmaja H, Kuzmin A, Shardt O, Silva G, Viggen EM. The Lattice Boltzmann Method: Principles and Practice. 1st ed. \textit{Springer Cham}. 2017. \href{https://doi.org/10.1007/978-3-319-44649-3}{DOI: 10.1007/978-3-319-44649-3}

\bibitem{Shan_Chen_Model}
Shan X,  Chen H. Lattice Boltzmann model for simulating flows with multiple phases and components. \textit{Physical review E}. 1993; 47(3), 1815. \href{https://doi.org/10.1103/PhysRevE.47.1815}{DOI: 10.1103/PhysRevE.47.1815}
    
\bibitem{Swift1996}
Swift MR, Orlandini E, Osborn WR,  Yeomans JM. Lattice Boltzmann simulations of liquid-gas and binary fluid systems. \textit{Physical Review E}. 1996; 54(5), 5041. \href{https://doi.org/10.1103/PhysRevE.54.5041}{DOI: 10.1103/PhysRevE.54.5041}

\bibitem{Gunstensen_1991}
Gunstensen AK, Rothman DH, Zaleski S, Zanetti G.  Lattice Boltzmann model of immiscible fluids. \textit{Physical review A}. 1991; 43(8), 4320.
\href{https://doi.org/10.1103/PhysRevA.43.4320}{DOI: 10.1103/PhysRevA.43.4320}

\bibitem{Zheng2006}
Zheng, HW, Shu, C,  Chew, YT. A lattice Boltzmann model for multiphase flows with large density ratio. \textit{Journal of computational physics}. 2006; 218(1), 353-371.
\href{https://doi.org/10.1016/j.jcp.2006.02.015}{DOI: 10.1016/j.jcp.2006.02.015}


\bibitem{Huang2015}
Huang H, Sukop M, Lu X. Multiphase lattice Boltzmann methods: Theory and application. 1st ed. Wiley-Blackwell; 2015. ISBN: 978-1-118-97134-5.

\bibitem{Huang2007}
Huang H, Thorne Jr DT, Schaap MG, Sukop MC. Proposed approximation for contact angles in Shan-and-Chen-type multicomponent multiphase lattice {B}oltzmann models. \textit{Physical Review E}. 2007 Dec 3; 76(6):066701. \href{https://doi.org/10.1103/PhysRevE.76.066701}{DOI: 10.1103/PhysRevE.76.066701}

\bibitem{Guo_forcing}
  Guo Z, Zheng C,  Shi B. Discrete lattice effects on the forcing term in the lattice Boltzmann method. \textit{Physical review E}. 2002; 65(4), 046308. \href{https://doi.org/10.1103/PhysRevE.65.046308}{DOI: 10.1103/PhysRevE.65.046308}

\bibitem[Alamian et al.(2024)]{Alamian2024}
Alamian R, Sawaf M, Stockinger C, Hadjadj A, Latt J, Shadloo MS. Modeling soot filter regeneration process through surface-reactive flow in porous media using iterative lattice Boltzmann method. \textit{Energy}. 2024; 289, 129980. \href{https://www.sciencedirect.com/science/article/pii/S0360544223033741?casa_token=x4J-4uSmFQIAAAAA:ce79eCxlHsbaO_OljjtINsYTV1K801Gb1RWtBXGUVJvCZ0Jso_-JWkFG6mMCRUdw3fQQLhIE4w}{DOI: 10.1016/j.energy.2023.129980}

\bibitem[Kang et al.(2010)]{Kang2010}
Kang Q, Lichtner PC, Viswanathan HS, Abdel-Fattah AI. Pore scale modeling of reactive transport involved in geologic CO$_2$ sequestration. \textit{Transport in porous media}. 2010; 82:197-213. \href{https://doi.org/10.1007/s11242-009-9443-9}{DOI: 10.1007/s11242-009-9443-9}

\bibitem[He et al.(2010)]{He1998}
He X, Chen S, Doolen GD. A novel thermal model for the lattice Boltzmann method in incompressible limit. \textit{Journal of computational physics}. 1998; 146:282-300. \href{https://doi.org/10.1006/jcph.1998.6057}{DOI: 10.1006/jcph.1998.6057}

\bibitem[Xu et al.(2018)]{Xu2018}
Xu Q, Long W, Jiang H, Zan C, Huang J, Chen X, et al. Pore-scale modelling of the coupled thermal and reactive flow at the combustion front during crude oil in-situ combustion. \textit{Chemical Engineering Journal}. 2018; 350:776-90. \href{https://doi.org/10.1016/j.cej.2018.04.114}{DOI: 10.1016/j.cej.2018.04.114}

\bibitem[Obrecht et al. (2013)]{Obrecht2013}
Obrecht C, Kuznik F, Tourancheau B, Roux JJ. Multi-GPU implementation of the lattice Boltzmann method. \textit{Computers \& Mathematics with Applications}. 2013; 65(2):252-61. \href{https://doi.org/10.1016/j.camwa.2011.02.020}{DOI: 10.1016/j.camwa.2011.02.020}

\bibitem[Xian and Takayuki (2011)]{Xian2011}
Xian W, Takayuki A. Multi-GPU performance of incompressible flow computation by lattice Boltzmann method on GPU cluster. \textit{Parallel Computing}. 2011;37(9):521-35. \href{https://doi.org/10.1016/j.parco.2011.02.007}{DOI: 10.1016/j.parco.2011.02.007}

\bibitem{Kreith1999}
Kreith F, editor. Fluid mechanics. CRC press; 1999. ISBN: 0-8493-0055-X.

\bibitem{stockinger2024lattice}
  Stockinger C, Raiolo A, Alamian R, Hadjadj A, Nieken U, 
  Shadloo MS. Lattice {B}oltzmann simulations of heterogeneous combustion reactions
  for application in porous media. \textit{Engineering Analysis with Boundary Elements}. 2024; 166:105817. \href{https://doi.org/10.1016/j.enganabound.2024.105817}{DOI: 10.1016/j.enganabound.2024.105817}

\bibitem{Abdelsamie2021}
Abdelsamie A, Lartigue G, Frouzakis CE, Thevenin D. The Taylor–Green vortex as a benchmark for high-fidelity combustion simulations using low-Mach solvers. \textit{Computers \& Fluids}. 2021; 223:104935. \href{https://doi.org/10.1016/j.compfluid.2021.104935}{DOI: 10.1016/j.compfluid.2021.104935}

\bibitem{Basit2010}
Basit R,  Basit MA. Simulation of phase separation process using lattice Boltzmann method. \textit{Canadian J. on Computing in Mathematics, Natural Sciences, Engineering \& Medicine}. 2010; 1, 71-76. 

\bibitem{Slinn1980}
Slinn, SA, Slinn, WGN. Predictions for particle deposition on natural waters. \textit{Atmospheric Environment}. 1980; (1967), 14(9), 1013-1016. \href{https://doi.org/10.1016/0004-6981(80)90032-3}{10.1016/0004-6981(80)90032-3}

\bibitem{Elimelech2013}
Elimelech M, Gregory J,  Jia X. Particle deposition and aggregation: measurement, modelling and simulation. \textit{Butterworth-Heinemann}. 2013.

\bibitem{Cavallo2023}
Cavallo DM, Chiavola O, Palmieri F, Mancaruso E,  Vaglieco BM. Experimental study on the effect of loading and regeneration for an optimized management of the DPF. \textit{Results in Engineering}. 2023; 18, 101048. \href{https://doi.org/10.1016/j.rineng.2023.101048}{10.1016/j.rineng.2023.101048}

\bibitem{yamamoto2009simulation}
Yamamoto K, Oohori S, Yamashita H, Daido S. Simulation on soot deposition and combustion in diesel particulate filter. \textit{Proceedings of the Combustion Institute}. 2009; 32(2):1965--1972. \href{https://doi.org/10.1016/j.proci.2008.06.081}{10.1016/j.proci.2008.06.081}

  



\end{thebibliography}

\appendix
\section{A Simple C++ Code Showing Implementation of \enquote{for\_each}}
\label{appendixA}
\begin{lstlisting}[frame=none]
// This is the STL way of implementing "for_each"
// Adding 1 to each element of the vector "vec"
#include <iostream>  //C++ input/output streams
#include <vector>    //Resizable array
#include <algorithm> //For std::for_each
#include <execution> // For parallel execution
using namespace std;

int main() {
    vector<int> vec = {1, 2, 3, 4, 5};
    //Apply parallel execution to each element using "Lambda function" on CPU
    for_each(execution::par, vec.begin(), vec.end(), [](int& m) {
        m += 1; // Add  1 to each element of vector
    });

    // Print the vector "vec"
    for_each(vec.begin(), vec.end(), [](int m) {
        cout << m << "\t";
    });

    return 0;
}
Output: 
2 3 4 5 6
\end{lstlisting}

\begin{lstlisting}[frame=none]
// This is the program without using "for_each"
// Adding 1 to each element of a vector "vec"
#include <iostream>      // C++ input/output streams
#include <vector>        // Resizable array

using namespace std;

int main() {
    std::vector<int> vec = {1, 2, 3, 4, 5};

    // Add 1 to each element of the vector "vec"
    for (int i = 0; i < vec.size(); ++i) {
        vec[i] += 1; // Add 1 to each element of the vector
    }

    // Print the vector "vec"
    for (int i = 0; i < vec.size(); ++i) {
        std::cout << vec[i] << "\t";  
    }

    return 0;
}

Output: 
2 3 4 5 6
\end{lstlisting}

\end{psfrags}
\end{document}